\documentclass[aps,pra,twocolumn,showpacs,amsmath,amssymb,superscriptaddress]{revtex4-1}

\usepackage{graphicx, amsmath, bm, braket, txfonts, color, ulem}

\usepackage[colorlinks]{hyperref}
\hypersetup{
colorlinks,
citecolor=blue,
filecolor=blue,
linkcolor=blue,
urlcolor=blue
}

\begin{document}

\title{Direct Rashba spin-orbit interaction in Si and Ge nanowires with different growth directions}

\author{Christoph Kloeffel}
\affiliation{Department of Physics, University of Basel, Klingelbergstrasse 82, CH-4056 Basel, Switzerland}

\author{Marko J. Ran\v{c}i\'{c}}
\affiliation{Department of Physics, University of Basel, Klingelbergstrasse 82, CH-4056 Basel, Switzerland}

\author{Daniel Loss}
\affiliation{Department of Physics, University of Basel, Klingelbergstrasse 82, CH-4056 Basel, Switzerland}
\affiliation{CEMS, RIKEN, Wako, Saitama 351-0198, Japan}

\date{\today}

\begin{abstract}
We study theoretically the low-energy hole states in Si, Ge, and Ge/Si core/shell nanowires (NWs). The NW core in our model has a rectangular cross section, the results for a square cross section are presented in detail. In the case of Ge and Ge/Si core/shell NWs, we obtain very good agreement with previous theoretical results for cylindrically symmetric NWs. In particular, the NWs allow for an unusually strong and electrically controllable spin-orbit interaction (SOI) of Rashba type. We find that the dominant contribution to the SOI is the ``direct Rashba spin-orbit interaction'' (DRSOI), which is an important mechanism for systems with heavy-hole-light-hole mixing. Our results for Si NWs depend significantly on the orientation of the crystallographic axes. The numerically observed dependence on the growth direction is consistent with analytical results from a simple model, and we identify a setup where the DRSOI enables spin-orbit energies of the order of millielectronvolts in Si NWs. Furthermore, we analyze the dependence of the SOI on the electric field and the cross section of the Ge or Si core. A helical gap in the spectrum can be opened with a magnetic field. We obtain the largest $g$~factors with magnetic fields applied perpendicularly to the NWs. 
\end{abstract}


\maketitle

\section{Introduction}
\label{sec:Introduction}

In recent years, there have been several novel trends toward spin-based quantum information processing with quantum dots (QDs) \cite{loss:pra98, kloeffel:annurev13}. One of these trends is the shift to \mbox{group-IV} materials such as Ge and Si \cite{maune:nat12, zwanenburg:rmp13, wu:pnas14, kawakami:nna14, veldhorst:nna14, veldhorst:nat15, reed:prl16, takeda:sciadv16, zajac:arX1708, mi:arX1710}. Both Ge and Si can be grown nuclear-spin-free, which is beneficial for implementing qubits with long dephasing times \cite{khaetskii:prl02, merkulov:prb02, coish:prb04, veldhorst:nna14, veldhorst:nat15, tyryshkin:nmat12}. Besides group-IV materials, qubits based on hole states, i.e., unfilled valence band states, have attracted a lot of attention \cite{bulaev:prl05, gundogdu:apl05, bulaev:prl07, heiss:prb07, fischer:prb08, trif:prl09, brunner:sci09, katsaros:nna10, katsaros:prl11, degreve:nphys11, greilich:nphot11, pingenot:prb11, godden:prl12, ares:prl13, warburton:nmat13, li:nlt15, vanbree:prb16, wang:nlt16}. Furthermore, various nanostructures such as nanowires (NWs) \cite{fasth:prl07, trif:prb08, nadjperge:nat10, schroer:prl11, petersson:nat12, vandenberg:prl13, pribiag:nna13} or hut wires \cite{zhang:prl12, watzinger:nlt16} are studied with great efforts. Semiconducting quantum wires are also promising platforms for, e.g., spin filters \cite{streda:prl03} and topological quantum computing with Majorana fermions \cite{alicea:rpp12, mourik:sci12, maier:prb14b, albrecht:nat16, lutchyn:arX1707}.

Hole states in Si- and Ge-based NWs, such as Ge/Si core/shell NWs \cite{lauhon:nat02, lu:pnas05, xiang:nat06, xiang:nna06, hu:nna07, roddaro:prl08, varahramyan:apl09, jiang:edl09, hao:nlt10, yan:nat11, nah:nlt12, hu:nna12, higginbotham:nlt14, higginbotham:prl14, brauns:prb16, conesaboj:nlt17}, comprise the mentioned trends and are interesting for many reasons. Besides the possibility to cancel the hyperfine interaction with nuclear spins by isotopic purification, there is no valley degree of freedom in the topmost valence band of bulk Ge and bulk Si. Furthermore, the spin-orbit interaction (SOI) of Dresselhaus type is absent because Ge and Si are bulk inversion symmetric \cite{winkler:book}. However, structure inversion asymmetry remains a source of SOI and can be controlled externally. In fact, an unusually strong ``direct Rashba spin-orbit interaction'' (DRSOI) \cite{kloeffel:prb11} has been predicted for the holes in Ge/Si core/shell NWs, which is an electric-field-induced mechanism that is not suppressed by the fundamental band gap of the semiconductor (in stark contrast to the standard Rashba SOI for electrons and holes, which is obtained in the third order of a multiband perturbation theory \cite{winkler:book}).  Thus far, the discovered DRSOI is consistent with experiments \cite{hao:nlt10, higginbotham:prl14, brauns:prb16}. The absence of Dresselhaus SOI and the presence of the DRSOI enable high electrical control, because the SOI can be switched ``on'' and ``off'' with moderate electric fields of only a few volts per micrometer \cite{kloeffel:prb11, kloeffel:prb13}. Control over the SOI is desirable because, on the one hand, SOI can be an unwanted source of relaxation and decoherence for spin qubits \cite{khaetskii:prb01, golovach:prl04, stano:prl06, hanson:rmp07, maier:prb13}, but on the other hand, just to name a few examples, SOI enables one-qubit operations via electric-dipole-induced spin resonance (EDSR) \cite{golovach:prb06, nowack:sci07, trif:prb08, nadjperge:nat10, schroer:prl11, petersson:nat12, vandenberg:prl13, pribiag:nna13, kloeffel:prb13, corna:arX1708}, long-distance two-qubit operations via superconducting resonators \cite{blais:pra04, wallraff:nat04, trif:prb08, petersson:nat12, kloeffel:prb13, nigg:prl17} or floating gates \cite{trifunovic:prx12, serina:prb17}, and the realization of the previously mentioned spin filters \cite{streda:prl03} and Majorana fermions \cite{alicea:rpp12, mourik:sci12, maier:prb14b, albrecht:nat16, lutchyn:arX1707}. 

Very recently, qubits have successfully been implemented with holes in Si NWs, using an industrial-level complementary metal-oxide semiconductor (CMOS) platform for the sample fabrication \cite{voisin:nlt16, maurand:ncomm16, crippa:arX1710}. Such Si NWs are now routinely fabricated \cite{coquand:ulisproc12, barraud:edl12, prati:nanotech12} and it has been demonstrated that even Ge/Si core/shell NWs with a compressively strained Ge core can be realized with a CMOS-compatible process \cite{jiang:edl09}. Depending on the details of the fabrication process, the cross sections of the NWs can, e.g., be approximately circular or rectangular \cite{coquand:ulisproc12, barraud:edl12, prati:nanotech12}. For instance, the Si NW in the setup of Ref.~\cite{voisin:nlt16} has an almost square cross section, with a side length of approximately $10\mbox{ nm}$. 

In this paper, we consider these recently fabricated nanostructures and study theoretically the spectrum and the SOI of holes in NWs with rectangular cross sections. For this, we use the Luttinger-Kohn (LK) Hamiltonian \cite{luttinger:pr55, luttinger:pr56}, rectangular hard-wall confinement, and a numerical approach to find the low-energy eigenstates. Since the Luttinger parameters $\gamma_2$ and $\gamma_3$ differ greatly in Si \cite{lawaetz:prb71}, a spherical approximation \cite{lipari:prl70, winkler:sst08} does not apply and it turns out that the results depend strongly on the orientation of the crystallographic axes. Remarkably, we find that the DRSOI in Si NWs allows for spin-orbit energies of several millielectronvolts, controllable with electric fields, provided that the growth direction is changed compared to recent setups \cite{coquand:ulisproc12, voisin:nlt16, maurand:ncomm16, crippa:arX1710}. We study the dependences on the NW dimensions, on the orientation of the crystallographic axes, and on the applied fields in detail and identify setups which are highly promising for applications that rely on a strong and electrically controllable SOI. Our numerical results for Si NWs are consistent with a simple analytical model that explains the significant dependence on the growth direction. Furthermore, we use the numerical approach to study Ge and Ge/Si core/shell NWs and find very good agreement with the effective model of Ref.~\cite{kloeffel:prb11}. For NWs with a thin Ge core, we show that a spin-orbit energy above 10~meV is feasible due to the DRSOI.     

The paper is organized as follows. In Sec.~\ref{sec:DRSOI}, we introduce and discuss the DRSOI, followed by the explanation of our model and the numerical approach in Sec.~\ref{sec:Model}. In Sec.~\ref{sec:Ge}, we consider hole states in Ge/Si core/shell NWs with square cross sections. We show that our model reproduces previous theoretical results for cylindrical Ge and Ge/Si core/shell NWs and study the dependence of the SOI on various parameters, such as the strength of the electric field. Our numerical and analytical results for Si NWs are analyzed in Secs.~\ref{sec:Si} and~\ref{sec:AnalyticalResultsSiNWs}, respectively, followed by a discussion about the accuracy in Sec.~\ref{sec:Accuracy} and concluding remarks in Sec.~\ref{sec:Conclusion}. Details of the theory are appended.

\section{Direct Rashba Spin-Orbit Interaction}
\label{sec:DRSOI}

We start by analyzing in detail the DRSOI~\cite{kloeffel:prb11} and explain why it is a pronounced feature of systems with strong heavy-hole-light-hole (HH-LH) mixing. For this, it is instructive to compare the standard regime of a two-dimensional (2D) system with that of NW quantum confinement considered in the present work. In the following, we assume that magnetic fields are absent (unless otherwise stated) and discuss the effect of an applied electric field. To set the notation for later use, we need to first review briefly the standard theory of electrons and holes in bulk and low-dimensional systems.

\subsection{Electrons}
\label{secsub:Electrons}

Many semiconductors with a zinc-blende lattice, such as GaAs and InAs, have a conduction band minimum at the $\Gamma$ point. Their leading-order Hamiltonian for bulk electrons near the $\Gamma$ point is 
\begin{equation}
H^{\rm el}_0 = \frac{\hbar^2}{2 m_{\rm eff}} \left( k_x^2 +  k_y^2 +  k_z^2 \right) ,
\label{eq:ElectronsH0}
\end{equation}  
where $\hbar k_j$ is the momentum operator for the $j$ axis and $m_{\rm eff}$ is the effective electron mass \cite{winkler:book}. An applied electric field $\bm{E}$ results in a force $-e \bm{E}$ on the electron, with 
\begin{equation}
H^{\rm el}_{\rm dir} = e \bm{E} \cdot \bm{r}
\end{equation}
as the associated Hamiltonian. The elementary positive charge is denoted by $e$ and $\bm{r}$ is the operator for the position of the electron. We note that $H^{\rm el}_0 + H^{\rm el}_{\rm dir}$ is independent of the spin, and so the direct coupling $H^{\rm el}_{\rm dir}$ to the electric field cannot lift the spin degeneracy.

However, in the presence of $\bm{E}$ there are higher-order corrections that provide a coupling to the spin. The most prominent mechanism is the standard Rashba SOI \cite{bychkov:jpcssp84, winkler:book}
\begin{equation}
H^{\rm el}_R = \alpha_{\rm el} \bm{E} \cdot \left( \bm{\sigma} \times \bm{k} \right) ,
\label{eq:RSOIElectronsGeneral}
\end{equation} 
where $\hbar \bm{k} = \hbar \left( k_x \bm{e}_x + k_y \bm{e}_y + k_z \bm{e}_z  \right)$ is the vector operator for the momentum, $\bm{\sigma} = \sigma_x \bm{e}_x + \sigma_y \bm{e}_y + \sigma_z \bm{e}_z$ is the vector of spin-1/2 Pauli matrices, and $\bm{e}_j$ is a unit vector along the $j$ axis. If we consider the special case of a NW along $z$ and an electric field $\bm{E} = E_x \bm{e}_x$ along $x$, the dominant effect of the Rashba SOI is described by  
\begin{equation}
H^{\rm el}_R \simeq \alpha_{\rm el} E_x \sigma_y k_z .
\label{eq:RSOIElectronsNW}
\end{equation} 
We note that the standard Rashba SOI of Eqs.~(\ref{eq:RSOIElectronsGeneral}) and (\ref{eq:RSOIElectronsNW}) enables the implementation of spin filters \cite{streda:prl03}, Majorana fermions \cite{alicea:rpp12}, EDSR \cite{golovach:prb06, nowack:sci07, trif:prb08, nadjperge:nat10, schroer:prl11, petersson:nat12, vandenberg:prl13}, and is the basis for many other useful effects and applications. Since it is a higher-order effect, however, the Rashba coefficient $\alpha_{\rm el}$ is relatively small in most materials. A perturbative analysis shows that $\alpha_{\rm el}$ depends strongly on the band structure parameters and is particularly small in semiconductors with a large fundamental band gap \cite{winkler:book}.

\subsection{Holes}
\label{secsub:Holes}

The situation is different in the valence band. In semiconductors such as GaAs or InAs (zinc-blende lattice) and Ge (diamond lattice), the hole spectrum near the $\Gamma$ point is well described by the LK Hamiltonian \cite{luttinger:pr55, luttinger:pr56}
\begin{equation}
H^{h}_{\rm 0} = \frac{\hbar^2}{2 m} \left[ \left( \gamma_1 + \frac{5}{2} \gamma_s \right) k^2 - 2 \gamma_s \left( \bm{k}\cdot\bm{J}  \right)^2 \right] , 
\label{eq:HolesH0}
\end{equation} 
where $m$ is the bare electron mass, $\gamma_1$ is a Luttinger parameter, $\bm{J} = J_x \bm{e}_x + J_y \bm{e}_y + J_z \bm{e}_z$ is the vector of spin-3/2 operators, and $k^2 = \bm{k}\cdot\bm{k} = k_x^2 + k_y^2 + k_z^2$. For simplicity, we used here the spherical approximation \cite{lipari:prl70, winkler:sst08}. That is, the original Luttinger parameters $\gamma_2$ and $\gamma_3$ were replaced by one parameter~$\gamma_s$, leading to invariance under arbitrary rotations of the coordinate system. In stark contrast to Eq.~(\ref{eq:ElectronsH0}) for electrons, the leading-order Hamiltonian for holes [Eq.~(\ref{eq:HolesH0})] already provides a coupling between the momentum and the spin.

The abovementioned coupling between the momentum and the spin has notable effects on the dispersion relation of holes in bulk. Considering $\ket{3/2}$, $\ket{1/2}$, $\ket{-1/2}$, and $\ket{-3/2}$ as the eigenstates of the spin operator $J_z$, with
\begin{eqnarray}
J_z \ket{\pm 3/2} &=& \pm \frac{3}{2} \ket{\pm 3/2} ,  \\
J_z \ket{\pm 1/2} &=& \pm \frac{1}{2} \ket{\pm 1/2} ,
\end{eqnarray}   
one finds
\begin{eqnarray}
H^{h}_{\rm 0} e^{i \tilde{k}_z z} \ket{\pm 3/2} &=& \frac{\hbar^2 \tilde{k}_z^2}{2 m_{\rm HH}} e^{i \tilde{k}_z z} \ket{\pm 3/2} , 
\label{eq:EigenstateHHBulk} \\
H^{h}_{\rm 0} e^{i \tilde{k}_z z} \ket{\pm 1/2} &=& \frac{\hbar^2 \tilde{k}_z^2}{2 m_{\rm LH}} e^{i \tilde{k}_z z} \ket{\pm 1/2} ,
\label{eq:EigenstateLHBulk}
\end{eqnarray}
where
\begin{equation}
m_{\rm HH} = \frac{m}{\gamma_1 - 2 \gamma_s} 
\label{eq:massHH}
\end{equation}
is the HH mass and 
\begin{equation}
m_{\rm LH} = \frac{m}{\gamma_1 + 2 \gamma_s} 
\label{eq:massLH}
\end{equation}
is the LH mass. We used here the $z$ axis as an example. Due to the spherical approximation, i.e., the rotational invariance of $H^{h}_{\rm 0}$ in Eq.~(\ref{eq:HolesH0}), analogous results are obtained when one considers an arbitrary spatial axis. That is, the effective mass of the hole depends strongly on the spin state and is large (HH) when the spin is parallel to the direction of motion. 

We wish to point out that the tilde of $\tilde{k}_z$ in Eqs.~(\ref{eq:EigenstateHHBulk}) and~(\ref{eq:EigenstateLHBulk}) was added because $\tilde{k}_z$ is a wave number, in contrast to the previously introduced $k_z$ which is an operator. For the sake of a simple notation, however, we will write $k_z$ for both the operator and the wave number in the remainder of this work (analogous for all axes).

\subsubsection{Holes in 2D-like systems}
\label{secsubsub:HolesIn2D}

Quantum wells \cite{vankesteren:prb90, gradl:prb14}, lateral QDs \cite{grbic:apl05, hanson:rmp07, kloeffel:annurev13}, and many self-assembled QDs \cite{katsaros:nna10, ares:prl13, warburton:nmat13} feature one special axis of very strong confinement and are therefore prominent examples for 2D-like quantum systems. Before we can discuss electric-field-induced effects, it is important that we remind us of several key properties of hole states in such 2D-like systems \cite{winkler:book, winkler:sst08}. Therefore, let us assume for simplicity that the confining potential $V(\bm{r}) = V_{\parallel}(x,y) + V_{\perp}(z)$ for the holes comprises a narrow hard-wall potential of width $L_z$ along the $z$ axis,
\begin{equation}
V_{\perp}(z) = \left\{ \begin{array}{ll}
0, & 0 < z < L_z, \\
\infty, & \mbox{otherwise} , \\
\end{array} \right.
\label{eq:ExampleVzHoles2D}
\end{equation}
and a much weaker in-plane confinement $V_{\parallel}(x,y)$ for the axes $x$ and $y$. In order to find the low-energy eigenstates of the Hamiltonian $H^{h}_{\rm 0} + V$, one can exploit the strong confinement along $z$ and focus first on the 1D Hamiltonian $H^{h}_{\rm 0}(k_z) + V_{\perp}(z)$, where  
\begin{equation}
H^{h}_{\rm 0}(k_z) =  \frac{\hbar^2}{2 m}  \left( \gamma_1 + \frac{5}{2} \gamma_s - 2 \gamma_s J_z^2 \right) k_z^2 
\label{eq:HolesH0kzOnly}
\end{equation}
is obtained from Eq.~(\ref{eq:HolesH0}) by omitting all terms with $k_x$ or $k_y$. As evident from Eq.~(\ref{eq:HolesH0kzOnly}), $H^{h}_{\rm 0}(k_z)$ simplifies to $\hbar^2 k_z^2/(2 m_{\rm HH})$ when the spin state is either $\ket{3/2}$ or $\ket{-3/2}$, and to $\hbar^2 k_z^2/(2 m_{\rm LH})$ when the spin state is either $\ket{1/2}$ or $\ket{-1/2}$. Consequently, given the example of $V_{\perp}(z)$ in Eq.~(\ref{eq:ExampleVzHoles2D}), it turns out that the states $\ket{\Phi_n^{\pm 3/2}}$ and $\ket{\Phi_n^{\pm 1/2}}$ with position-space representations
\begin{eqnarray}
\Braket{z | \Phi_n^{\pm 3/2}} &=& \left\{ \begin{array}{cl} 
\sqrt{\frac{2}{L_z}} \sin\biggl( \frac{n \pi z}{L_z} \biggr) \ket{\pm 3/2} , & 0 < z < L_z , \\
0, & \mbox{otherwise},  
\end{array} \right. 
\label{eq:PhinPm3ov2PSRepr} \\
\Braket{z | \Phi_n^{\pm 1/2}} &=& \left\{ \begin{array}{cl} 
\sqrt{\frac{2}{L_z}} \sin\biggl( \frac{n \pi z}{L_z} \biggr) \ket{\pm 1/2} , & 0 < z < L_z , \\
0, & \mbox{otherwise},  
\end{array} \right. 
\end{eqnarray}
are the eigenstates of $H^{h}_{\rm 0}(k_z) + V_{\perp}(z)$ with
\begin{eqnarray}
E_n^{\rm HH} &=& n^2 \frac{\hbar^2 \pi^2}{2 m_{\rm HH} L_z^2} , 
\label{eq:HHEnergies1D} \\ 
E_n^{\rm LH} &=& n^2 \frac{\hbar^2 \pi^2}{2 m_{\rm LH} L_z^2} 
\label{eq:LHEnergies1D}
\end{eqnarray}
as the respective eigenenergies. The $n \in \{1, 2, \cdots \}$ in Eqs.~(\ref{eq:PhinPm3ov2PSRepr}) to (\ref{eq:LHEnergies1D}) is a quantum number. We note that the HH-LH splitting
\begin{equation}
\Delta_{\rm HH-LH} = \frac{\hbar^2 \pi^2}{2 L_z^2} \left( \frac{1}{m_{\rm LH}} - \frac{1}{m_{\rm HH}} \right)
\end{equation}
is a large energy in our 2D-like system because $L_z$ is relatively small.  

Product states that consist of $\ket{\Phi_n^{\pm 3/2}}$ or $\ket{\Phi_n^{\pm 1/2}}$ and suitable orbital parts for the $x$-$y$ plane form a set of basis states that can be used to analyze $H^{h}_{\rm 0} + V$. As the HH-LH splitting $\Delta_{\rm HH-LH}$ provides a relatively large energy gap between the basis states with $\ket{\Phi_n^{\pm 3/2}}$ and those with $\ket{\Phi_n^{\pm 1/2}}$, it turns out that the low-energy eigenstates of $H^{h}_{\rm 0} + V$ feature almost exclusively the spin states $\ket{3/2}$ and $\ket{-3/2}$. We wish to emphasize that the tiny admixtures of basis states with spin $\ket{\pm 1/2}$ can nevertheless have substantial effects on characteristic properties such as the $g$ factors \cite{ares:prl13, watzinger:nlt16}.

The above-discussed example highlights two major features of low-energy hole states in 2D-like systems, and we note that these features are not restricted to our specific example of the confining potential. First, the involved spin states are almost exclusively $\ket{3/2}$ and $\ket{-3/2}$, which are the two spin states parallel to the axis of strong confinement. Second, there is a strong connection between the spin states $\ket{\pm 3/2}$ ($\ket{\pm 1/2}$) and the HH mass $m_{\rm HH}$ (LH mass $m_{\rm LH}$). Consequently, the hole states with spin $\ket{3/2}$ or $\ket{-3/2}$ in 2D-like systems are commonly referred to as HH states, whereas those with spin $\ket{1/2}$ or $\ket{-1/2}$ are commonly referred to as LH states. This nomenclature is equivalent to the bulk case. We note, however, that the spin projections $\pm 3/2$ and $\pm 1/2$ refer here to the axis of strong confinement instead of the direction of motion in bulk.

The fact that the low-energy hole states of a 2D-like system with strong confinement along $z$ contain nearly exclusively the spin states $\ket{3/2}$ and $\ket{-3/2}$ has remarkable consequences. Mathematically, these consequences are evident from the identities
\begin{eqnarray}
\bra{\pm 3/2} J_z \ket{\pm 3/2} &=& \pm \frac{3}{2} , \\
\bra{\pm 3/2} J_{x,y} \ket{\pm 3/2} &=& 0 ,
\end{eqnarray}
and
\begin{eqnarray}
\bra{3/2} J_\mu \ket{- 3/2} &=& 0 , \label{eq:JMatrixElementHHstates} \\
\bra{3/2} J_\mu  J_\nu  \ket{- 3/2} &=& 0,
\label{eq:JSquaredMatrixElementHHstates}
\end{eqnarray}
which hold for any $\mu, \nu \in \left\{x,y,z\right\}$. For instance, the Zeeman term $2 \kappa \mu_B \bm{B}\cdot \bm{J}$ \cite{luttinger:pr56, winkler:book} becomes inefficient when the magnetic field $\bm{B}$ lies in the $x$-$y$ plane. Indeed, the measured $g$ factors usually exhibit a pronounced anisotropy, where small (large) values are observed when $\bm{B}$ is applied in-plane (out-of-plane) \cite{vankesteren:prb90, nichele:prl14, katsaros:nna10, watzinger:nlt16}. Besides the $g$ factors, also the SOI is affected by the HH character of the low-energy states. For an electric field $\bm{E} = E_z \bm{e}_z$ along the axis of strong confinement, the standard Rashba SOI for holes \cite{winkler:prb00, winkler:book}
\begin{equation}
H^h_R = \alpha_h \bm{E} \cdot \left( \bm{k} \times \bm{J} \right) 
\label{eq:RSOIHolesGeneral}
\end{equation} 
reduces to 
\begin{equation}
H^h_R = \alpha_h E_z \left( k_x J_y - k_y J_x \right) .
\label{eq:RSOIHoles2DwithEz}
\end{equation}
Because of Eq.~(\ref{eq:JMatrixElementHHstates}), one finds
\begin{equation}
\bra{3/2}  \left( k_x J_y - k_y J_x \right) \ket{- 3/2} = 0 ,
\end{equation}
and so Eq.~(\ref{eq:RSOIHoles2DwithEz}) does not provide a $k$-linear coupling between states of pure-HH type. The SOI for holes in 2D-like systems therefore requires a detailed analysis, and many interesting and insightful results have already been obtained \cite{winkler:book, winkler:prb00, bulaev:prl07, winkler:sst08, chesi:prl11, nichele:prb14, nichele:prl14, srinivasan:prl17, hung:prb17, liu:arX1708}. For a comparison with Sec.~\ref{secsubsub:HolesInNWs}, it is essential to note that 
\begin{equation}
\bra{3/2} H^{h}_{\rm 0} \ket{- 3/2} = 0 
\label{eq:32Hh0Minus32}
\end{equation}
due to Eq.~(\ref{eq:JSquaredMatrixElementHHstates}). Hence, we can conclude that the DRSOI, which will be explained in detail below, is strongly suppressed in 2D-like systems because of the large HH-LH splitting.

\subsubsection{Holes in NWs and elongated NW QDs}
\label{secsubsub:HolesInNWs}

When there are two axes of strongest confinement, which, in particular, is the case for NWs with square or circular cross sections, a simple separation between HH and LH states as in Sec.~\ref{secsubsub:HolesIn2D} is no longer possible, i.e., even the low-energy eigenstates of such 1D-like hole systems may exhibit a strong mixing of HH and LH states \cite{sercel:prb90, csontos:prb09}. This fact can enable novel effects that are negligible in 2D-like systems. In the following, we want to recall key elements of the effective model of Ref.~\cite{kloeffel:prb11} for low-energy hole-states in Ge/Si core/shell NWs, because Ref.~\cite{kloeffel:prb11} showed that the combination of an applied electric field and a potential with two (or three, see Sec.~\ref{secsubsub:HolesIn0DlikeQDs}) axes of strongest hole confinement results in an unusually strong SOI that is not suppressed by the fundamental band gap, referred to as the DRSOI.

When the NW axis is $z$ and an electric field $\bm{E} = E_x \bm{e}_x$ is applied along $x$, the low-energy hole spectrum in Ge/Si core/shell NWs is well described by the 4$\times$4 Hamiltonian 
\begin{equation}
H_{\rm 4x4}^{\rm eff} = 
\begin{pmatrix}
\frac{\hbar^2 k_z^2}{2 m_g}  &  0  &  e U E_x  &  - i C k_z \\
0  &  \frac{\hbar^2 k_z^2}{2 m_g}  &  - i C k_z  &  - e U E_x \\
e U E_x  &  i C k_z  &  \frac{\hbar^2 k_z^2}{2 m_e} + \Delta  &  0 \\
i C k_z  &  - e U E_x  &  0  &  \frac{\hbar^2 k_z^2}{2 m_e} + \Delta
\end{pmatrix} ,
\label{eq:RecallPRB2011Matrix}
\end{equation}
where $m_g = 0.043 m$ and $m_e = 0.054 m$ are effective masses, $C = 7.26\hbar^2/(m R)$ and $U = 0.15R$ are inversely and directly proportional to the core radius $R$, and $\Delta = \Delta_{\rm BP} + 0.73 \hbar^2/(m R^2)$ is an energy gap that results from the shell-induced strain and the confinement, for which a cylindrically symmetric hard wall was assumed at the core-shell interface,
\begin{equation}
V(x,y) = \left\{ \begin{array}{ll}
0 , & \sqrt{x^2 + y^2} < R, \\
\infty , & \mbox{ otherwise}. 
\end{array}
\right. 
\label{eq:VcylindricalHardWall}
\end{equation}
Equation~(\ref{eq:RecallPRB2011Matrix}) is obtained when the Hamiltonian $H^{h}_{\rm 0} + V(x,y) - e E_x x$ is projected onto the low-energy subspace spanned by $\ket{g_{+}}$, $\ket{g_{-}}$, $\ket{e_{+}}$, and $\ket{e_{-}}$, which are the four basis states of the shown matrix. Static strain caused by the shell, if present, is accounted for by the Bir-Pikus (BP) Hamiltonian \cite{birpikus:book} and simply rescales the energy gap $\Delta$ via $\Delta_{\rm BP}$ \cite{kloeffel:prb14}. The four basis states are eigenstates of the Hamiltonian $H^{h}_{\rm 0}(k_x, k_y) + V(x,y)$, where $H^{h}_{\rm 0}(k_x, k_y)$ corresponds to Eq.~(\ref{eq:HolesH0}) with $k_z = 0$. The states $\ket{g_{+}}$ and $\ket{g_{-}}$ are the two degenerate ground states of $H^{h}_{\rm 0}(k_x, k_y) + V(x,y)$, whereas $\ket{e_{+}}$ and $\ket{e_{-}}$ are the two degenerate excited states with second-lowest eigenenergy. The subscript ``$+$'' (``$-$'') refers to a spin block, meaning here that the state contains the two spin states $\ket{3/2}$ and $\ket{-1/2}$ ($\ket{-3/2}$ and $\ket{1/2}$), and we recall that the $\ket{j_z}$ with $j_z \in \{3/2, 1/2, -1/2, -3/2\}$ are the eigenstates of $J_z$ and satisfy $J_z \ket{j_z} = j_z \ket{j_z}$. The basis states $\ket{g_{+}}$, $\ket{g_{-}}$, $\ket{e_{+}}$, and $\ket{e_{-}}$ were derived with an approach similar to that of Refs.~\cite{sercel:prb90, csontos:prb09}, and details on their explicit form are provided in Ref.~\cite{kloeffel:prb11} and the Supplementary Information (SI) of Ref.~\cite{kloeffel:prb13}. If one studies an infinitely long NW, the wave function that accounts for the $z$ direction is simply a phase factor of type $e^{i k_z z}$, where $k_z$ is a wave number, and so the operator $k_z$ in Eq.~(\ref{eq:RecallPRB2011Matrix}) becomes a continuous parameter \cite{footnote:kzOperatorWavenumber}.

\begin{figure}[tb]
\begin{center}
\includegraphics[width=0.90\linewidth]{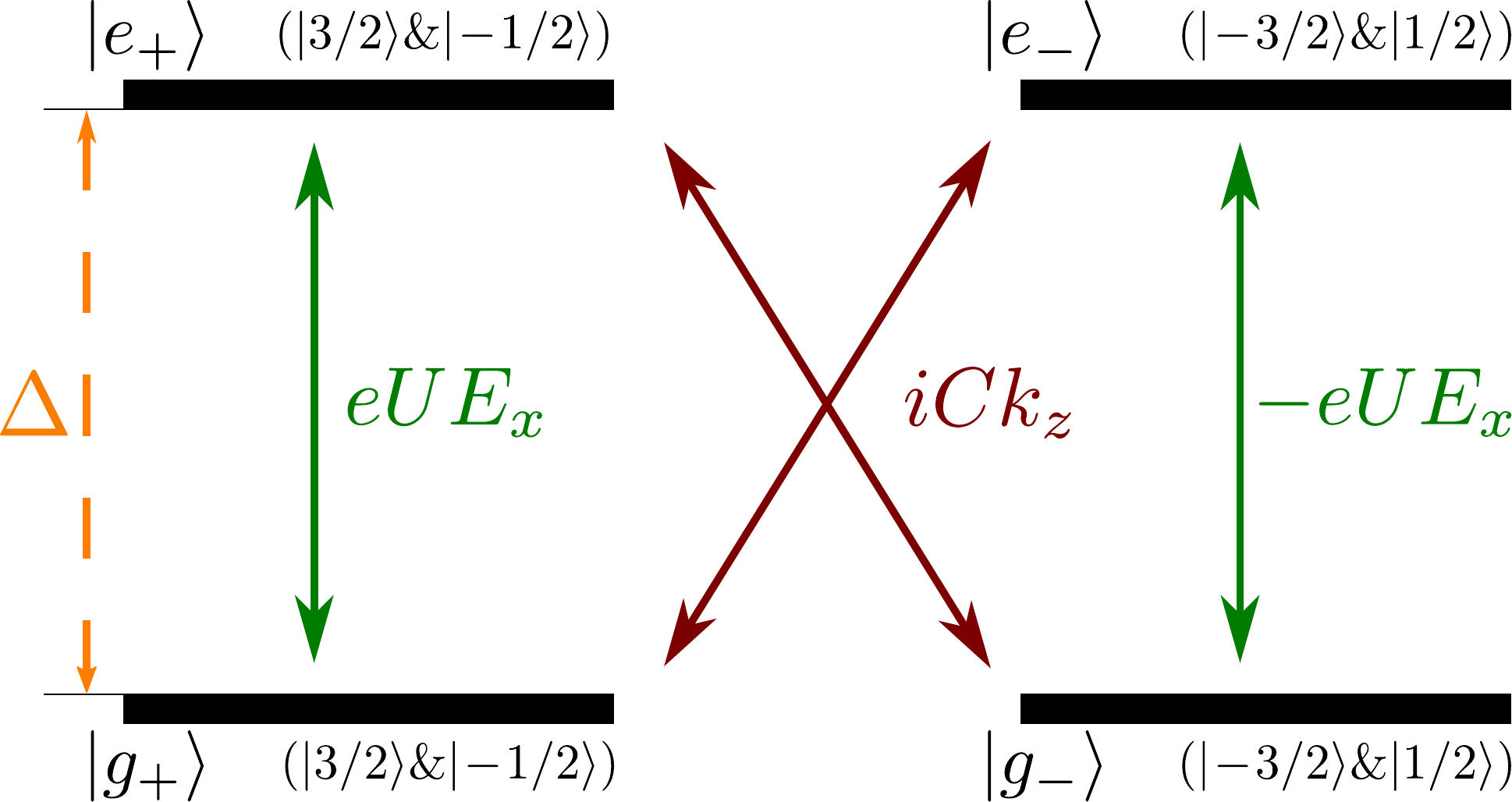}
\caption{The basis states of Eq.~(\ref{eq:RecallPRB2011Matrix}) and their couplings. At $k_z = 0$ and $E_x = 0$, the states $\ket{g_\pm}$ and $\ket{e_\pm}$ are the two ground states and first excited states, respectively, of the Hamiltonian. They differ in energy by $\Delta$ and the contained spin states are shown in parentheses. The couplings proportional to $E_x$ result directly from the electric-field-induced shift $- e E_x x$ (spin conserving) in the potential energy and therefore cannot occur between states that have different spins. The cross couplings proportional to $k_z$ result from the LK Hamiltonian. The combination of the shown couplings within the low-energy subspace results in an unusually strong SOI of Rashba type, the DRSOI \cite{kloeffel:prb11}. The cross couplings would be absent if the spin states for ``$+$'' and ``$-$'' were solely $\ket{3/2}$ and $\ket{-3/2}$, and so the DRSOI in 2D-like systems is strongly suppressed by the HH-LH splitting. The additional energies $\hbar^2 k_z^2/(2 m_{g,e})$ in Eq.~(\ref{eq:RecallPRB2011Matrix}) are not shown in the sketch. Details are provided in the text and in Refs.~\cite{kloeffel:prb11, kloeffel:prb13}. }
\label{fig:DRSOI}
\end{center}
\end{figure}

In Eq.~(\ref{eq:RecallPRB2011Matrix}), the coupling terms proportional to $E_x$ are obtained when the operator
\begin{equation}
H_{\rm dir}^{h} = - e \bm{E} \cdot \bm{r} ,
\label{eq:potentialGradientHoles}
\end{equation}
which reduces to $H_{\rm dir}^{h} = - e E_x x$ given $\bm{E} = E_x \bm{e}_x$, is projected onto the low-energy subspace. The effect of this electric-field-induced coupling is that the hole is pushed along the electric field, i.e., towards the boundary of the Ge-core cross section. Furthermore, $H_{\rm dir}^{h}$ preserves the spin and therefore cannot couple a basis state of type ``$+$'' with one of type~``$-$''. Remarkably, such a coupling between ``$+$'' and ``$-$'' is caused by the terms in $H^{h}_{\rm 0}$ that are linear in $J_x$ or $J_y$. For instance, $J_x J_z k_x k_z$ can couple $\ket{g_\pm}$ with $\ket{e_\mp}$. This is possible since $\bra{1/2} J_x \ket{3/2}$, $\bra{-1/2} J_x \ket{1/2}$, and $\bra{-3/2} J_x \ket{-1/2}$ are nonzero and because the basis states of type ``$+$'' (``$-$'') contain both $\ket{3/2}$ and $\ket{-1/2}$ ($\ket{-3/2}$ and $\ket{1/2}$) as a consequence of the confinement potential of the NW. Therefore, it is important to note that the off-diagonal elements proportional to $k_z$ in Eq.~(\ref{eq:RecallPRB2011Matrix}) result from the LK Hamiltonian $H^{h}_{\rm 0}$. The discussed couplings caused by $H_{\rm dir}^{h}$ and $H^{h}_{\rm 0}$ are illustrated in Fig.~\ref{fig:DRSOI}.

The Hamiltonian of Eq.~(\ref{eq:RecallPRB2011Matrix}) is solely based on the LK Hamiltonian [Eq.~(\ref{eq:HolesH0})], the confinement of the hole to the NW [Eq.~(\ref{eq:VcylindricalHardWall})], the strain (if a shell is present), and the potential gradient that is caused by an electric field [Eq.~(\ref{eq:potentialGradientHoles})]. It turns out that this Hamiltonian already features a strong SOI of Rashba type, the so-called DRSOI \cite{kloeffel:prb11}, even though standard terms for Rashba SOI, in particular $H^h_R$ [Eq.~(\ref{eq:RSOIHolesGeneral})], were not yet included. The DRSOI becomes particularly evident when we consider the special case where $|e U E_x / \Delta|$ and $|C k_z / \Delta|$ can be treated as small parameters in a perturbative analysis. In this case, a Schrieffer-Wolff transformation (quasi-degenerate perturbation theory \cite{winkler:book}) of Eq.~(\ref{eq:RecallPRB2011Matrix}) yields the effective Hamiltonian
\begin{equation}
H_{\rm 2x2}^{\rm eff} = \left( \frac{\hbar^2}{2 m_g} - \frac{C^2}{\Delta} \right) k_z^2 + \frac{2 e C U}{\Delta} E_x \tilde{\sigma}_y k_z
\label{eq:H2x2effWithDRSOI}
\end{equation}   
for the two subbands of lowest energy, where $\tilde{\sigma}_y$ corresponds to a spin-1/2 Pauli matrix. This is equivalent to the well-known effective Hamiltonian
\begin{equation}
H_{\rm 1D}^{\rm el} =  \frac{\hbar^2}{2 m_{\rm eff}} k_z^2 + \alpha_{\rm el} E_x \sigma_y k_z
\label{eq:H1DForElectronsWithRSOI}
\end{equation}
for electrons in a NW with Rashba SOI (see Sec.~\ref{secsub:Electrons}). Hence, in the regime where $|e U E_x / \Delta| \ll 1$ and $|C k_z / \Delta| \ll 1$ we can identify $\alpha_{\rm DR} = 2 e C U / \Delta$ as the effective Rashba coefficient of the DRSOI. Using the parameters $\gamma_1 = 13.35$ and $\gamma_s = 5.114$ for Ge \cite{footnote:GeLKparams}, one obtains the aforementioned values for $C$ and $U$ and therefore $C U = 1.1 \hbar^2/m$. Considering $\Delta = 20\mbox{ meV}$ for instance, which is a realistic subband splitting for typical Ge/Si core/shell NWs \cite{kloeffel:prb11, kloeffel:prb14, lu:pnas05}, one finds $\alpha_{\rm DR} = 8.4\mbox{ nm}^2 e$. This value is much greater than the calculated Rashba coefficient $\alpha_{\rm el} = 0.05\mbox{ nm}^2 e$ for electrons in GaAs and even exceeds $\alpha_{\rm el} = 1.2\mbox{ nm}^2 e$ and $\alpha_{\rm el} = 5.2\mbox{ nm}^2 e$ for electrons in InAs and InSb, respectively \cite{winkler:book}. Furthermore, in stark contrast to GaAs, InAs, or InSb, Dresselhaus SOI is absent in Ge and Si because of bulk inversion symmetry. Therefore, the DRSOI in Ge/Si core/shell NWs results in a strong SOI that is highly controllable with moderate electric fields, and a few volts per micrometer are sufficient to achieve spin-orbit energies of the order of millielectronvolts \cite{kloeffel:prb11}. Indeed, recent experimental studies have reported a strong and electrically tunable SOI for holes in such NWs \cite{hao:nlt10, higginbotham:prl14}. 

While we focused in our discussion on the case $|e U E_x / \Delta| \ll 1$ and $|C k_z / \Delta| \ll 1$ for illustration purposes, we wish to emphasize that this regime is not crucial in order to achieve a strong DRSOI. However, we also wish to point out that the effective Rashba coefficient of the DRSOI (see $\alpha_{\rm DR}$ in the example above) is not always independent of the applied electric field and may even decrease rapidly once the electric field exceeds a certain value. Details will be provided in Secs.~\ref{sec:Ge}, \ref{sec:Si}, and \ref{sec:AnalyticalResultsSiNWs}. We recall that the effective Rashba coefficient of the DRSOI can be understood as the coefficient $\alpha_{\rm DR}$ in an effective SOI term of type $\alpha_{\rm DR} E_x \tilde{\sigma}_y k_z$ that relies directly on $H^h_{\rm dir}$ [Eq.~(\ref{eq:potentialGradientHoles})] rather than on higher-order (i.e., involving the other bands of the semiconductor) corrections for the valence band that are also generated by the electric field, such as $H^h_R$ [Eq.~(\ref{eq:RSOIHolesGeneral})].

For the holes in Ge/Si core/shell NWs, it turns out that the standard Rashba SOI $H^h_R$ [Eq.~(\ref{eq:RSOIHolesGeneral})] has essentially the same effect on the spectrum as $H^h_{\rm dir}$ [Eq.~(\ref{eq:potentialGradientHoles})], even though $H^h_R$ and $H^h_{\rm dir}$ differ greatly. In particular, the latter does not contain any spin operators. While $H^h_{\rm dir}$ is independent of the semiconductor and therefore independent of the fundamental band gap $E_0$, $H^h_R$ is a third-order correction and its coefficient $\alpha_h$ is approximately proportional to $E_0^{-2}$ (see, e.g., the derivation in Ref.~\cite{winkler:book}). As a consequence, the DRSOI dominates for typical Ge/Si core/shell NWs and $H^h_R$ is completely negligible. 

On the one hand, the discovered DRSOI is unusually strong and provides a remarkable degree of external control. On the other hand, the effect resembles that of a standard Rashba SOI [compare, e.g., Eqs.~(\ref{eq:H2x2effWithDRSOI}) and (\ref{eq:H1DForElectronsWithRSOI})]. Therefore, the DRSOI has a wide range of applications. For instance it is an especially useful tool for the implementation of Majorana fermions \cite{alicea:rpp12, maier:prb14b} and for hole-spin qubits in NW QDs \cite{kloeffel:prb13, nigg:prl17}. The predicted anisotropy and electrical tunability of the hole $g$-factor in elongated Ge/Si NW QDs \cite{maier:prb13} has already been observed experimentally \cite{brauns:prb16}. 

Of course, the discussed model for low-energy hole states applies not only to Ge/Si core/shell NWs, but also to arbitrary semiconducting NWs, provided that the approximations made in the derivation of the model are still well justified. For all these details of the model and additional information, such as magnetic-field-induced effects, we refer to Refs.~\cite{kloeffel:prb11, maier:prb13, kloeffel:prb13, kloeffel:prb14} and the SI of Ref.~\cite{kloeffel:prb13}. 

We want to conclude this Sec.~\ref{secsubsub:HolesInNWs} by highlighting two major differences compared with the 2D-like systems in Sec.~\ref{secsubsub:HolesIn2D}. First, it is evident that the terms proportional to $C k_z$ in the low-energy Hamiltonian of Eq.~(\ref{eq:RecallPRB2011Matrix}) are crucial for the DRSOI. These terms originate from the LK Hamiltonian $H^{h}_{\rm 0}$ and provide a coupling between basis states of different spin type (Fig.~\ref{fig:DRSOI}). In a similar model for 2D-like systems, $H^{h}_{\rm 0}$ cannot cause such a strong coupling between low-energy basis states with different spin, because the low-energy basis states in a 2D-like system are of HH-type, see Sec.~\ref{secsubsub:HolesIn2D}. That is, if we consider $z$ as the axis of strongest confinement, the low-energy states of the 2D-like system contain almost exclusively the spin states $\ket{3/2}$ and $\ket{-3/2}$. The LK Hamiltonian $H^{h}_{\rm 0}$ [Eq.~(\ref{eq:HolesH0})] features products $J_\mu J_\nu$ of two but not of three or more spin operators, which would be needed to couple $\ket{3/2}$ with $\ket{-3/2}$. Consequently, Eq.~(\ref{eq:32Hh0Minus32}) applies and the DRSOI in 2D-like systems is suppressed by the HH-LH splitting. 

The second difference affects the terminology. In 2D-like systems (Sec.~\ref{secsubsub:HolesIn2D}), eigenstates with a relatively small in-plane momentum can almost exclusively be formed with basis states of either HH or LH type (weak HH-LH mixing). Therefore, these eigenstates are themselves often referred to as HH or LH states. More precisely, when $z$ is the direction of strong confinement, the eigenstates with a large contribution of the spin states $\ket{3/2}$ and $\ket{-3/2}$ ($\ket{1/2}$ and $\ket{-1/2}$) are simply referred to as HH (LH) states because of the close connection between these spin states and the HH mass $m_{\rm HH}$ (LH mass $m_{\rm LH}$). In stark contrast to Sec.~\ref{secsubsub:HolesIn2D}, in systems with two axes of strongest confinement (Sec.~\ref{secsubsub:HolesInNWs}) even the low-energy eigenstates can have relatively large contributions of both HH- and LH-type basis states (strong HH-LH mixing) \cite{sercel:prb90, csontos:prb09}. Furthermore, one should always be aware of the type of basis states that were considered when HH and LH contributions are discussed, because there is more variety in the literature than in the case of 2D-like systems. For instance, there may be basis states with a HH-like dispersion relation only for the motion along the NW axis or only for the motion in a transverse direction, all of which might be referred to as HH states. Thus, the names HH and LH states in the context of NWs can be ambiguous without a precise specification. In the upcoming analysis of the hole states in Si, Ge, and Ge/Si core/shell NWs, this will be taken into account.

\subsubsection{Holes in QDs with similar confinement for all directions}
\label{secsubsub:HolesIn0DlikeQDs}

We are currently not aware of an analysis of the DRSOI for QDs that exhibit a very similar confinement along all three directions. It is, however, reasonable to assume that a relatively strong DRSOI is also feasible in these systems, since HH-LH mixing is usually inevitable for the holes in such QDs \cite{sercel:prb90, csontos:prb09}. In such systems, the DRSOI would most likely lead to an effective SOI term of type $\alpha_{\rm DR} \bm{E} \cdot \left( \bm{k} \times \bm{J} \right)$, where $\alpha_{\rm DR}$ is the effective Rashba coefficient of the DRSOI. The value of $\alpha_{\rm DR}$ may be determined with a model that comprises only the LK Hamiltonian, the BP Hamiltonian (if strain is present), the confining potential, and the direct coupling to the electric field [Eq.~(\ref{eq:potentialGradientHoles})].

\section{Model}
\label{sec:Model}

\subsection{Hamiltonian}
\label{secsub:ModelHamiltonian}

The Hamiltonian of our model for low-energy hole states in NWs is
\begin{equation}
H = H_{\rm LK} + H_{\rm BP} + H^h_{\rm dir} + H^h_R + H^h_Z + V .
\label{eq:HamiltonianOurModel}
\end{equation}
In the following, the contributions to this Hamiltonian are explained. We want to point out that our Hamiltonian contains a global minus sign compared with that for valence band electrons, since holes are unfilled valence band states.

\subsubsection{Luttinger-Kohn Hamiltonian}
\label{secsubsub:ModelHamLK}

The LK Hamiltonian without Zeeman terms (separately discussed in Sec.~\ref{secsubsub:ModelHamElectrMagnFields}) is \cite{luttinger:pr55, luttinger:pr56}
\begin{eqnarray}
H_{\rm LK} = 
\frac{\hbar^2}{2m}\Biggl[ & &
\left(\gamma_1 + \frac{5 \gamma_2}{2}\right)k^2 
- 2 \gamma_2 \left( k_{x^\prime}^2 J_{x^\prime}^2 + k_{y^\prime}^2 J_{y^\prime}^2 + k_{z^\prime}^2 J_{z^\prime}^2 \right) \nonumber \\
& & - 4 \gamma_3 \left( \{k_{x^\prime}, k_{y^\prime}\}\{J_{x^\prime}, J_{y^\prime}\} \mbox{ + c.p.} \right)
\Biggr] ,
\label{eq:LKfullMainaxes} 
\end{eqnarray}
where ``c.p.''\ stands for cyclic permutations and $\{A, B\} = (AB + BA)/2$. We recall that $m$ is the free electron mass, $\gamma_{1,2,3}$ are the Luttinger parameters, and $J_i$ are spin-3/2 operators obeying $[J_{x^\prime}, J_{y^\prime}] = J_{x^\prime} J_{y^\prime} - J_{y^\prime} J_{x^\prime} = i J_{z^\prime}$ (and analogous for cyclic permutations). It is important to note that the $\hbar k_i$ in Eq.~(\ref{eq:LKfullMainaxes}) correspond to the kinetic electron momenta, i.e., 
\begin{equation}
\bm{k} = - i \nabla + \frac{e}{\hbar} \bm{A}, 
\end{equation}  
where $-e$ is the electron charge, $\nabla$ is the Nabla operator, and $\bm{A}$ is the vector potential with $\bm{B} = \nabla \times \bm{A}$ \cite{luttinger:pr56}. Consequently, one finds $\bm{k}\times\bm{k} = - i e \bm{B}/\hbar$ \cite{winkler:book}, i.e., the components $k_i$ no longer commute in the presence of a magnetic field $\bm{B}$.
Furthermore, we note that the axes $x^\prime$, $y^\prime$, and $z^\prime$ in Eq.~(\ref{eq:LKfullMainaxes}) correspond to the main crystallographic axes.

\subsubsection{Bir-Pikus Hamiltonian}

Strain-based effects on the hole states are described by the BP Hamiltonian \cite{birpikus:book} 
\begin{eqnarray}
H_{\rm BP} &=& 
- \left(a + \frac{5 b}{4} \right)  \left( \varepsilon_{x'x'} + \varepsilon_{y'y'} + \varepsilon_{z'z'} \right) \nonumber \\
& & + b \left( \varepsilon_{x'x'} J_{x'}^2 + \varepsilon_{y'y'} J_{y'}^2 + \varepsilon_{z'z'} J_{z'}^2 \right) \nonumber \\
& & + \frac{2 d}{\sqrt{3}} \left( \varepsilon_{x'y'} \{J_{x^\prime}, J_{y^\prime}\} \mbox{ + c.p.} \right) ,
\label{eq:BirPikusFull}
\end{eqnarray}
where $a$, $b$, and $d$ are the deformation potentials, $\varepsilon_{ij} = \varepsilon_{ji}$ are the strain tensor elements, and $x^\prime$, $y^\prime$, $z^\prime$ are again the main crystallographic axes.

\subsubsection{Electric and magnetic fields}
\label{secsubsub:ModelHamElectrMagnFields}

An applied electric field $\bm{E}$ is accounted for by the direct coupling $H^h_{\rm dir}$ [Eq.~(\ref{eq:potentialGradientHoles})] and by the standard Rashba SOI $H^h_R$ [Eq.~(\ref{eq:RSOIHolesGeneral})]. An applied magnetic field $\bm{B}$ enters the calculation via the aforementioned vector potential $\bm{A}$. The used gauge and further details are provided in Appendix~\ref{app:OrbitalContribMagnField}. In addition, we include the Zeeman term \cite{luttinger:pr56, winkler:book}  
\begin{equation}
H^h_Z = 2 \kappa \mu_B \bm{B}\cdot\bm{J} 
\end{equation}
with $\mu_B$ as the Bohr magneton. The anisotropic Zeeman term $2 q \mu_B \bm{B}\cdot\bm{\mathcal{J}}$ \cite{luttinger:pr56, winkler:book}, where $\bm{\mathcal{J}} = \bm{e}_{x'}J_{x'}^3 + \bm{e}_{y'}J_{y'}^3 + \bm{e}_{z'}J_{z'}^3$, is omitted in our model since $|q| \ll |\kappa|$ for Si and Ge \cite{lawaetz:prb71}.

\subsubsection{Confinement}

\begin{figure}[tb]
\begin{center}
\includegraphics[width=0.90\linewidth]{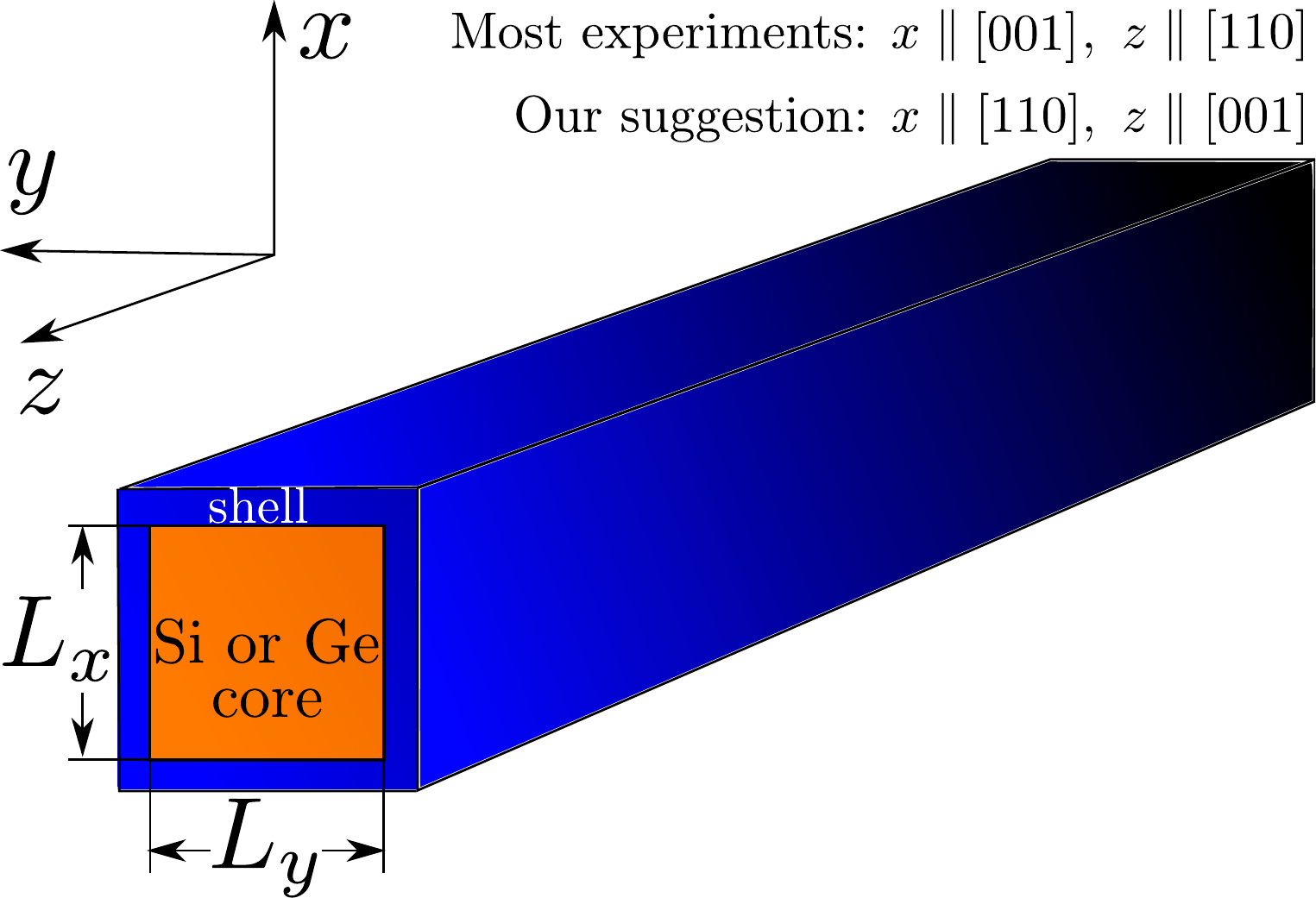}
\caption{Sketch of the NWs and the coordinate system considered in this work. The core of the NW has a rectangular cross section which lies in the $x$-$y$ plane. The NW axis is the $z$ axis. In the case of a Ge/Si core/shell NW, the Ge core is compressively strained by the Si shell. Since standard silicon-on-insulator (or bulk Si \cite{jiang:edl09}) wafers have a (100) surface, the $x$ axis in the sketch usually corresponds to a main crystallographic axis when a CMOS-compatible NW is fabricated as in Refs.~\cite{voisin:nlt16, maurand:ncomm16, crippa:arX1710, coquand:ulisproc12, barraud:edl12, jiang:edl09, prati:nanotech12}. We find that a much stronger SOI can be induced in Si NWs when $x \parallel [110]$ and $z \parallel [001]$ (see Secs.~\ref{sec:Si} and~\ref{sec:AnalyticalResultsSiNWs}). }
\label{fig:SetupWireAndQD}
\end{center}
\end{figure}

In the present work, we consider NWs with rectangular cross sections. More generally, we consider core-shell NWs whose cores have rectangular cross sections, as illustrated in Fig.~\ref{fig:SetupWireAndQD}. The height of the core is $L_x$, the width is $L_y$, and the cross section lies in the $x$-$y$ plane with $|x| < L_x / 2$ and $|y| < L_y / 2$. The wire axis corresponds to the $z$ axis, analogous to Sec.~\ref{secsubsub:HolesInNWs}. We assume hard-wall confinement at the core-shell interface, and so the confining potential is
\begin{equation}
V = V(x,y) = \left\{ \begin{array}{ll} 0, & |x| < \frac{L_x}{2} \mbox{ and } |y| < \frac{L_y}{2} ,   \\ 
\infty, & \mbox{otherwise}.
\end{array} \right. 
\label{eq:ConfinementRectangularNWxy}
\end{equation}
The same confining potential is used to model a bare NW of height $L_x$ and width $L_y$. We note that the bare NW can be considered as a core-shell NW in the limit of a vanishing shell.

\newpage

\subsection{Basis states and numerical diagonalization}
\label{secsub:BasisStatesNumDiag}

The functions \cite{csontos:prb09, harada:prb06}
\begin{equation}
f_{n_x , n_y}(x,y) = \frac{2 \sin\Bigl[n_x \pi \Bigl(\frac{x}{L_x} + \frac{1}{2} \Bigr)\Bigr] \sin\Bigl[n_y \pi \Bigl(\frac{y}{L_y} + \frac{1}{2} \Bigr)\Bigr] }{\sqrt{L_x L_y}}  
\end{equation}
with $n_{x,y} \in \{1,2,\cdots\}$ satisfy the relations
\begin{gather}
0 = f_{n_x , n_y}(-L_x/2,y) =  f_{n_x , n_y}(L_x/2 ,y) , \\
0 = f_{n_x , n_y}(x,-L_y/2) =  f_{n_x , n_y}(x,L_y/2) , \\
\int_{- L_x/2}^{L_x / 2} dx \int_{- L_y / 2}^{L_y / 2} dy \hspace{0.05cm} f_{n_x, n_y}(x,y) f_{n_x^\prime,n_y^\prime}(x,y) = \delta_{n_x , n_x^\prime} \delta_{n_y , n_y^\prime} , 
\end{gather}
where $\delta_{ij}$ is the Kronecker delta. Thus, they are consistent with the hard-wall boundary conditions and form a complete set of orthonormal basis functions for the transverse orbital part of a wave function. Consequently, the hole states may be written as linear combinations of basis states $\ket{n_x , n_y , k_z , j_z}$ whose position-space representation is 
\begin{equation}
\Braket{x,y,z | n_x , n_y , k_z , j_z} = f_{n_x , n_y}(x,y) e^{i k_z z} \ket{j_z} 
\label{eq:BasisStatesNWinside}
\end{equation}
if $|x| < L_x / 2$ and $|y| < L_y / 2$, or
\begin{equation}
\Braket{x,y,z | n_x , n_y , k_z , j_z} = 0
\label{eq:BasisStatesNWoutside}
\end{equation}
otherwise. In Eq.~(\ref{eq:BasisStatesNWinside}), the factor $e^{i k_z z}$ with wave number $k_z$ results from the model assumption of an infinitely long NW, i.e., from the translational invariance along the $z$ axis. The spin states $\ket{j_z}$ are eigenstates of $J_z$ with eigenvalues $j_z \in \{3/2$, $1/2,$ $-1/2,$ $-3/2\}$, i.e., $J_z \ket{j_z} = j_z \ket{j_z}$. The spin-3/2 operators $J_x$, $J_y$, and $J_z$ are implemented in our calculations via the standard matrix representation which is shown, e.g., in Appendix~C of Ref.~\cite{winkler:book} or in Eqs.~(A1) to (A3) of Ref.~\cite{kloeffel:prb11}. 

In order to analyze the low-energy hole states in the NWs, we project the Hamiltonian $H$ [Eq.~(\ref{eq:HamiltonianOurModel})] onto the subspace that is spanned by the 36 basis states $\ket{n_x , n_y , k_z , j_z}$ [Eqs.~(\ref{eq:BasisStatesNWinside}) and (\ref{eq:BasisStatesNWoutside})] with $n_{x, y} \leq 3$. Having chosen the desired values for all input parameters, such as the wave number $k_z$ and the applied electric and magnetic fields, the resulting 36$\times$36 matrix is diagonalized numerically. We note that calculations with only 16 basis states, namely those with $n_{x, y} \leq 2$, yielded results that are similar to those plotted here with $n_{x, y} \leq 3$. Since the eigenenergy of a particle in a hard-wall potential increases quadratically with the quantum number, rather than, e.g., linearly as in the case of harmonic confinement, only minor quantitative corrections to the low-energy hole states can be expected from basis states with large quantum numbers $n_x$ and $n_y$. We therefore conclude that our subspace with $n_{x, y} \leq 3$ is on the one hand large enough to feature the most important couplings and to provide reasonably accurate results, and on the other hand small enough to enable fast computation \cite{watzinger:nlt16}.

\pagebreak

\subsection{Nanowire fabrication}
\label{secsub:ModelCoordinateSystems}

The details of the NW fabrication enter our model via the relations between the axes $x',y',z'$ (main crystallographic axes) and $x,y,z$ (NW and setup, see Fig.~\ref{fig:SetupWireAndQD}). As already mentioned in Sec.~\ref{sec:DRSOI}, the unit vector that points along the axis $j$ is referred to as $\bm{e}_{j}$. Furthermore, we use $\bm{e}_{x'}$, $\bm{e}_{y'}$, and $\bm{e}_{z'}$ for the crystallographic directions [100], [010], and [001], respectively. In the following, we consider the two cases where the NW axis $z$ coincides with the [001] and the [110] direction.

\subsubsection{Nanowire axis along $[001]$}
\label{secsubsub:ModelAxisZalong001}

When the $z$ direction coincides with the [001] direction, one obtains
\begin{eqnarray}
\bm{e}_x &=& \bm{e}_{x^\prime} \cos\phi + \bm{e}_{y^\prime} \sin\phi , \label{eq:RelationsAxesMainTextFor001start} \\
\bm{e}_y &=& - \bm{e}_{x^\prime} \sin\phi + \bm{e}_{y^\prime} \cos\phi , \\
\bm{e}_z &=& \bm{e}_{z^\prime} , \label{eq:RelationsAxesMainTextFor001end}
\end{eqnarray}
where the angle $\phi$ depends again on the details of the NW fabrication and determines the orientation of the crystallographic axes with respect to the transverse directions. As described in Appendix~\ref{appsub:TransformationFor001}, we can use the shown relations between the unit vectors to rewrite, among other things, the LK Hamiltonian of Eq.~(\ref{eq:LKfullMainaxes}). By doing so, we find 
\begin{widetext}
\begin{eqnarray}
H_{\rm LK}^{\rm [001]}(\phi) =  \frac{\hbar^2}{2m} & \Biggl[ & 
\left(\gamma_1 + \frac{5 \gamma_2}{2}\right)k^2 - 2 \gamma_2 \left( k_x^2 J_x^2 + k_y^2 J_y^2 \right) \left(\cos^4\phi + \sin^4\phi \right) - 2 \gamma_2 k_z^2 J_z^2 \nonumber \\
& & - \gamma_2 \left[ \left( k_y^2 - k_x^2 \right) \{ J_x , J_y \} + \{ k_x , k_y \} \left(J_y^2 - J_x^2 \right) \right] \sin(4\phi) 
- \gamma_2 \left( k_x^2 J_y^2 + k_y^2 J_x^2 + 4 \{ k_x , k_y \} \{ J_x , J_y \} \right) \sin^2(2\phi) \nonumber \\
& & - \gamma_3 \left[ \left(k_x^2 -k_y^2 \right) \sin(2\phi) + 2 \{ k_x , k_y \} \cos(2\phi) \right] \left[ \left(J_x^2 - J_y^2 \right) \sin(2\phi) + 2 \{ J_x , J_y \} \cos(2\phi) \right] \nonumber \\
& & - 4 \gamma_3 \left( \{ k_y , k_z \} \{ J_y , J_z \} + \{ k_z , k_x \} \{ J_z , J_x \} \right) \Biggr]
\label{eq:LKfull001}
\end{eqnarray}
\end{widetext}
and we mention again that $\{A, B\} = (AB + BA)/2$. For details of the calculation, see the SI \cite{supplement}. As expected from symmetry considerations, Eq.~(\ref{eq:LKfull001}) features a $\pi/2$-periodicity, i.e., 
\begin{equation}
H_{\rm LK}^{\rm [001]}(\phi \pm \pi/2) = H_{\rm LK}^{\rm [001]}(\phi). 
\label{eq:PeriodicityHLK001}
\end{equation}
In this work, we are particularly interested in two special cases. If $\phi = 0$, the axes $x, y, z$ of the NW coincide with the crystallographic directions [100], [010], [001], and the LK Hamiltonian of Eq.~(\ref{eq:LKfullMainaxes}) is equivalent to
\begin{eqnarray}
H_{\rm LK}^{\rm [001]}(0) = 
\frac{\hbar^2}{2m}\Biggl[ & &
\left(\gamma_1 + \frac{5 \gamma_2}{2}\right)k^2 
- 2 \gamma_2 \left( k_{x}^2 J_{x}^2 + k_{y}^2 J_{y}^2 + k_{z}^2 J_{z}^2 \right) \nonumber \\
& & - 4 \gamma_3 \left( \{k_{x}, k_{y}\}\{J_{x}, J_{y}\} \mbox{ + c.p.} \right)
\Biggr] .
\label{eq:LKx100y010z001} 
\end{eqnarray}
However, if $\phi = \pi/4$, the axes $x$ and $y$ correspond to the directions [110] and $[\bar{1}10]$, respectively, and Eq.~(\ref{eq:LKfullMainaxes}) is equivalent to
\begin{eqnarray}
H_{\rm LK}^{\rm [001]}(\pi/4) =  \frac{\hbar^2}{2m} & \Biggl[ & 
\left(\gamma_1 + \frac{5 \gamma_2}{2}\right)k^2 - \gamma_2 \left( k_x^2 J_x^2 + k_y^2 J_y^2 + 2 k_z^2 J_z^2 \right)  \nonumber \\
& & - \gamma_2 \left( k_x^2 J_y^2 + k_y^2 J_x^2 + 4 \{ k_x , k_y \} \{ J_x , J_y \} \right) \nonumber \\
& & - 4 \gamma_3 \left( \{ k_y , k_z \} \{ J_y , J_z \} + \{ k_z , k_x \} \{ J_z , J_x \} \right) \nonumber \\
& & - \gamma_3  \left(k_x^2 -k_y^2 \right) \left(J_x^2 - J_y^2 \right) \Biggr] . 
\label{eq:LKx110yM110z001} 
\end{eqnarray}

\subsubsection{Nanowire axis along $[110]$}
\label{secsubsub:ModelAxisZalong110}

When the NW axis $z$ corresponds to the [110] direction, the relations between the basis vectors are
\begin{eqnarray}
\bm{e}_x &=& \bm{e}_{x^\prime} \frac{\sin\xi}{\sqrt{2}} - \bm{e}_{y^\prime} \frac{\sin\xi}{\sqrt{2}} + \bm{e}_{z^\prime} \cos\xi , \label{eq:RelationsAxesMainTextFor110start} \\
\bm{e}_y &=& \bm{e}_{x^\prime} \frac{\cos\xi}{\sqrt{2}} - \bm{e}_{y^\prime} \frac{\cos\xi}{\sqrt{2}} - \bm{e}_{z^\prime} \sin\xi , \\
\bm{e}_z &=& \bm{e}_{x^\prime} \frac{1}{\sqrt{2}} + \bm{e}_{y^\prime} \frac{1}{\sqrt{2}} . \label{eq:RelationsAxesMainTextFor110end}
\end{eqnarray}     
The arbitrary angle is denoted here by $\xi$ in order to avoid confusion with the previously introduced angle $\phi$. The Hamiltonian $H_{\rm LK}^{\rm [110]}(\xi)$, which we obtain by following Appendix~\ref{appsub:TransformationFor110} and rewriting Eq.~(\ref{eq:LKfullMainaxes}), is relatively lengthy and shown explicitly in the SI \cite{supplement}. We wish to mention that the $\pi$-periodicity 
\begin{equation}
H_{\rm LK}^{\rm [110]}(\xi \pm \pi) = H_{\rm LK}^{\rm [110]}(\xi) ,
\end{equation}
which is expected from symmetry considerations, is indeed satisfied and contrasts the $\pi/2$-periodicity of $H_{\rm LK}^{\rm [001]}(\phi)$ for the NWs with $z$ along [001] [see Sec.~\ref{secsubsub:ModelAxisZalong001}, Eqs.~(\ref{eq:LKfull001}) and (\ref{eq:PeriodicityHLK001})]. 

If $\xi = 0$, the axes $x,y,z$ coincide with the directions [001], $[1\bar{1}0]$, [110] and Eq.~(\ref{eq:LKfullMainaxes}) is equivalent to
\begin{eqnarray}
H_{\rm LK}^{\rm [110]}(0) 
= \frac{\hbar^2}{2m} & \Biggl[ & 
\left(\gamma_1 + \frac{5 \gamma_2}{2}\right)k^2 
- \gamma_2 \left( 2 k_x^2 J_x^2 + k_y^2 J_y^2 + k_z^2 J_z^2  \right)  \nonumber \\ & & 
- \gamma_2  \left( k_y^2 J_z^2 + k_z^2 J_y^2 + 4 \{ k_y , k_z \} \{ J_y , J_z \} \right)  \nonumber \\ & &
- 4 \gamma_3 \left( \{ k_x , k_y \} \{ J_x , J_y \} +  \{ k_z , k_x \} \{ J_z , J_x \} \right) \nonumber \\ & &
- \gamma_3 \left( k_y^2 - k_z^2 \right) \left( J_y^2 - J_z^2 \right) 
\Biggr]  .
\label{eq:LKx001y1M10z110}
\end{eqnarray} 
We note that this special case, where $x$ is parallel to a main crystallographic axis and the NW axis is oriented along the [110] direction, is of particular relevance for our discussion because it applies to recently fabricated Si NWs that are based on silicon-on-insulator technology \cite{coquand:ulisproc12, barraud:edl12, voisin:nlt16, maurand:ncomm16}.

\section{Ge and Ge/Si Core/Shell Nanowires}
\label{sec:Ge}

In this section, we present the results for Ge NWs and Ge/Si core/shell NWs.

\subsection{Parameters and static strain}
\label{secsub:GeParameters}

The valence band parameters for Ge are \cite{lawaetz:prb71} $\gamma_1 = 13.35$, $\gamma_2 = 4.25$, $\gamma_3 = 5.69$, and $\kappa = 3.41$. For the coefficient $\alpha_h$ of the standard Rashba SOI, we use \cite{kloeffel:prb11} $\alpha_h = - 0.4\mbox{ nm$^2 e$}$ based on Refs.~\cite{winkler:book, richard:prb04} (see also Appendix~\ref{app:TermsCausedByEFields}). We are particularly interested in NWs with square cross sections, and so
\begin{equation}
L_x = L_y = s 
\end{equation}
is used for the plots in this Sec.~\ref{sec:Ge}, which means that the Ge core has a square cross section with side length $s$. Considering square cross sections is interesting for two reasons. First, the DRSOI is expected to be very strong because the holes are equally confined in two directions, in stark contrast to holes in 2D-like systems (Sec.~\ref{secsubsub:HolesIn2D}). Second, the special case $L_x = L_y$ allows for a reasonable comparison between the newly obtained results and those from previous theoretical studies with circular cross sections \cite{sercel:prb90, csontos:prb09, kloeffel:prb11, kloeffel:prb13, maier:prb13}. More specifically, we will compare previous results for a given core radius $R$ with those for $s = 2 R$ in our model, i.e., $s/2 = R$.

If a Si shell is present, the resulting strain in the Ge core must be taken into account. In the case of cylindrical Ge/Si core/shell NWs, the strain in the Ge core was found to be constant, with the strain tensor elements $\varepsilon_{zz}$ and $\varepsilon_{\perp} = \varepsilon_{xx} = \varepsilon_{yy}$ depending on the relative shell thickness and with $0 = \varepsilon_{xy} = \varepsilon_{xz} = \varepsilon_{yz}$ \cite{menendez:anphbe11, kloeffel:prb14}. Numerical simulations of the strain field profile in core-shell NWs revealed that the core strain remains approximately position-independent (particularly near the core center) when the cross section is hexagonal instead of circular \cite{groenqvist:jap09, hestroffer:nanotech10, hocevar:apl13}. Since NWs with square cross sections also feature a high degree of symmetry and since we study low-energy hole states, which are mostly located near the core center, we believe that the core strain in our model can also be considered as constant, provided that $L_x = L_y$. For the strain tensor elements $\epsilon_{ij}$ in our model, we therefore use the results from Ref.~\cite{kloeffel:prb14}. In the SI \cite{supplement}, we provide detailed information on how the BP Hamiltonian of Eq.~(\ref{eq:BirPikusFull}), which is based on the main crystallographic axes $x^\prime , y^\prime , z^\prime$, is rewritten such that it refers to the axes $x,y,z$ (see Fig.~\ref{fig:SetupWireAndQD}). We note that when the core strain is constant, all spin-independent terms in the BP Hamiltonian only provide a global energy shift in our model and therefore cannot affect the results. Hence, the hydrostatic deformation potential $a$ drops out when we consider a square cross section. The two remaining deformation potentials in the BP Hamiltonian for Ge are \cite{birpikus:book} $b \simeq -2.5\mbox{ eV}$ and $d \simeq -5.0\mbox{ eV}$. 

After an extensive analysis of our results for the two cases $z \parallel [001]$ and $z \parallel [110]$ described in Sec.~\ref{secsub:ModelCoordinateSystems}, using various values for the angles $\phi$ and $\xi$, respectively, we conclude that the orientation of the crystallographic axes has only minor effects on the low-energy hole states in Ge/Si core/shell NWs. This finding is not very surprising, because the small value $(\gamma_3 - \gamma_2)/\gamma_1 = 10.8\%$ indicates that the spherical approximation applies well to Ge \cite{lipari:prl70, winkler:sst08, lawaetz:prb71}. Moreover, the spherical approximation also applies to the BP Hamiltonian, since $d = \sqrt{3}b$ is almost satisfied for the deformation potentials of Ge \cite{birpikus:book}. 

Since we obtain similar results for all orientations of the crystallographic axes, we choose the spherical approximation for the plots in this Sec.~\ref{sec:Ge}. That is, Figs.~\ref{fig:GeSpectrumNoStrainComparison} to \ref{fig:GeSiCoreShellESOmax} are independent of the orientation of the crystallographic axes and they can be interpreted as averaged results that closely resemble the various data sets calculated with nonspherical (cubic) corrections. For Figs.~\ref{fig:GeSpectrumNoStrainComparison} to \ref{fig:GeSiCoreShellESOmax}, we set $b = -2.5\mbox{ eV}$, $d = \sqrt{3}b$, and $\gamma_2 = \gamma_3 = \gamma_s = 5.114$ \cite{footnote:GeLKparams}.

\subsection{Hole spectrum without applied fields}
\label{secsub:GeNWsNoFields}

\begin{figure}[tb]
\begin{center}
\includegraphics[width=0.90\linewidth]{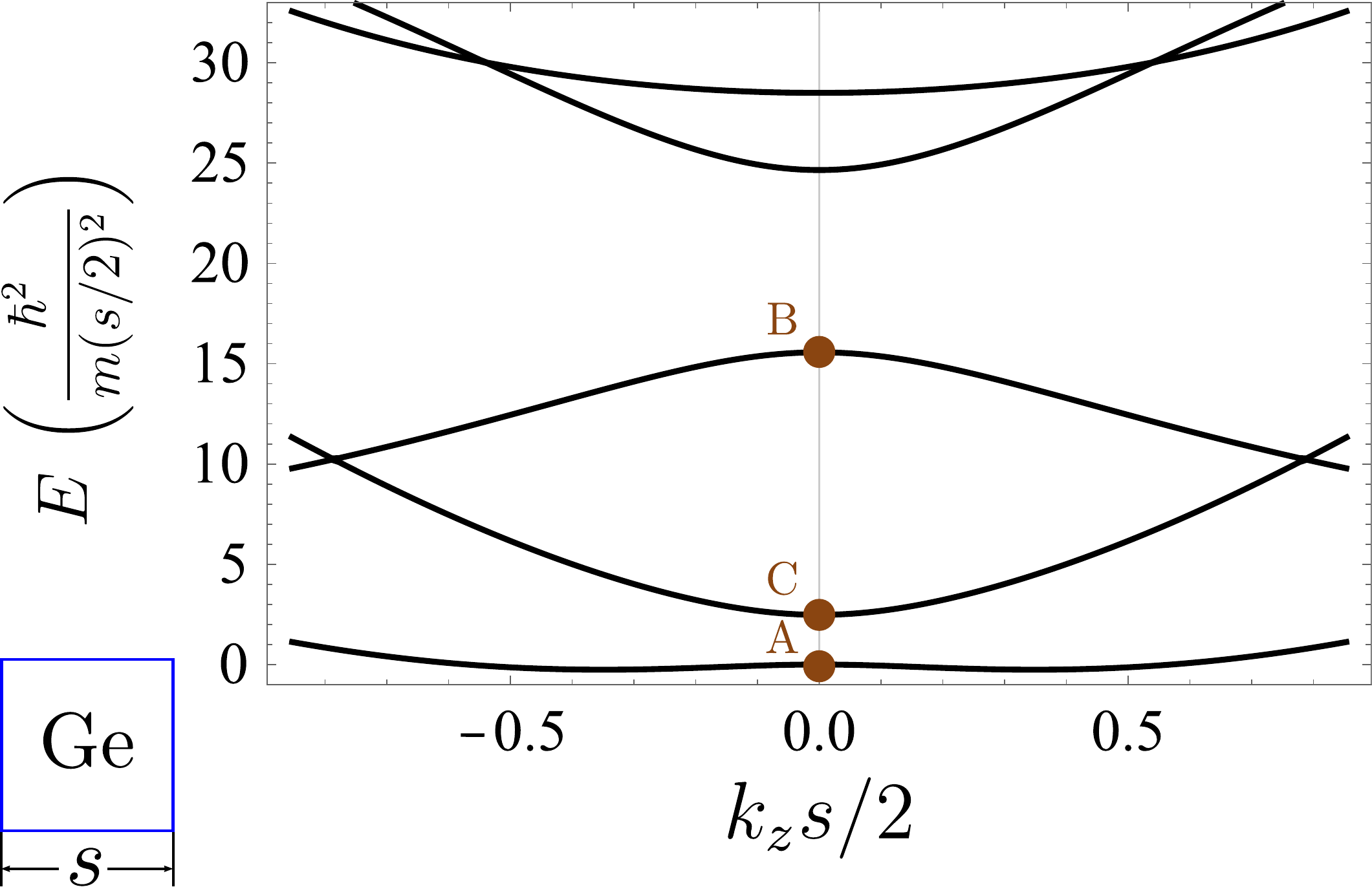} \\ \mbox{ } \\
\includegraphics[width=0.90\linewidth]{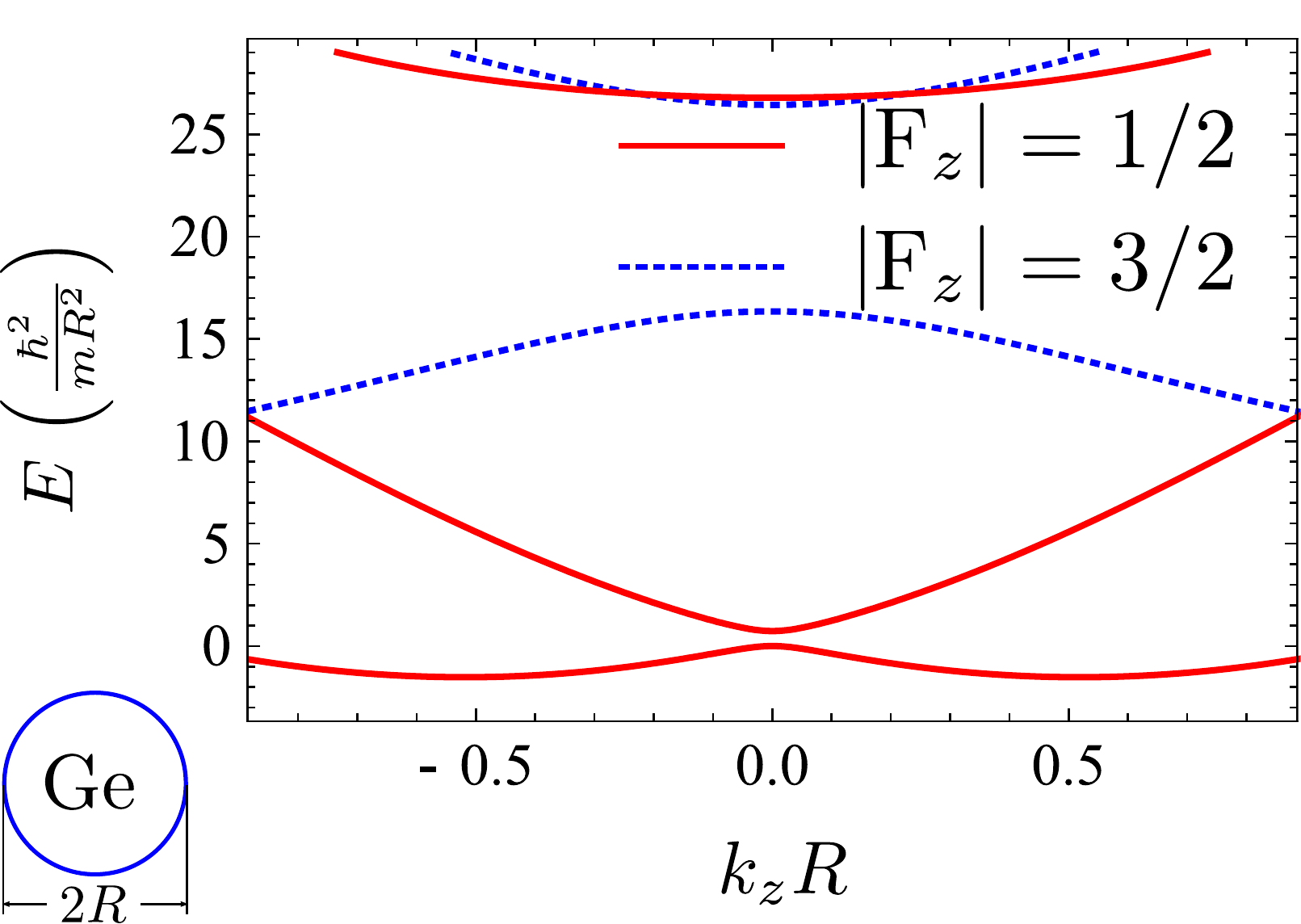}
\caption{Low-energy hole spectrum of an unstrained Ge NW. Each line corresponds to two degenerate subbands. (Top) Spectrum for a square cross section, calculated with the model of Sec.~\ref{sec:Model}. The two eigenstates of type~A consist predominantly of the basis states $\ket{1,1,0,\pm\frac{1}{2}}$ ($96.4\%$). Those of type~B have $\ket{1,1,0,\pm\frac{3}{2}}$ ($51.6\%$) as their largest contribution. For more information on the properties of A, B, and C, we refer to Sec.~\ref{secsub:SiNWsNoFields}. (Bottom) Spectrum for a circular cross section, adapted from Ref.~\cite{kloeffel:prb11}. Due to the cylindrical symmetry, the subbands can be classified via the total angular momentum $F_z$ along the NW axis \cite{sercel:prb90, csontos:prb09}. Despite the different cross sections, the spectra in the upper and lower panel closely resemble each other. Around $k_z = 0$, the effective mass for the degenerate subbands of lowest energy is negative. We note that $\hbar^2/(m R^2) = 3.05\mbox{ meV}$ for $R = 5\mbox{ nm}$. The label $E$ at the vertical axis stands for ``Energy''. }
\label{fig:GeSpectrumNoStrainComparison}
\end{center}
\end{figure}

The upper panel of Fig.~\ref{fig:GeSpectrumNoStrainComparison} shows our simulated hole spectrum of an unstrained Ge NW. The plot is independent of the side length $s$ of the square cross section. In the idealized case of cylindrical symmetry, the hole spectrum of a NW can be calculated exactly \cite{sercel:prb90, csontos:prb09, kloeffel:prb11}. The result for a cylindrical Ge NW, taken from Ref.~\cite{kloeffel:prb11}, is displayed in the lower panel of Fig.~\ref{fig:GeSpectrumNoStrainComparison}. A comparison between the two spectra reveals very good agreement. We note that each line in Fig.~\ref{fig:GeSpectrumNoStrainComparison} is twofold degenerate. In particular, one finds a relatively small energy gap $\Delta$ at $k_z = 0$ between the two ground states and the two excited states with second-lowest energy. Moreover, for small $k_z$ the two degenerate subbands of lowest energy feature a dispersion with negative effective mass. 

In the presence of a Si shell, the abovementioned splitting $\Delta$ increases because of the compressive strain in the Ge core. When $\Delta$ increases, the effective mass for the degenerate subbands of lowest energy changes from negative to positive. This can also be seen in Eq.~(\ref{eq:H2x2effWithDRSOI}), where the term $C^2 / \Delta$ that leads to a negative effective mass decreases with increasing $\Delta$. Thus, one obtains electron-like parabolic spectra in the low-energy regime when the Ge core is sufficiently strained due to a Si shell. This transition is illustrated in the SI \cite{supplement}, where the spectrum for a bare Ge NW is shown next to those for a Ge/Si core/shell NW with increasing shell thickness. 

\begin{figure}[tb]
\begin{center}
\includegraphics[width=0.90\linewidth]{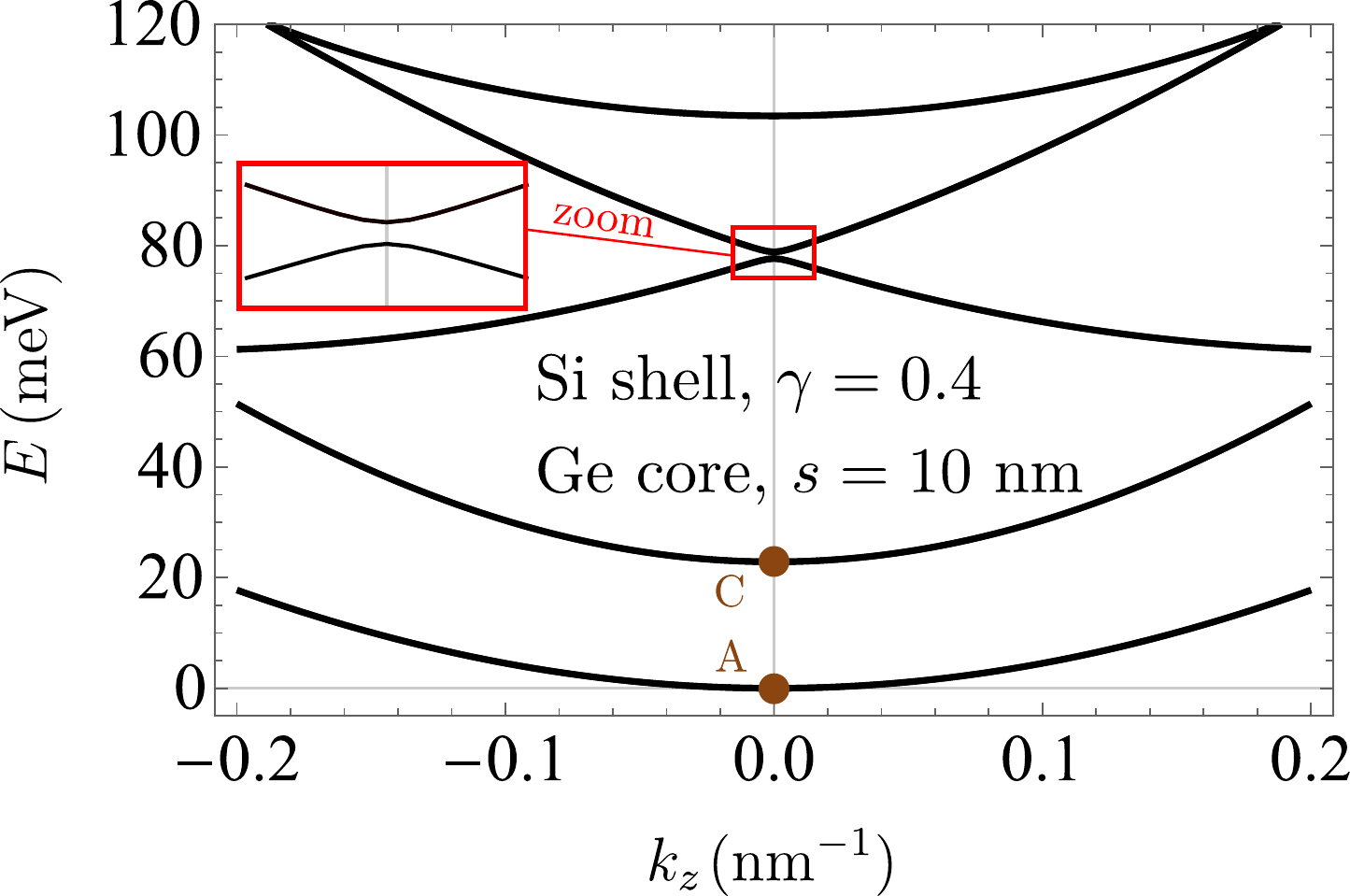}
\caption{Low-energy hole spectrum of a Ge/Si core/shell NW. The cross section of the Ge core is a square with side length $s = 10\mbox{ nm}$ and the relative shell thickness is $\gamma = 0.4$. Each line represents two degenerate subbands. Compared with the case of a bare Ge NW in Fig.~\ref{fig:GeSpectrumNoStrainComparison} (top), the contribution of $\ket{1,1,0,\pm\frac{1}{2}}$ ($97.8\%$) to the A-type eigenstates is greater, which is a consequence of the compressive strain in the Ge core that is caused by the Si shell. Moreover, the strain significantly increases the gap (referred to as $\Delta$ in the text) between the eigenstates of type A and C, leading to an electron-like parabolic dispersion relation with positive effective mass for the subbands of lowest energy. At $k_z = 0$, the two energetically lower states in the enlarged frame do not contain the basis states $\ket{1,1,0,\pm\frac{3}{2}}$ at all. The two energetically higher states originate from the B-type states of Fig.~\ref{fig:GeSpectrumNoStrainComparison}, but the contribution of $\ket{1,1,0,\pm\frac{3}{2}}$ decreased to only $14.5\%$. }
\label{fig:GeSpectrumWithStrain}
\end{center}
\end{figure}

The continuous increase of the energy gap $\Delta$ with increasing Si shell can easily be understood by analyzing the BP Hamiltonian in the spherical approximation $d = \sqrt{3}b$. Because of the strain field profile of the Ge core, the BP Hamiltonian has the simple, effective form \cite{kloeffel:prb11, kloeffel:prb14} $H_{\rm BP} = |b| \left[\varepsilon_{\perp}(\gamma) - \varepsilon_{zz}(\gamma) \right] J_z^2$, where the parameter $\gamma$ is the relative shell thickness of the Ge/Si core/shell NW and where we exploited that $b$ is negative. In the presence of a Si shell ($\gamma > 0$), the difference $\varepsilon_{\perp}(\gamma) - \varepsilon_{zz}(\gamma)$ of the strain tensor elements is positive and increases with increasing $\gamma$. Thus, holes with spin states $\ket{\pm 1/2}$ are energetically favored by $H_{\rm BP}$. Analyzing the ground states at $k_z = 0$ in Fig.~\ref{fig:GeSpectrumNoStrainComparison} reveals that the holes are predominantly found near the core center and that they feature the spin states $\ket{\pm 1/2}$ with very high probability, with only small corrections that involve other spin states \cite{sercel:prb90, csontos:prb09, kloeffel:prb11}. The first excited states at $k_z = 0$ have a higher contribution of $\ket{\pm 3/2}$ than the ground states. This explains why the gap $\Delta$ between the two ground states and the two first excited states is increased by $H_{\rm BP}$ as a result of the shell-induced strain, leading to a positive effective mass in the subbands of lowest energy when the Si shell is sufficiently thick (see also the SI \cite{supplement}). An example is provided in Fig.~\ref{fig:GeSpectrumWithStrain}, where we display the calculated spectrum for a Ge/Si core/shell NW with $s = 10\mbox{ nm}$ and relative shell thickness $\gamma = 0.4$. The latter yields $\varepsilon_{zz} = -21.8 \times 10^{-3}$ and $\varepsilon_{\perp} = -5.8 \times 10^{-3}$ for the core strain \cite{kloeffel:prb14}.     

\subsection{Hole spectrum with applied fields}
\label{secsub:GeNWsWithFields}

\begin{figure}[tb]
\begin{center}
\includegraphics[width=0.90\linewidth]{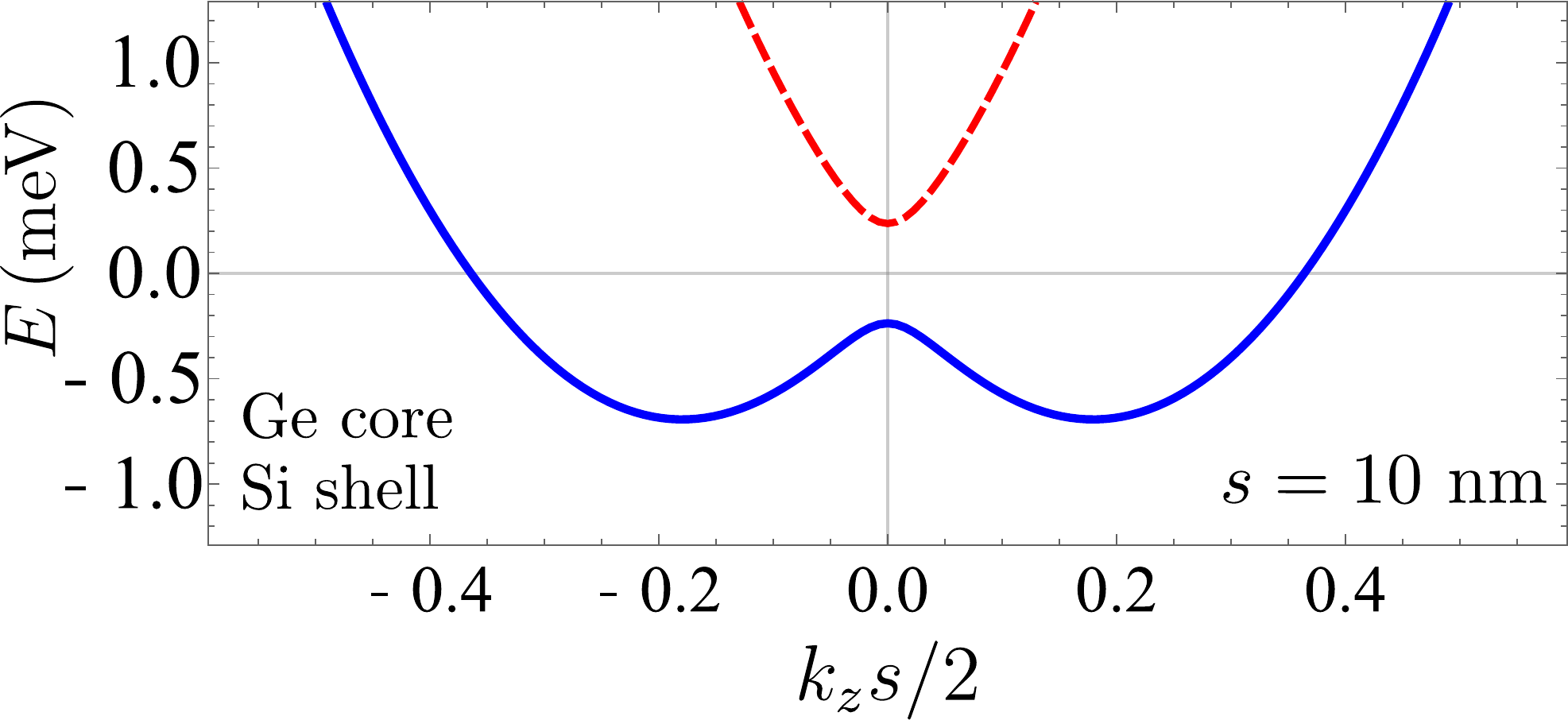} \\ \mbox{ } \\
\includegraphics[width=0.90\linewidth]{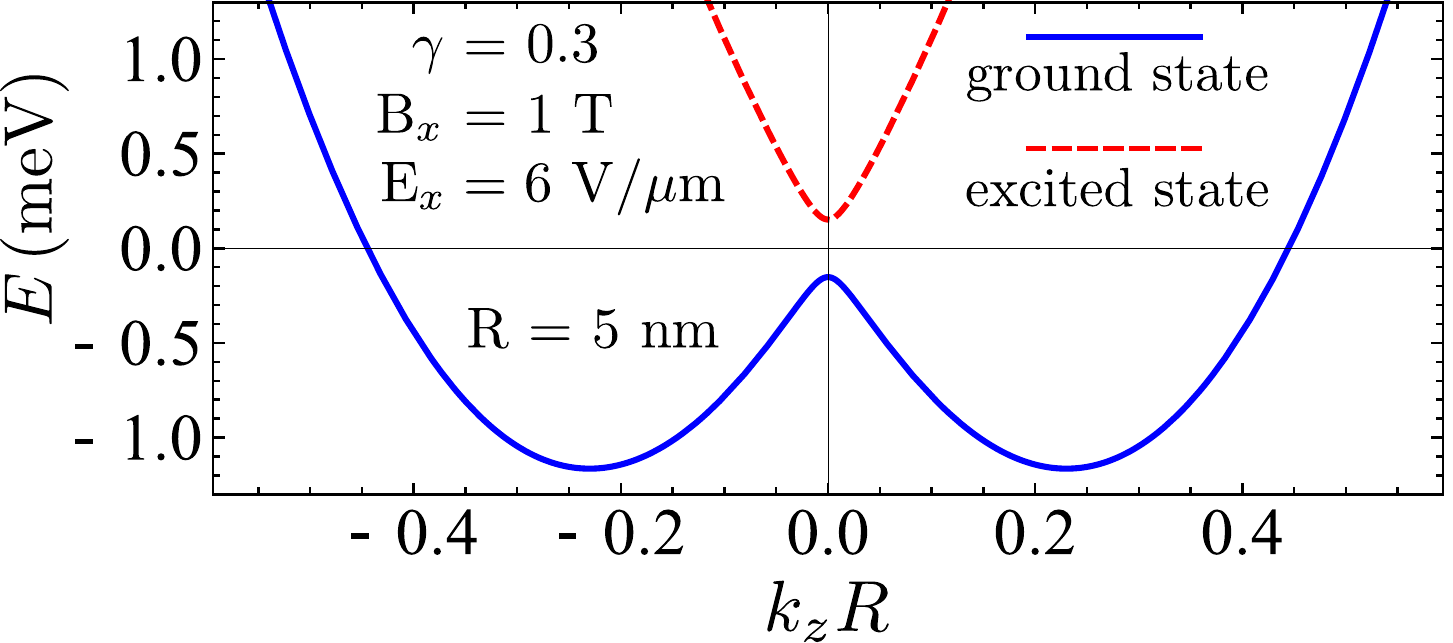}
\caption{Subbands of lowest energy in a Ge/Si core/shell NW with relative shell thickness $\gamma = 0.3$. Each line represents a single subband, because the degeneracy is lifted by an electric field $E_x = 6\mbox{ V/$\mu$m}$ and a magnetic field $B_x = 1\mbox{ T}$ applied along the $x$ axis (see Fig.~\ref{fig:SetupWireAndQD}). The spectrum in the upper panel was calculated with the model of Sec.~\ref{sec:Model} using $s = 10\mbox{ nm}$ (square cross section). The lower panel was adapted from Ref.~\cite{kloeffel:prb11} (circular cross section). In both cases, the spin-orbit energy is approximately one millielectronvolt. The Zeeman gap at $k_z = 0$ corresponds to a $g$ factor of 8.3 (top) and 5.2 (bottom), respectively. When the magnetic field is applied along $z$ instead of $x$, our calculation for the square cross section yields a $g$ factor of 1.8. As expected \cite{kloeffel:prb11, maier:prb13, brauns:prb16}, this value is smaller than the one for the perpendicularly applied magnetic field. }
\label{fig:GeSiCoreShellDRSOIComparison}
\end{center}
\end{figure}

In Figs.~\ref{fig:GeSpectrumNoStrainComparison} and \ref{fig:GeSpectrumWithStrain}, we have not yet included any electric or magnetic fields. In the upper panel of Fig.~\ref{fig:GeSiCoreShellDRSOIComparison}, we show our simulated hole spectrum for the two subbands of lowest energy in a Ge/Si core/shell NW with $s = 10\mbox{ nm}$ and $\gamma = 0.3$. Furthermore, the electric field $E_x = 6\mbox{ V/$\mu$m}$ and the magnetic field $B_x = 1\mbox{ T}$ are applied in the $x$ direction. We observe very good agreement with the lower panel of Fig.~\ref{fig:GeSiCoreShellDRSOIComparison}, which was obtained with the effective model developed in Ref.~\cite{kloeffel:prb11} for cylindrical Ge/Si core/shell NWs, using $R = 5\mbox{ nm}$ and otherwise exactly the same parameters as above. This agreement is of great importance, because it confirms that the DRSOI discovered in Ref.~\cite{kloeffel:prb11} also occurs in NWs with approximately square cross sections and that the derived effective model is consistent with the numerical approach of the present work.

We note that $\gamma$ was defined in Refs.~\cite{kloeffel:prb11, kloeffel:prb14} as the ratio between the shell thickness and the core radius $R$. Thus, $\gamma = 0.3$ corresponds to a rather thin Si shell of $1.5\mbox{ nm}$ thickness when $R = s/2 = 5\mbox{ nm}$ as in Fig.~\ref{fig:GeSiCoreShellDRSOIComparison}. Nevertheless, the core strain $\varepsilon_{zz} = -18.6 \times 10^{-3}$ and $\varepsilon_{\perp} = -4.8 \times 10^{-3}$ associated with $\gamma = 0.3$ \cite{kloeffel:prb14} is already sufficient at $R = s/2 = 5\mbox{ nm}$ to change the effective mass from negative to positive for the subbands of lowest energy, leading to an electron-like parabolic spectrum. The SOI leads to a shift of the two parabolas along the $k_z$ axis, and the associated spin-orbit energy in Fig.~\ref{fig:GeSiCoreShellDRSOIComparison} is approximately $1\mbox{ meV}$, despite the moderate electric field of only a few volts per micrometer. We verified that this strong SOI results from the DRSOI, because the spectra in Fig.~\ref{fig:GeSiCoreShellDRSOIComparison} remain unchanged (apart from negligibly small, quantitative corrections) when we set $\alpha_h = 0$. The magnetic field lifts the degeneracy at $k_z = 0$, with a $g$ factor greater than~5 in Fig.~\ref{fig:GeSiCoreShellDRSOIComparison} \cite{footnote:gfactor}, and so the low-energy hole spectrum in Ge/Si core/shell NWs is very useful, among other things, for the implementation of Majorana fermions \cite{alicea:rpp12, maier:prb14b} and spin filters \cite{streda:prl03}. As we illustrate in the following, even spin-orbit energies that clearly exceed $1\mbox{ meV}$ can be achieved with these NWs.            

In Sec.~\ref{secsubsub:HolesInNWs}, we discussed the 4$\times$4 Hamiltonian of Eq.~(\ref{eq:RecallPRB2011Matrix}), which was derived in Ref.~\cite{kloeffel:prb11} for Ge/Si core/shell NWs with cylindrical symmetry. Starting from this Hamiltonian and considering $|e U E_x / \Delta| \ll 1$ and $|C k_z / \Delta| \ll 1$, i.e., the regime where the splitting $\Delta$ is relatively large, we used perturbation theory and obtained the 2$\times$2 Hamiltonian in Eq.~(\ref{eq:H2x2effWithDRSOI}) for the two subbands of lowest energy. Analogously, we can derive
\begin{equation}
H_{\rm 2x2}^{\rm eff} = \frac{\hbar^2}{2 m_{\rm avg}} k_z^2 + C \tilde{\sigma}_y k_z
\label{eq:H2x2effWithDRSOIStrongEx}
\end{equation}
for the case of a very strong electric field, meaning that $|e U E_x|$ is much greater than all other energies in Eq.~(\ref{eq:RecallPRB2011Matrix}), apart from global energy shifts on the diagonal. In this case, the low-energy eigenstates contain almost equal superpositions of the two basis states $\ket{g_{+}}$ and $\ket{e_{+}}$ or $\ket{g_{-}}$ and $\ket{e_{-}}$. In the derivation of Eq.~(\ref{eq:H2x2effWithDRSOIStrongEx}), we exploited that the masses $m_g$ and $m_e$ are similar for Ge, and replaced both of them by an average effective mass $m_{\rm avg}$ that satisfies
\begin{equation}
\frac{1}{m_{\rm avg}} = \frac{1}{2} \left( \frac{1}{m_g} + \frac{1}{m_e} \right).
\end{equation} 
When we compare Eq.~(\ref{eq:H2x2effWithDRSOIStrongEx}) with the well-known effective Hamiltonian for electrons in Rashba NWs [Eq.~(\ref{eq:H1DForElectronsWithRSOI})], we can identify $\alpha_{\rm DR} = C/E_x$ as the effective Rashba coefficient of the DRSOI. For the spin-orbit length $l_{\rm SO}$ and the spin-orbit energy $E_{\rm SO}$, we obtain \cite{kloeffel:prb11}
\begin{eqnarray}
l_{\rm SO} &=& \frac{\hbar^2}{m_{\rm avg} C} , \\
E_{\rm SO} &=& \frac{m_{\rm avg} C^2}{2 \hbar^2} . 
\label{eq:ESOGeSiEffModelLimitOfStrongEx}
\end{eqnarray}
The latter suggests that
\begin{equation}
E_{\rm SO} = \left(\frac{10\mbox{ nm}}{R}\right)^2 \times 0.96\mbox{ meV}
\label{eq:EstimateESOGeSiNWsLargeE}  
\end{equation}
can be achieved with Ge/Si core/shell NWs, where we recall that $R$ is the radius of the Ge core. 

Equations~(\ref{eq:H2x2effWithDRSOIStrongEx}) to (\ref{eq:EstimateESOGeSiNWsLargeE}) are remarkably simple and depend neither on the splitting $\Delta$ nor on the electric field $E_x$. However, they require that $|e U E_x / \Delta| \gg 1$. It is therefore important to note that the 4$\times$4 Hamiltonian of Eq.~(\ref{eq:RecallPRB2011Matrix}) is reliable only when the subspace spanned by the four basis states $\ket{g_{\pm}}$ and $\ket{e_{\pm}}$ is sufficiently separated from other states. As explained in the SI of Ref.~\cite{kloeffel:prb13}, we estimate that this subspace can be considered as isolated when $|E_x| \lesssim (10\mbox{ nm}/ R)^3 \times 5 \mbox{ V/$\mu$m}$. The decay of this upper bound with $R^{-3}$ can be understood from the $R^{-2}$-type decrease of the level spacings and the proportionality to $E_x R$ of the couplings that are caused by the term $- e E_x x$. With the parameters for Ge/Si core/shell NWs, we consequently find that $|e U E_x / \Delta| > 1$ is accessible within the allowed parameter regime when $R \lesssim 5\mbox{ nm}$ \cite{footnote:estimateStrongEx}. 

\begin{figure}[tb]
\begin{center}
\includegraphics[width=0.895\linewidth]{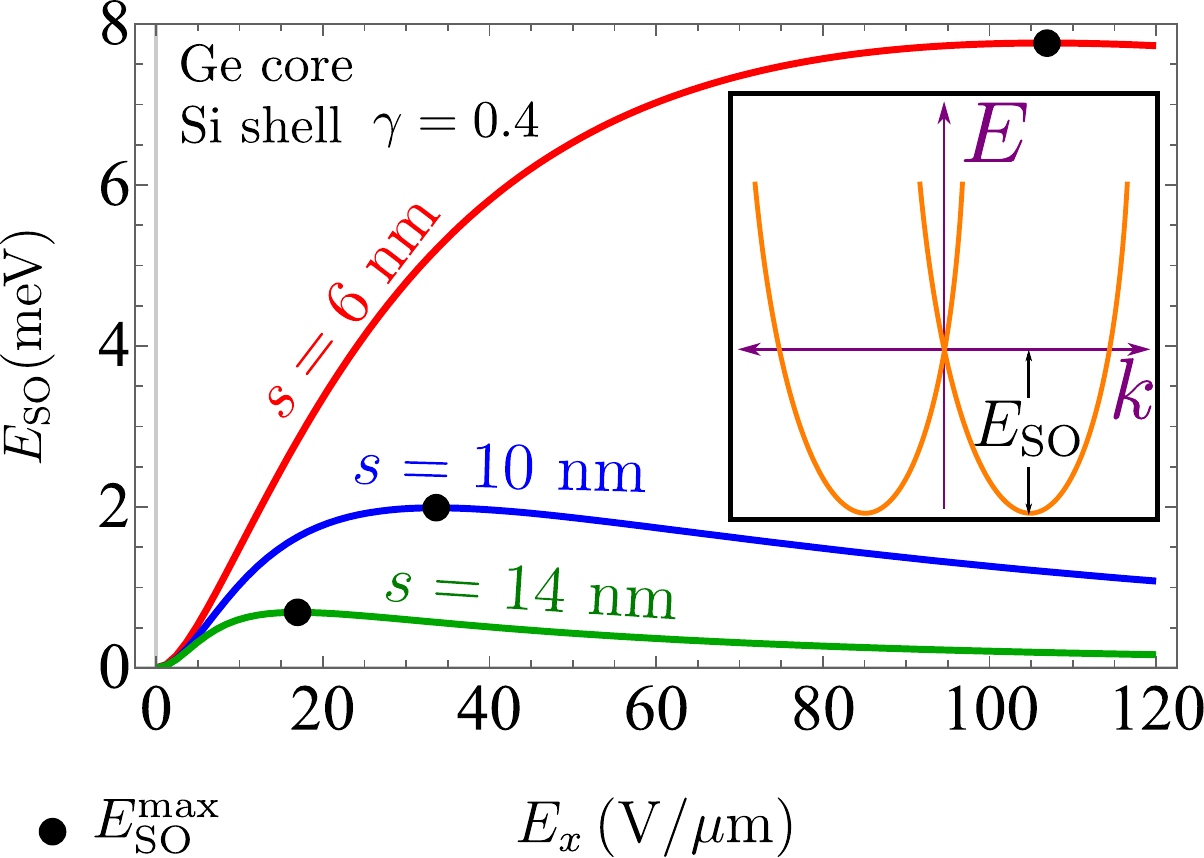}
\caption{Spin-orbit energy $E_{\rm SO}$ as a function of the electric field $E_x$ for three Ge/Si core/shell NWs. The NWs differ in the side length~$s$ of the Ge core, the relative shell thickness is always $\gamma = 0.4$. In each of the three cases, a black dot marks the point where the maximal spin-orbit energy $E_{\rm SO}^{\rm max}$ is reached. The sketched $E$-$k$ diagram in the inset illustrates how we extract the spin-orbit energy $E_{\rm SO}$ from a low-energy hole spectrum. At a given set of parameters, we obtain $E_{\rm SO}$ from the numerically calculated spectrum, similar to that of Fig.~\ref{fig:GeSiCoreShellDRSOIComparison} (top), setting all magnetic fields to zero. In the absence of magnetic fields, there is a degeneracy at $k_z = 0$, as sketched in the diagram. }
\label{fig:GeSiCoreShellESOvsEx}
\end{center}
\end{figure}

\begin{figure}[tb]
\begin{center}
\includegraphics[width=0.895\linewidth]{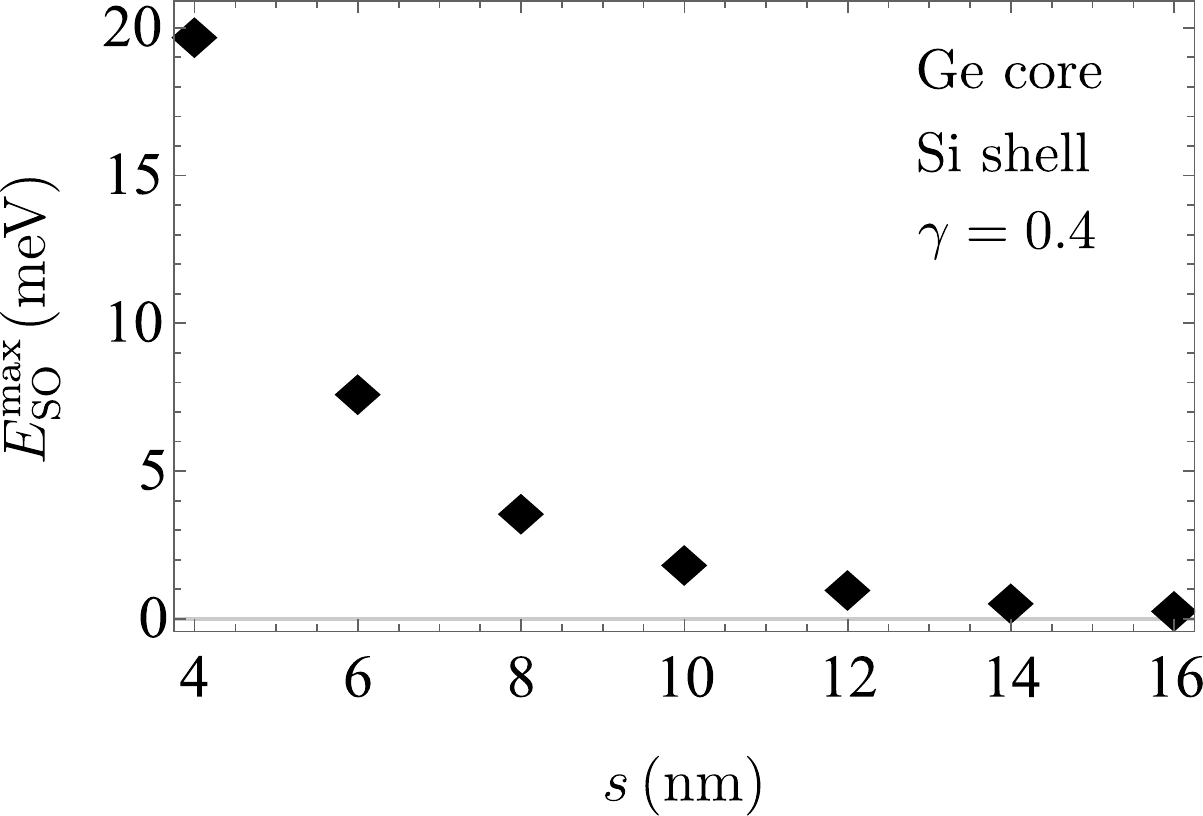}
\caption{Maximal spin-orbit energy $E_{\rm SO}^{\rm max}$ as a function of the side length $s$ for Ge/Si core/shell NWs with relative shell thickness $\gamma = 0.4$. Details for the three data points at $s = 6\mbox{ nm}$, $s = 10\mbox{ nm}$, and $s = 14\mbox{ nm}$ are shown in Fig.~\ref{fig:GeSiCoreShellESOvsEx}. }
\label{fig:GeSiCoreShellESOmax}
\end{center}
\end{figure}

Transferring the abovementioned results to Ge/Si core/shell NWs with square cross sections suggests that for small side lengths $s \lesssim 10\mbox{ nm}$, the spin-orbit energy is approximately constant at strong electric fields, with a value around $E_{\rm SO} = (20\mbox{ nm}/ s)^2 \times 0.96 \mbox{ meV}$. In Fig.~\ref{fig:GeSiCoreShellESOvsEx}, we plot the numerically calculated spin-orbit energy $E_{\rm SO}$ as a function of the applied electric field $E_x$ for the three examples $s = 6\mbox{ nm}$, $s = 10\mbox{ nm}$, and $s = 14\mbox{ nm}$. The relative shell thickness is always $\gamma = 0.4$. In every case, $E_{\rm SO}$ first increases rapidly with increasing $E_x$, then reaches a maximum value $E_{\rm SO}^{\rm max}$, and finally decays slowly when $E_x$ is further increased. However, as evident from Fig.~\ref{fig:GeSiCoreShellESOvsEx}, reaching $E_{\rm SO}^{\rm max}$ in thin NWs (small $s$) requires a stronger electric field than in thicker NWs (larger $s$). Moreover, also the achievable spin-orbit energies depend strongly on the size of the NW. Figure~\ref{fig:GeSiCoreShellESOmax} shows the obtained values for $E_{\rm SO}^{\rm max}$ as a function of $s$. 

The comparison with the effective model for cylindrically symmetric Ge/Si core/shell NWs reveals good agreement. First, we find from Fig.~\ref{fig:GeSiCoreShellESOvsEx} that $E_{\rm SO}^{\rm max}$ is approximately reached when $E_x$ is chosen such that the ratio $|e U E_x / \Delta|$ is of the order of one, using $R = s/2$ for the estimate. Second, the decay after $E_{\rm SO}^{\rm max}$ is relatively slow, particularly for NWs with a small $s$. Third, the values of $E_{\rm SO}^{\rm max}$ at small $s$ in Fig.~\ref{fig:GeSiCoreShellESOmax} scale approximately with $s^{-2}$. When we consider a Ge core with $s = 6\mbox{ nm}$, the abovementioned term $(20\mbox{ nm}/ s)^2 \times 0.96 \mbox{ meV}$ yields a spin-orbit energy of $10.7\mbox{ meV}$, which agrees well with the simulated $E_{\rm SO}^{\rm max} = 7.8\mbox{ meV}$.        

We note that the decrease of $E_{\rm SO}$ in Fig.~\ref{fig:GeSiCoreShellESOvsEx} towards zero at very strong $E_x$ is not reproduced by the effective model for cylindrical Ge/Si core/shell NWs. According to our estimates, more than the four basis states $\ket{g_{\pm}}$ and $\ket{e_{\pm}}$ would have to be included in this regime. Similarly, also the model of Sec.~\ref{sec:Model} will lose validity when the electric field becomes so strong that the considered number of basis states in our numerical approach is insufficient. It is therefore not surprising that, in the regime of strong electric fields, our numerical results and the 4$\times$4 model for cylindrical NWs eventually deviate from each other. At these electric fields, more research about the hole states and their SOI is needed for reliable predictions, as explained in Sec.~\ref{sec:Accuracy}.

\subsection{Results beyond the spherical approximation}
\label{secsub:GeNWsBeyondSpherApprox}

Finally, we want to comment on effects of the growth direction. When we consider a bare Ge NW with square cross section and calculate the hole spectrum with $\gamma_2 = 4.25$ and $\gamma_3 = 5.69$ \cite{lawaetz:prb71} instead of the spherical approximation $\gamma_2 = \gamma_3$, the key features in the low-energy regime of Fig.~\ref{fig:GeSpectrumNoStrainComparison} (top) are preserved. That is, the splitting $\Delta$ between the eigenstates of type~A and~C at $k_z = 0$ is relatively small and the effective mass for the two degenerate subbands of lowest energy is negative. Varying the orientation of the crystallographic axes leads here to rather large quantitative differences. For the three special cases mentioned in Sec.~\ref{secsub:ModelCoordinateSystems}, i.e., $x \parallel \mbox{[100]}$ and $z \parallel \mbox{[001]}$, $x \parallel \mbox{[110]}$ and $z \parallel \mbox{[001]}$, $x \parallel \mbox{[001]}$ and $z \parallel \mbox{[110]}$, we obtain the effective masses $-0.45 m$, $-0.034 m$, $-0.13 m$, respectively. The results for $\Delta$ in units of $\hbar^2/[m (s/2)^2]$ are 5.6, 2.8, and 2.0. Next, we consider the Ge/Si core/shell NWs. We list the results that we obtain by recalculating the data of Fig.~\ref{fig:GeSiCoreShellDRSOIComparison} (top), which shows the lowest-energy subbands of a Ge/Si core/shell NW with $\gamma = 0.3$, $s = 10\mbox{ nm}$, $B_x = 1\mbox{ T}$, and $E_x = 6\mbox{ V/$\mu$m}$. For the $g$~factor at $k_z = 0$, we get 7.3, 8.8, and 6.3, respectively, when we use $\gamma_{2,3}$ as above and otherwise unchanged parameters. The effective mass, calculated at $B_x = E_x = 0$, is $0.11 m$, $0.21 m$, and $0.081 m$. The spin-orbit energy is $0.34\mbox{ meV}$, $1.3\mbox{ meV}$, and $0.31\mbox{ meV}$. Thus, although these spin-orbit energies differ by less than a factor of five, we note that the largest of the three is obtained when $x \parallel \mbox{[110]}$ and $z \parallel \mbox{[001]}$. In Secs.~\ref{sec:Si} and~\ref{sec:AnalyticalResultsSiNWs}, we will show that this orientation is a particularly promising choice for Si NWs.

\section{Si Nanowires}
\label{sec:Si}

Given the agreement in Sec.~\ref{sec:Ge} between our new approach and previous theoretical results for Ge and Ge/Si core/shell NWs, we now use our model from Sec.~\ref{sec:Model} to analyze the hole states in Si NWs. Importantly, a model as in Ref.~\cite{kloeffel:prb11} no longer applies, because in Si the Luttinger parameter $\gamma_3$ is approximately four times greater than $\gamma_2$ \cite{lawaetz:prb71}. That is, the hole spectrum around the $\Gamma$ point is highly anisotropic for bulk Si, and so a spherical approximation \cite{lipari:prl70, winkler:sst08} does not apply.

\subsection{Parameters}
\label{secsub:SiParameters}

In our calculations for Si NWs, we use \cite{lawaetz:prb71} $\gamma_1 = 4.22$, $\gamma_2 = 0.39$, $\gamma_3 = 1.44$, and we note that these numbers agree well with those provided elsewhere \cite{winkler:book, adachi:properties, richard:prb04}. The reported values for $\kappa$ (e.g., $\kappa = -0.42$ \cite{winkler:book}) differ more than those for the $\gamma_{i}$, and we set $\kappa = -0.26$ \cite{lawaetz:prb71} in our simulations. We recall that the choice for $\kappa$ only matters when the effects of a magnetic field are studied. As described in Appendix~\ref{app:TermsCausedByEFields}, we use $\alpha_h = 0.002\mbox{ nm$^2 e$}$ based on Refs.~\cite{winkler:book, richard:prb04}, and so the standard Rashba coefficient for holes in Si is extremely small. 

Unless stated otherwise we use $L_x = L_y = s$ in this section. Hence, analogous to the case of Ge/Si core/shell NWs (Sec.~\ref{sec:Ge}), we focus here on setups where the Si core has a square cross section. However, in contrast to Ge/Si core/shell NWs, the materials that surround recently fabricated Si NWs \cite{voisin:nlt16, maurand:ncomm16, crippa:arX1710, coquand:ulisproc12, barraud:edl12, prati:nanotech12} usually do not lead to considerable strain in the Si core, and so we treat the Si NWs as unstrained.

\subsection{Hole spectrum without applied fields}
\label{secsub:SiNWsNoFields}

\begin{figure}[htbp]
\begin{center}
\includegraphics[width=0.90\linewidth]{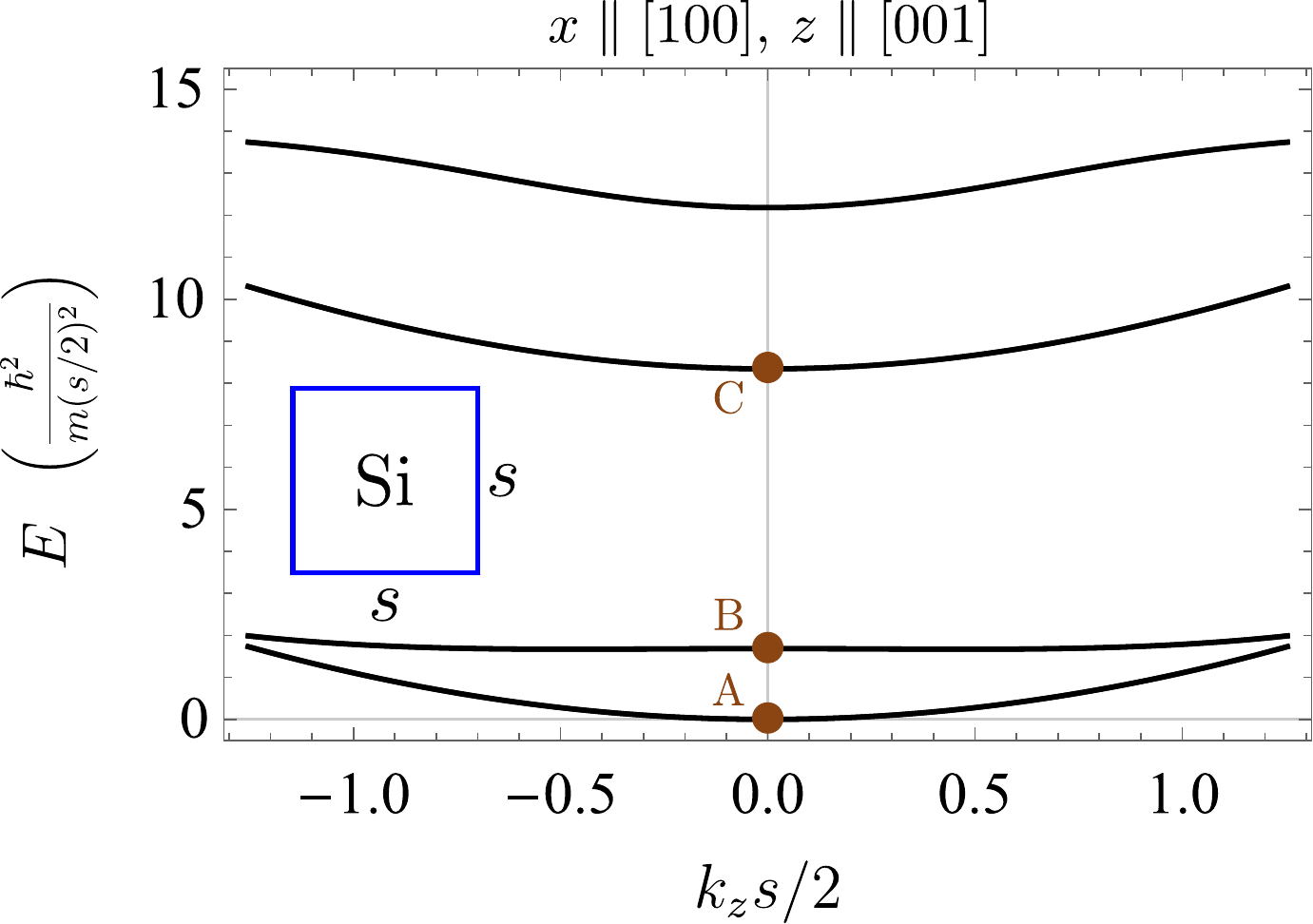} \\ \mbox{ } \\
\includegraphics[width=0.90\linewidth]{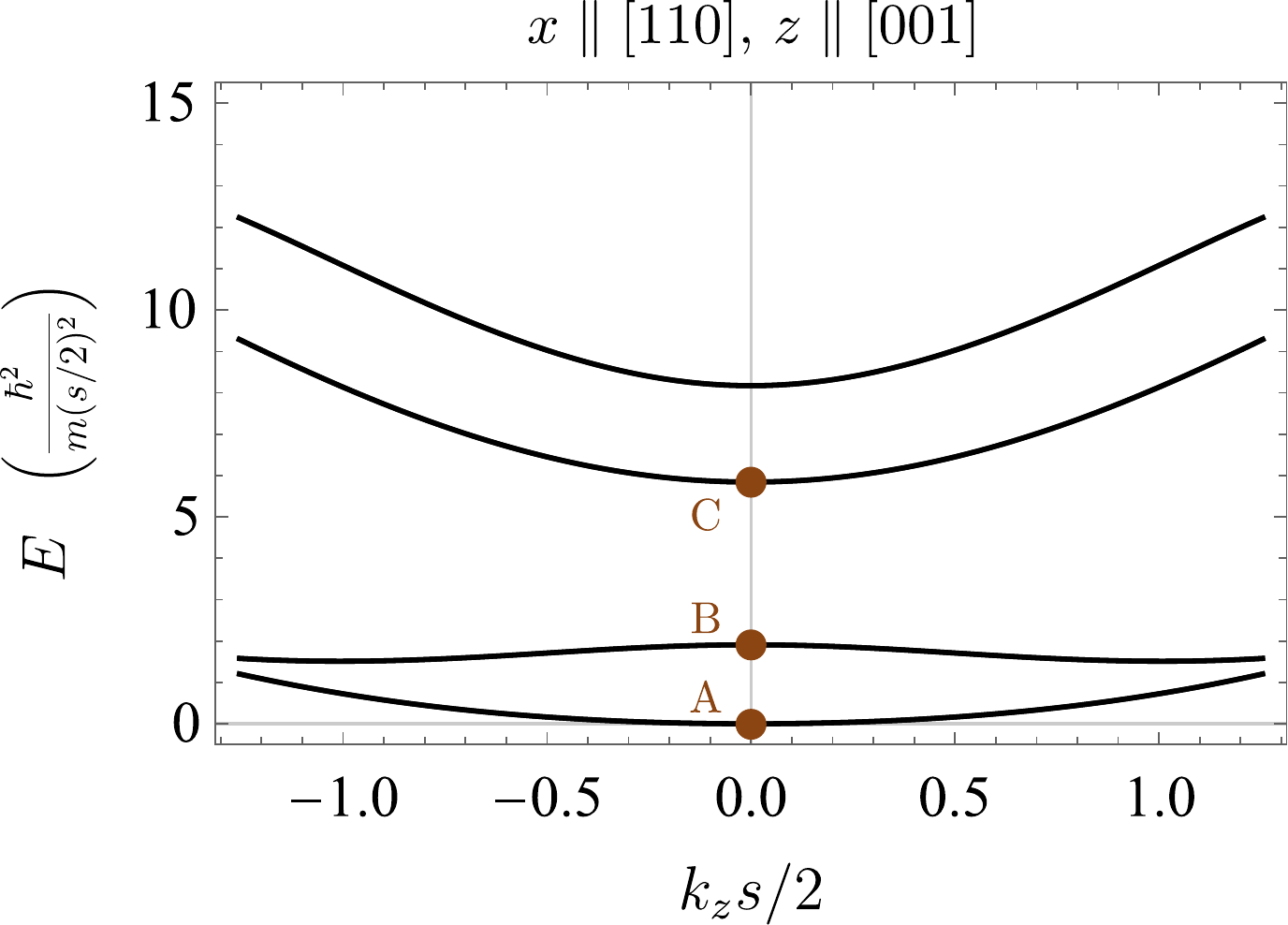} \\ \mbox{ } \\
\includegraphics[width=0.90\linewidth]{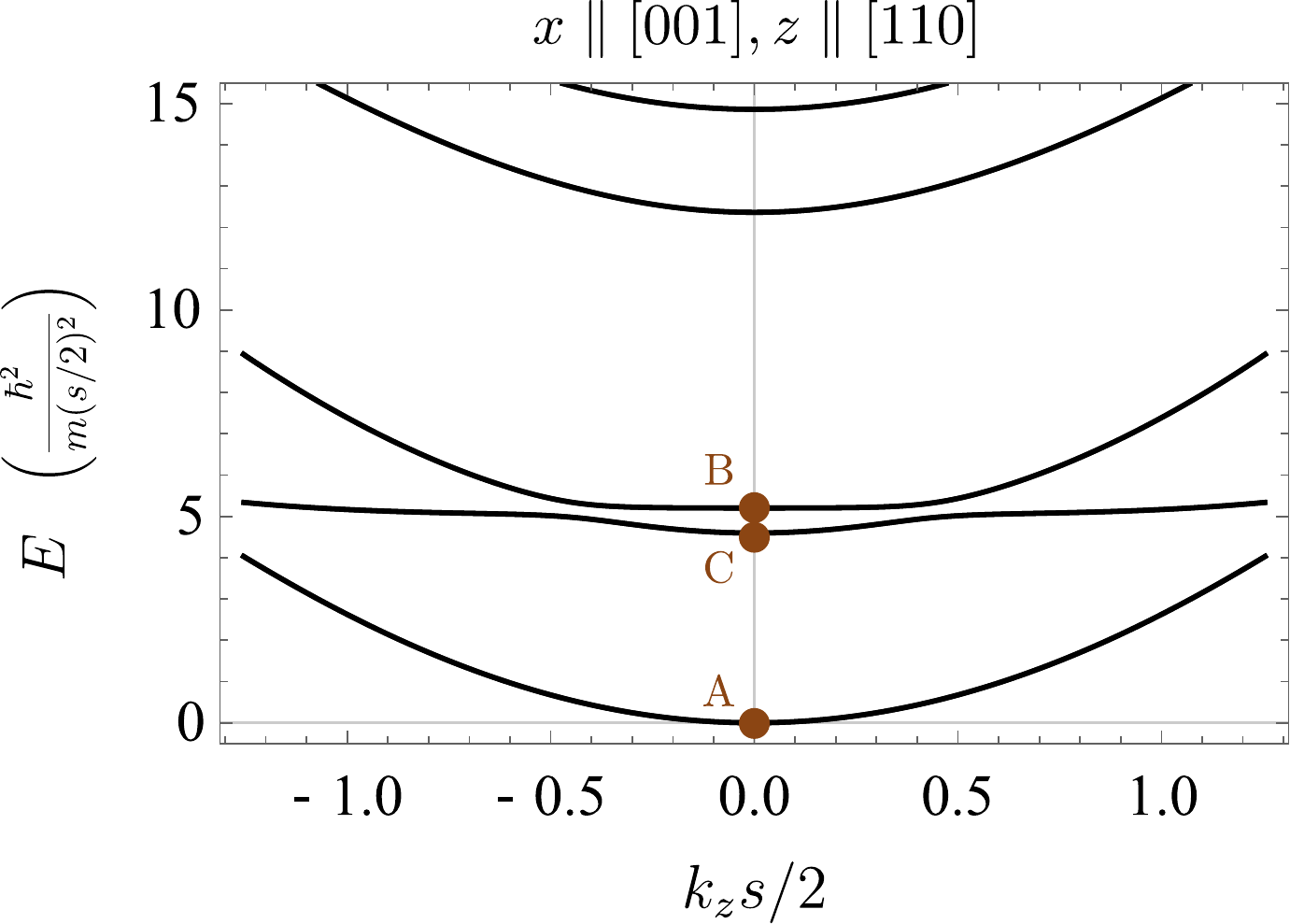}
\caption{Low-energy hole spectra of Si NWs with different orientations of the crystallographic axes (see labels at the top of each panel and Fig.~\ref{fig:SetupWireAndQD} for an illustration of the axes $x$ and $z$). In each case, the cross section of the NW is a square with side length $s$. The spectra in the top, middle and bottom panel were calculated with $\phi = 0$, $\phi = \pi/4$, and $\xi = 0$, respectively (see Secs.~\ref{secsubsub:ModelAxisZalong001} and \ref{secsubsub:ModelAxisZalong110}). Every line corresponds to two degenerate subbands. The marked eigenstates of type A, B, and C are discussed in Sec.~\ref{secsub:SiNWsNoFields}. The relative contribution of the basis states $\ket{1,1,0,\pm \frac{1}{2}}$ to the two eigenstates of type A is $97.7\%$ (top), $99.9\%$ (middle), and $95.0\%$ (bottom), respectively. That of $\ket{1,1,0,\pm \frac{3}{2}}$ to the two eigenstates of type B is $95.9\%$ (top), $99.8\%$ (middle), and $85.5\%$ (bottom). Despite the absence of strain, the effective mass of the lowest-energy subbands is always positive. }
\label{fig:SiSpectraNoFields}
\end{center}
\end{figure}

Since $\gamma_2$ and $\gamma_3$ differ greatly in Si, the orientation of the crystallographic axes with respect to the setup (Fig.~\ref{fig:SetupWireAndQD}) strongly affects the hole spectrum in the NW. This is illustrated in Fig.~\ref{fig:SiSpectraNoFields}, where we plot the low-energy hole spectrum for three different cases. Each line in Fig.~\ref{fig:SiSpectraNoFields} is twofold degenerate. Despite the substantial differences between these three spectra, they all exhibit an important common feature. In stark contrast to unstrained Ge NWs (Fig.~\ref{fig:GeSpectrumNoStrainComparison}), the two degenerate subbands of lowest energy always show an electron-like, parabolic dispersion relation with a positive effective mass, even though the Si NW is unstrained. However, the value of this effective mass strongly depends on the orientation of the crystallographic axes. 

We obtain the abovementioned effective mass by fitting the function $\hbar^2 k_z^2 / (2 m_{\rm eff})$ to the degenerate subbands of lowest energy. When the $z$ axis, i.e., the NW axis, corresponds to the $[001]$ direction, we refer to the fitted mass as $m_{\rm fit}^{[001]}(\phi)$, where $\phi$ is the angle described in Sec.~\ref{secsubsub:ModelAxisZalong001}. Analogously, we use the notation $m_{\rm fit}^{[110]}(\xi)$ when $z$ coincides with $[110]$, with $\xi$ as in Sec.~\ref{secsubsub:ModelAxisZalong110}. For the three cases shown in Fig.~\ref{fig:SiSpectraNoFields}, we obtain $m_{\rm fit}^{[001]}(0) = 0.446 m$ (top panel, $x \parallel [100]$, $z \parallel [001]$), $m_{\rm fit}^{[001]}(\pi/4) = 0.794 m$ (middle panel, $x \parallel [110]$, $z \parallel [001]$), and $m_{\rm fit}^{[110]}(0) = 0.184 m$ (bottom panel, $x \parallel [001]$, $z \parallel [110]$).

The low-energy eigenstates at $k_z = 0$ in the spectra of Fig.~\ref{fig:SiSpectraNoFields} may be grouped into three types, which we briefly describe in the following. The two eigenstates of type~A closely resemble the basis states $\ket{1,1,0,\pm \frac{1}{2}}$ (see Sec.~\ref{secsub:BasisStatesNumDiag} for the definition of the basis states). Thus, with a high probability the holes are found near the core center and with the spin states $\ket{\pm 1/2}$. Similarly, the two eigenstates of type~B consist predominantly of $\ket{1,1,0,\pm\frac{3}{2}}$. The two eigenstates of type~C mainly contain basis states of type $\ket{1,2,0,j_z}$ and $\ket{2,1,0,j_z}$. At least approximately, they can be considered as eigenstates with total angular momenta $\pm 1/2$ (in units of $\hbar$) and nonzero orbital angular momenta along~$z$. 

In our calculated spectra for bare Ge NWs (Fig.~\ref{fig:GeSpectrumNoStrainComparison}, top) and Si NWs with $z \parallel [110]$ (Fig.~\ref{fig:SiSpectraNoFields}, bottom), the two ground states at $k_z = 0$ are of type~A, the states with second-lowest energy are of type~C, and those with third-lowest energy are of type~B. These properties were also found previously in a calculation for GaAs NWs with hard-wall confinement \cite{csontos:prb09}. We note in passing that the basis states $\ket{g_\pm}$ and $\ket{e_\pm}$ of the effective model for Ge/Si core/shell NWs \cite{kloeffel:prb11, kloeffel:prb13} (Sec.~\ref{secsubsub:HolesInNWs}) resemble states of type A and C, respectively. Remarkably, the orientation of the crystallographic axes in Si NWs even affects the order of the subbands. While in Fig.~\ref{fig:SiSpectraNoFields} the ground states are always of type~A, it turns out that the eigenstates of second-lowest (third-lowest) energy at $k_z = 0$ are of type B (C) when $z \parallel [001]$, which contrasts the spectrum in the bottom panel for $z \parallel [110]$ where B and C are swapped. The observed change in the order of the subbands can be understood as follows. The energy gap at $k_z = 0$ between the eigenstates of type A and B is largely determined by the splitting $\bra{1,1,0,\pm \frac{3}{2}} H_{\rm LK} \ket{1,1,0,\pm \frac{3}{2}} - \bra{1,1,0,\pm \frac{1}{2}} H_{\rm LK} \ket{1,1,0,\pm \frac{1}{2}}$, which yields $2 \gamma_2 \hbar^2 \pi^2 / (s^2 m)$ if $z \parallel [001]$. If $z \parallel [110]$, this splitting changes by the factor $(3 \gamma_3 + \gamma_2)/(4 \gamma_2)$. For the parameters of Si, this corresponds to an approximately threefold increase of the gap between eigenstates of type A and B, which is consistent with the numerical data of Fig.~\ref{fig:SiSpectraNoFields}.

\subsection{Hole spectrum with applied fields}
\label{secsub:SiNWsWithFields}

\begin{figure}[tb]
\begin{center}
\includegraphics[width=0.90\linewidth]{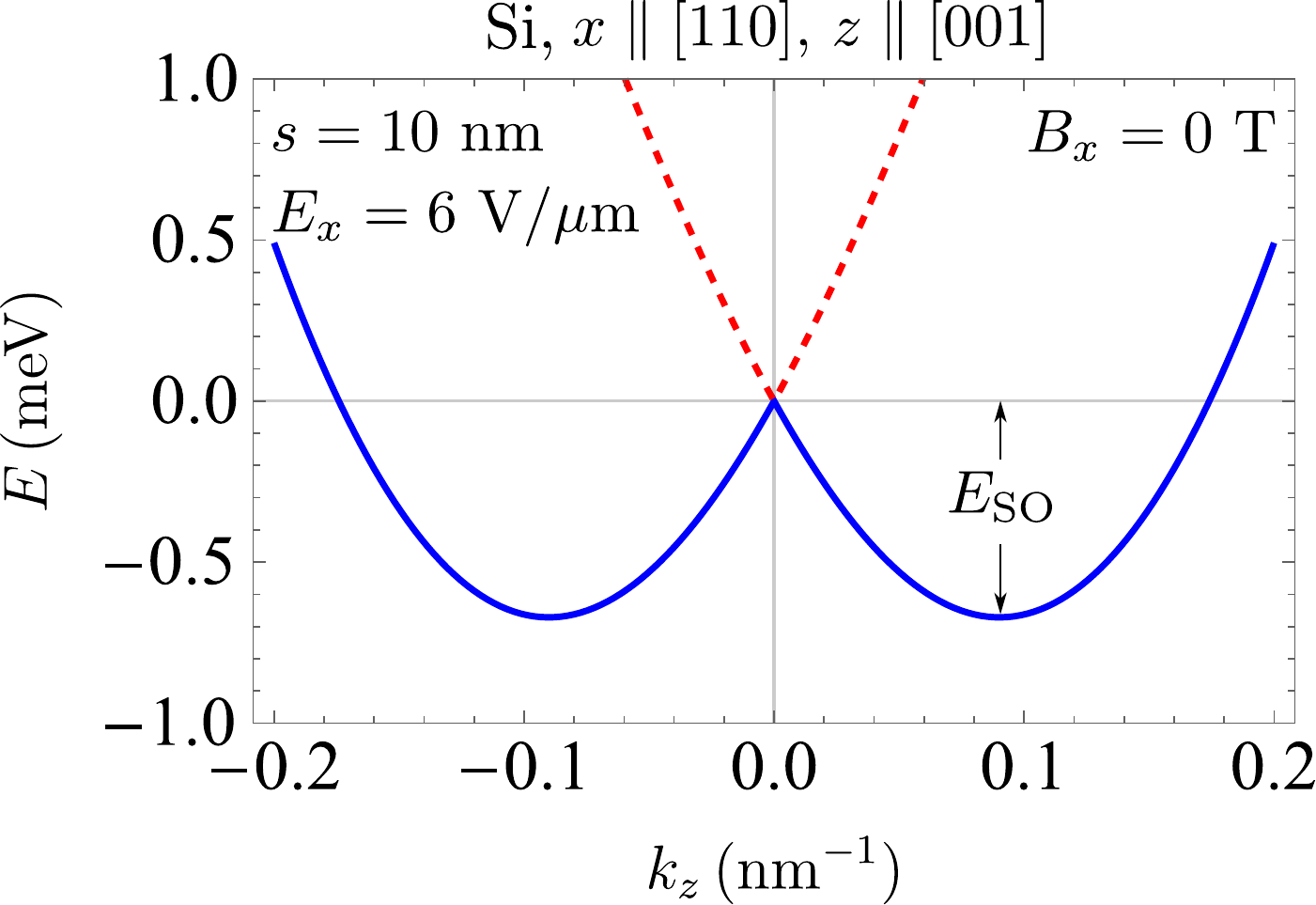} \\ \mbox{ } \\
\includegraphics[width=0.90\linewidth]{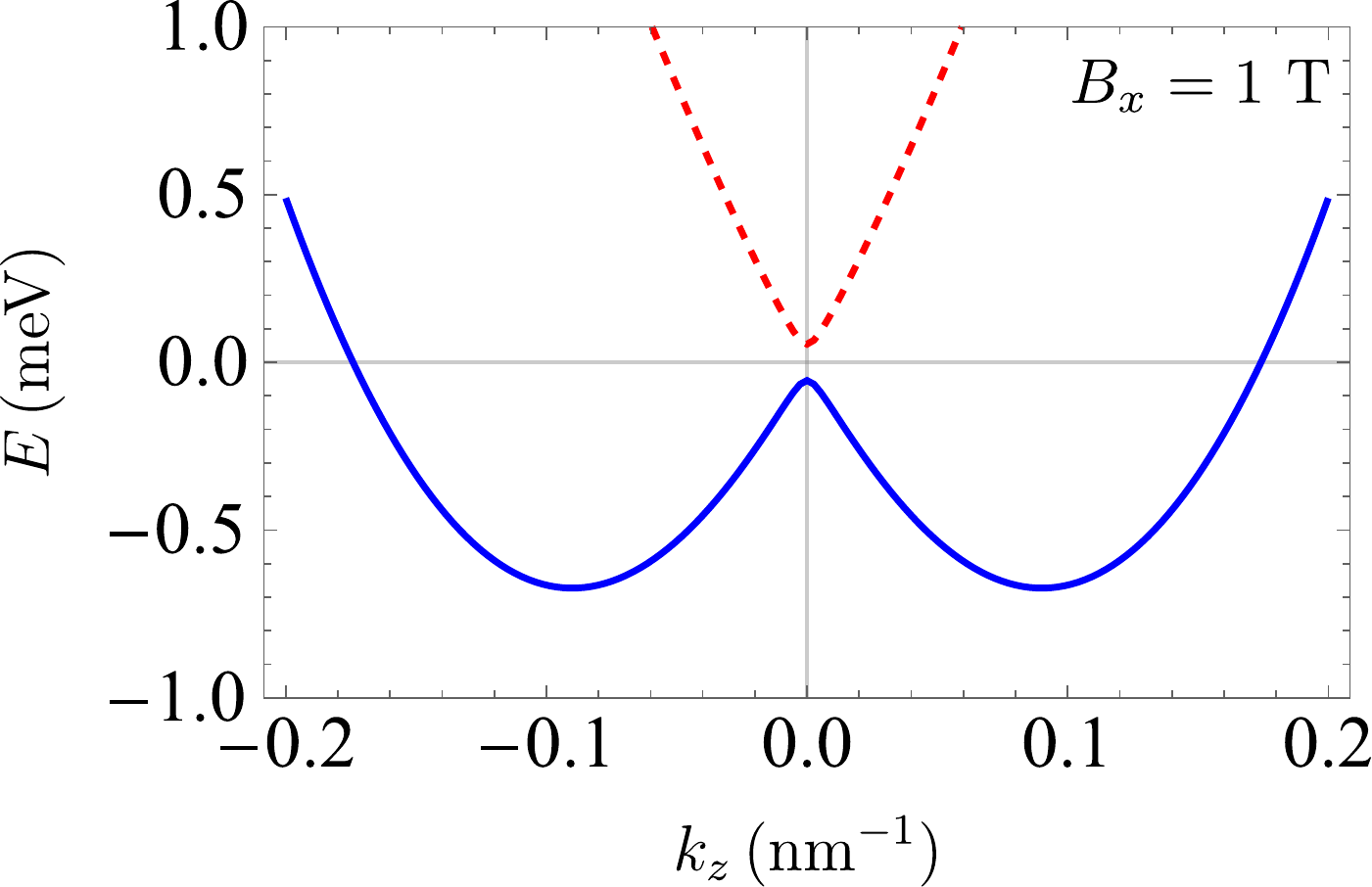}
\caption{Subbands of lowest energy in a Si NW with side length $s = 10\mbox{ nm}$. An electric field $E_x = 6\mbox{ V/$\mu$m}$ is applied along the $x$ axis (see Fig.~\ref{fig:SetupWireAndQD}). The plotted spectra were calculated with $\phi = \pi/4$, i.e., the $x$ axis corresponds to the [110] direction and the NW axis $z$ corresponds to the [001] direction. (Top) The electric field leads to a SOI and shifts the two originally degenerate subbands in opposite directions along the $k_z$ axis. This lifts the degeneracy, except at $k_z = 0$. Ground (excited) states are represented by the solid blue (dashed red) lines. The spin-orbit energy $E_{\rm SO}$ is of the order of one millielectronvolt. (Bottom) A magnetic field $B_x$ along $x$ opens a Zeeman gap at $k_z = 0$, with a $g$ factor of approximately 2.   }
\label{fig:SiExamplesSpectrum}
\end{center}
\end{figure}

The orientation of the crystallographic axes is of particular importance for the SOI. In fact, we find that the SOI in Si NWs can be strong. As an example, the upper panel of Fig.~\ref{fig:SiExamplesSpectrum} shows the calculated spectrum for the two subbands of lowest energy in a Si NW with $x \parallel [110]$, $z \parallel [001]$, and $s = 10\mbox{ nm}$. A moderate electric field $E_x = 6\mbox{ V/$\mu$m}$ is sufficient for a spin-orbit energy close to one millielectronvolt. As illustrated in the lower panel of Fig.~\ref{fig:SiExamplesSpectrum}, an additional magnetic field $B_x = 1\mbox{ T}$ leads to a Zeeman gap of \mbox{0.11 meV} at $k_z = 0$, which corresponds to an effective $g$ factor of~1.9 \cite{footnote:gfactor}. When the magnetic field is applied along $z$ instead of $x$, we obtain a $g$~factor of about~0.6. (With $\kappa = -0.42$ \cite{winkler:book} instead of $\kappa = -0.26$ \cite{lawaetz:prb71}, the results are $g \simeq 2.7$ and $g \simeq 0.8$, respectively.) The possibility to open a gap at $k_z = 0$ in the spectrum is important, e.g., for the implementation of spin filters \cite{streda:prl03} and Majorana fermions~\cite{alicea:rpp12, lutchyn:arX1707}.    

\begin{figure}[tb]
\begin{center}
\includegraphics[width=0.90\linewidth]{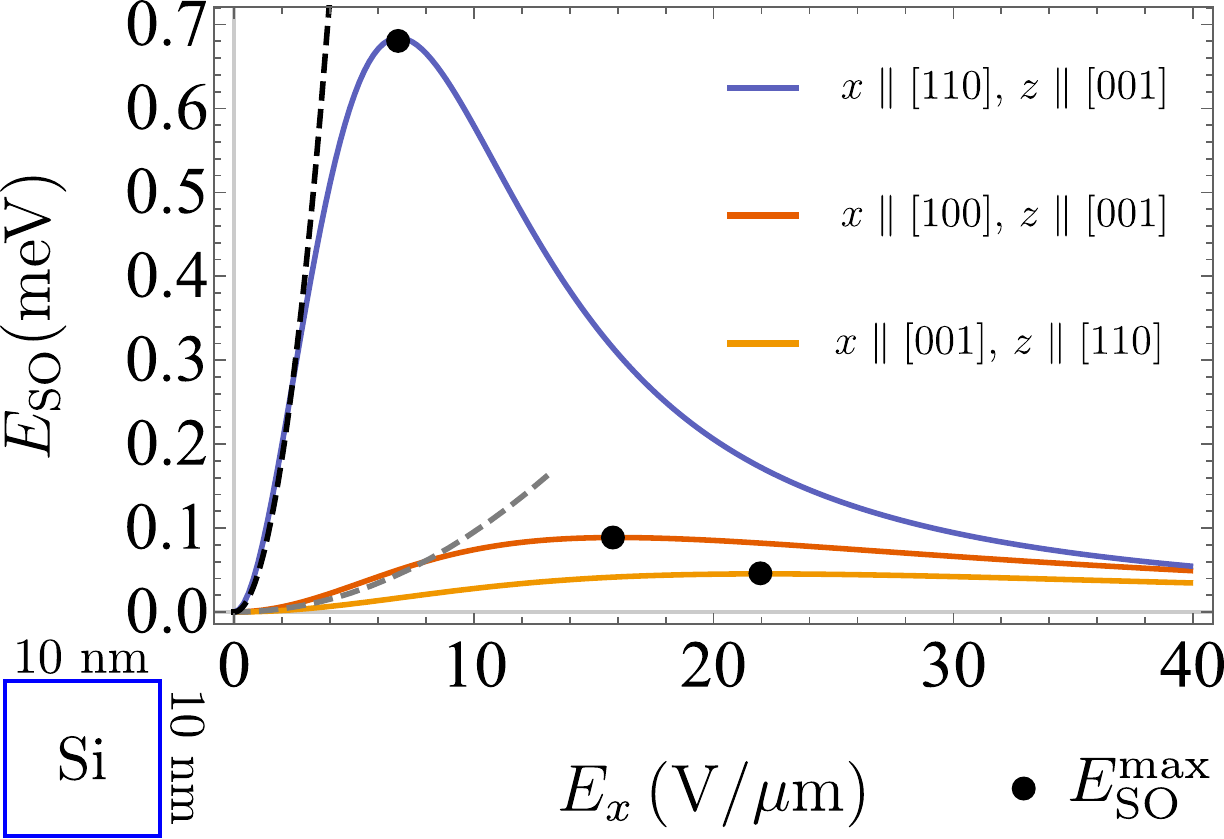}
\caption{Dependence of the spin-orbit energy $E_{\rm SO}$ on the electric field $E_x$ for three Si NWs with identical cross sections ($s = 10\mbox{ nm}$). For each of the three NWs, the orientation of the crystallographic axes is different and provided by the labels of the respective lines. The blue, red, and orange solid lines were calculated analogously to Fig.~\ref{fig:GeSiCoreShellESOvsEx}, using $\phi = \pi/4$, $\phi = 0$, and $\xi = 0$, respectively. Black dots mark the points with maximal spin-orbit energy $E_{\rm SO}^{\rm max}$. The dashed lines are plots of Eq.~(\ref{eq:AnalyticalModelSiESOforPlots}) with $\phi = \pi/4$ (black) and $\phi = 0$ (gray). In the regime of small $E_x$, they almost coincide with the numerically calculated solid lines. For details on Eq.~(\ref{eq:AnalyticalModelSiESOforPlots}) and the underlying model, see Sec.~\ref{sec:AnalyticalResultsSiNWs}. }
\label{fig:SiESO}
\end{center}
\end{figure}

\begin{figure}[tb]
\begin{center}
\includegraphics[width=0.90\linewidth]{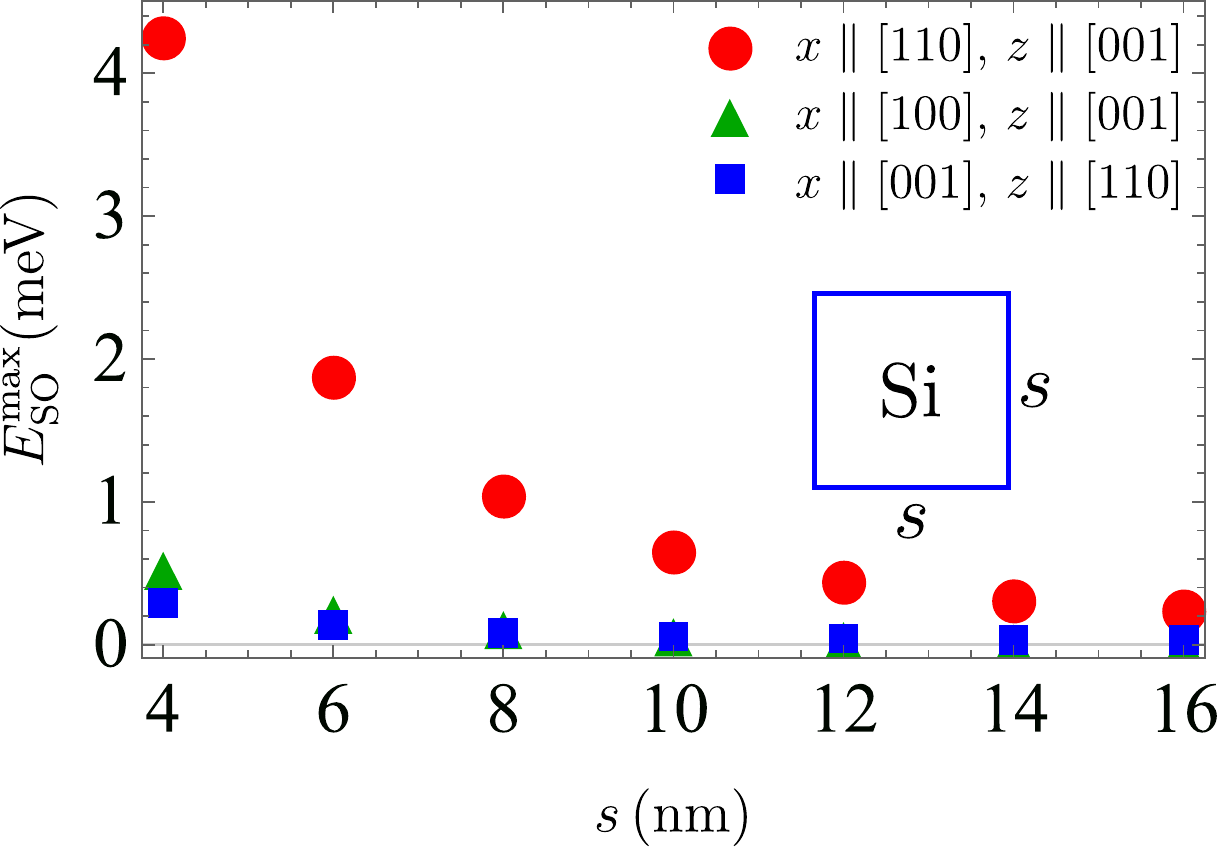}
\caption{Maximal spin-orbit energy $E_{\rm SO}^{\rm max}$ as a function of the side length $s$ for Si NWs with different orientations of the crystallographic axes (see explanations at the top right of the figure). Details for the three data points at $s = 10\mbox{ nm}$ are shown in Fig.~\ref{fig:SiESO}, the other data were calculated analogously. We note that $E_{\rm SO}^{\rm max} \propto s^{-2}$ in good approximation. }
\label{fig:SiESOmax}
\end{center}
\end{figure}

Similarly to Sec.~\ref{secsub:GeNWsWithFields}, we plot in Fig.~\ref{fig:SiESO} the spin-orbit energy $E_{\rm SO}$ as a function of $E_x$. The three solid curves correspond all to a Si NW with $s = 10\mbox{ nm}$, but the orientation of the crystallographic axes is different. Remarkably, the curves in Fig.~\ref{fig:SiESO} and the maximally achievable spin-orbit energies $E_{\rm SO}^{\rm max}$ differ greatly. This can also be seen in Fig.~\ref{fig:SiESOmax}, where we show $E_{\rm SO}^{\rm max}$ as a function of the side length $s$.       

A key result which is evident from Figs.~\ref{fig:SiESO} and \ref{fig:SiESOmax} is that the case with $x \parallel [110]$ and $z \parallel [001]$ leads to significantly larger spin-orbit energies than the cases where $x$ coincides with a main crystallographic axis. This suggests that the SOI of holes in Si NWs can be much increased compared with recent experiments. In fact, the smallest values for $E_{\rm SO}^{\rm max}$ in Figs.~\ref{fig:SiESO} and \ref{fig:SiESOmax} are obtained when $x$ is parallel to a main crystallographic axis and $z \parallel [110]$, which is the case in many recent devices with Si NWs \cite{coquand:ulisproc12, barraud:edl12, voisin:nlt16, maurand:ncomm16}. 

We recall that the spin-orbit energy $E_{\rm SO}$ and the spin-orbit length $l_{\rm SO}$ are \cite{kloeffel:prb11}
\begin{eqnarray}
E_{\rm SO} &=& \frac{m_{\rm eff} \alpha^2 E_x^2}{2 \hbar^2} , 
\label{eq:SpinOrbitEnergySiSection}  \\
l_{\rm SO} &=& \frac{\hbar^2}{m_{\rm eff} \left| \alpha E_x \right|}
\label{eq:SpinOrbitLengthDepAlpha} 
\end{eqnarray}
for an electron-like Hamiltonian of the form $\hbar^2 k_z^2/(2 m_{\rm eff}) + \alpha E_x \sigma_y k_z$. Hence, the effective Rashba coefficient $\alpha$ and the spin-orbit length can be calculated with the data from Figs.~\ref{fig:SiESO} and \ref{fig:SiESOmax} via
\begin{eqnarray}
|\alpha| &=& \frac{\hbar}{|E_x|} \sqrt{\frac{2 E_{\rm SO}}{m_{\rm eff}}} , 
\label{eq:AlphaSiSection}\\
l_{\rm SO} &=& \frac{\hbar}{\sqrt{2 m_{\rm eff} E_{\rm SO}}} .
\label{eq:SpinOrbitLengthDepESO}  
\end{eqnarray} 
A reasonable choice for the effective mass $m_{\rm eff}$ is the fitted mass obtained in Sec.~\ref{secsub:SiNWsNoFields} for the spectrum without applied fields. That is, for the three cases plotted in Figs.~\ref{fig:SiESO} and \ref{fig:SiESOmax}, $m_{\rm eff}$ should be replaced by $m_{\rm fit}^{[001]}(0) = 0.446 m$, $m_{\rm fit}^{[001]}(\pi/4) = 0.794 m$, and $m_{\rm fit}^{[110]}(0) = 0.184 m$, respectively. 

When $x \parallel [110]$ and $z \parallel [001]$, we obtain both the largest effective mass and the largest spin-orbit energies compared with the other two cases. Although $E_{\rm SO} \propto m_{\rm eff}$ [Eq.~(\ref{eq:SpinOrbitEnergySiSection})], it is important to note that the greatest ratios $E_{\rm SO}/m_{\rm eff}$ and, therefore, the greatest $|\alpha|$ are obtained for this orientation of the crystallographic axes as well. Moreover, the fact that $m_{\rm fit}^{[001]}(\pi/4)$ is large is advantageous for achieving a short spin-orbit length [Eqs.~(\ref{eq:SpinOrbitLengthDepAlpha}) and (\ref{eq:SpinOrbitLengthDepESO})]. For instance, the data in Fig.~\ref{fig:SiESOmax} reveal that for each of the considered values of $s$, the calculated $E_{\rm SO}^{\rm max}$ in the case of $x \parallel [110]$ and $z \parallel [001]$ exceeds the $E_{\rm SO}^{\rm max}$ for $x \parallel [001]$ and $z \parallel [110]$ by a factor of about 15. Furthermore, $m_{\rm fit}^{[001]}(\pi/4)/m_{\rm fit}^{[110]}(0) = 4.3$. We therefore conclude that, at any fixed side length $s$, changing the orientation of the crystallographic axes could reduce the shortest possible spin-orbit length in recently fabricated Si NWs \cite{coquand:ulisproc12, barraud:edl12, prati:nanotech12, voisin:nlt16, maurand:ncomm16, crippa:arX1710} by a factor of eight.

Another possibility to decrease the shortest achievable spin-orbit length is to decrease the side length $s$ of the Si NW. As in the case of Ge/Si core/shell NWs (Sec.~\ref{secsub:GeNWsWithFields}), we find from the data in Fig.~\ref{fig:SiESOmax} that $E_{\rm SO}^{\rm max}$ scales approximately with $s^{-2}$, and so the minimal spin-orbit length is approximately proportional to~$s$. We wish to point out, however, that reaching $E_{\rm SO}^{\rm max}$ in a thin NW requires a stronger electric field than in thicker ones, as shown before for Ge/Si core/shell NWs (see also Fig.~\ref{fig:GeSiCoreShellESOvsEx}).  

Thus far, we have focused on electric fields $\bm{E}$ that are applied along the $x$ axis. Remarkably, for the case with $x \parallel [110]$ and $z \parallel [001]$, we find that even larger spin-orbit energies can be obtained when the electric field is parallel to the diagonal of the cross section. For instance, while $E_{\rm SO}^{\rm max} \simeq 0.7\mbox{ meV}$ when $s = 10\mbox{ nm}$ and $\bm{E} \parallel x$, we obtain $E_{\rm SO} = 1.5\mbox{ meV}$ with the same Si NW when $E_x = E_y = 10\mbox{ V/$\mu$m}$. However, in the case of $x \parallel [001]$ and $z \parallel [110]$, we did not find a noteworthy enhancement of $E_{\rm SO}^{\rm max}$ by changing the direction of $\bm{E}$.            

Setting $\alpha_h = 0$ in the simulations does not affect the presented results, apart from negligible quantitative deviations. Therefore, we conclude that our results for the SOI of low-energy hole states in Si NWs (Sec.~\ref{secsub:SiNWsWithFields}) and Ge/Si core/shell NWs (Sec.~\ref{secsub:GeNWsWithFields}) are based on the DRSOI. As a consequence, the values for $\alpha$ that may be calculated with Eq.~(\ref{eq:AlphaSiSection}) and our simulated results correspond to the values of $\alpha_{\rm DR}$, which is the effective Rashba coefficient of the DRSOI.

\pagebreak

\section{Analytical Results for Si Nanowires}
\label{sec:AnalyticalResultsSiNWs}

In order to explain why a surprisingly large DRSOI can be achieved with Si NWs when $x \parallel [110]$ and $z \parallel [001]$, we consider a simple model. In this section, we use the Hamiltonian
\begin{equation}
H = H_{\rm LK}^{\rm [001]}(\phi) + V(x,y) - e E_x x  
\end{equation} 
for a NW with $z \parallel [001]$. The LK Hamiltonian $H_{\rm LK}^{\rm [001]}(\phi)$ and the confining potential $V(x,y)$ are displayed in Eqs.~(\ref{eq:LKfull001}) and~(\ref{eq:ConfinementRectangularNWxy}), respectively. For simplicity, we assume that the NW has a square cross section, i.e., $L_x = L_y = s$, where $s$ is the side length.

\subsection{Subspace and projected Hamiltonian}
\label{secsub:AnalyticalModelSiSubspace}

The matrix 
\begin{widetext}
\begin{equation}
H_{\rm proj} = \begin{pmatrix}
\frac{\hbar^2 k_z^2}{2 m_{\rm LH}^{\prime}} & 0 & p_1 E_x & 0 & 0 & - i p_2 k_z \\
0 & \frac{\hbar^2 k_z^2}{2 m_{\rm LH}^{\prime}} & 0 & i p_2 k_z & p_1 E_x & 0 \\
p_1 E_x & 0 & \Delta_1 + \frac{\hbar^2 k_z^2}{2 m_{\rm LH}^{\prime}} & \nu(\phi) & 0 & 0 \\
0 & - i p_2 k_z & \nu^{*}(\phi) & \Delta_1 + \Delta_2 + \frac{\hbar^2 k_z^2}{2 m_{\rm HH}^{\prime}} & 0 & 0 \\
0 & p_1 E_x & 0 & 0 & \Delta_1 + \frac{\hbar^2 k_z^2}{2 m_{\rm LH}^{\prime}} & \nu^{*}(\phi) \\
i p_2 k_z & 0 & 0 & 0 & \nu(\phi) & \Delta_1 + \Delta_2 + \frac{\hbar^2 k_z^2}{2 m_{\rm HH}^{\prime}} 
\end{pmatrix} 
\label{eq:Matrix6x6ModelSi}
\end{equation}
\end{widetext}
shows the projection of $H$ onto a subspace with the six basis states $\ket{1,1,k_z,\frac{1}{2}}$, $\ket{1,1,k_z,-\frac{1}{2}}$, $\ket{2,1,k_z,\frac{1}{2}}$, $\ket{2,1,k_z,-\frac{3}{2}}$, $\ket{2,1,k_z,-\frac{1}{2}}$, and $\ket{2,1,k_z, \frac{3}{2}}$. For details on the basis states, we refer to Sec.~\ref{secsub:BasisStatesNumDiag}. The two states $\ket{1,1,k_z,\pm\frac{1}{2}}$ are used in the model because we are interested in the lowest-energy subbands and the expectation value $\bra{n_x , n_y, 0, j_z} H \ket{n_x , n_y, 0, j_z}$ at $k_z = 0$ is minimal for these two states. In Eq.~(\ref{eq:Matrix6x6ModelSi}), the energy $\bra{1,1,0,\pm\frac{1}{2}} H \ket{1,1,0,\pm\frac{1}{2}} = \hbar^2 \pi^2 (\gamma_1 - \gamma_2)/(s^2 m)$ was subtracted from the diagonal, as it corresponds to a global offset. The two states $\ket{2,1,k_z,\pm\frac{1}{2}}$ are included because among all basis states with a reasonably small $n_x \leq 3$, these are the only states that are coupled to $\ket{1,1,k_z,\pm\frac{1}{2}}$ via the electric field $E_x$. Finally, the two states $\ket{2,1,k_z,\pm\frac{3}{2}}$ are taken into account because these are coupled to both $\ket{1,1,k_z,\pm\frac{1}{2}}$ and $\ket{2,1,k_z,\mp\frac{1}{2}}$ due to the LK Hamiltonian. 

In Eq.~(\ref{eq:Matrix6x6ModelSi}), the prefactors are
\begin{eqnarray}
p_1 &=& \frac{16 e s}{9 \pi^2} , 
\label{eq:p1AnalyticalModelSi} \\
p_2 &=& \frac{8 \gamma_3 \hbar^2}{\sqrt{3} m s} .
\end{eqnarray}
The prime at the effective masses
\begin{eqnarray}
m_{\rm HH}^{\prime} &=& \frac{m}{\gamma_1 - 2 \gamma_2} , \\
m_{\rm LH}^{\prime} &=& \frac{m}{\gamma_1 + 2 \gamma_2} 
\end{eqnarray}
was added to avoid confusion with Eqs.~(\ref{eq:massHH}) and (\ref{eq:massLH}), which are based on the spherical approximation. The expressions for the two splittings read
\begin{eqnarray}
\Delta_1 &=& \frac{3 \hbar^2 \pi^2 (\gamma_1 - \gamma_2)}{2 s^2 m} , 
\label{eq:Delta1AnalyticalModelSi} \\
\Delta_2 &=& \frac{5 \hbar^2 \pi^2 \gamma_2}{s^2 m} . 
\label{eq:Delta2AnalyticalModelSi}
\end{eqnarray} 
We wish to emphasize that the orientation of the crystallographic axes enters the model via the coupling
\begin{equation}
\nu(\phi) = - \frac{3 \sqrt{3} \hbar^2 \pi^2 e^{2 i \phi} \left[ \gamma_2 \cos(2 \phi) - i \gamma_3 \sin(2 \phi) \right] }{2 s^2 m} .
\end{equation}
At this stage, we can already see why $x \parallel [110]$ is favorable for the DRSOI. When one compares the case $\phi = \pi/4$, i.e., $x \parallel [110]$, with the case $\phi = 0$, i.e., $x \parallel [100]$, one finds
\begin{equation}
\frac{\nu(\pi/4)}{\nu(0)} = \frac{\gamma_3}{\gamma_2} ,
\end{equation} 
and so the coupling at $\phi = \pi/4$ is four times stronger because $\gamma_3 / \gamma_2 \approx 4$ in Si \cite{lawaetz:prb71, winkler:book}.

\subsection{Results}
\label{secsub:AnalyticalModelSiResults}

The Hamiltonian $H_{\rm proj}$ in Eq.~(\ref{eq:Matrix6x6ModelSi}) can be used to derive analytical results for the low-energy states. First, we find a unitary transformation that exactly diagonalizes $H_{\rm proj}$ when $k_z = 0$ and $E_x = 0$. Second, we apply this transformation to $H_{\rm proj}$. Third, by considering terms that contain $k_z$ or $E_x$ as perturbations, we perform a second-order Schrieffer-Wolff transformation (quasi-degenerate perturbation theory \cite{winkler:book}) and obtain an effective 2$\times$2 Hamiltonian for the subbands of lowest energy. Neglecting the corrections to the effective mass, this approach yields
\begin{equation}
H_{\rm 2x2}^{\rm eff, [001]} = \frac{\hbar^2 k_z^2}{2 m_{\rm LH}^{\prime}} + \alpha_{\rm DR}(\phi) E_x \tilde{\sigma}_y k_z  + \delta_{\rm DR}(\phi) E_x \tilde{\sigma}_x k_z , 
\label{eq:AnalyticalSiModel2x2Ham}
\end{equation}
where $\tilde{\sigma}_x$ and $\tilde{\sigma}_y$ refer to spin-1/2 Pauli matrices. The coefficient
\begin{equation}
\delta_{\rm DR}(\phi) = \chi(\phi) (\gamma_2 - \gamma_3) \sin(4 \phi) ,
\label{eq:AnalyticalResultSiDelta}
\end{equation}
where
\begin{eqnarray}
\chi(\phi) &=&  \frac{\gamma_3}{7 \gamma_2^2 - 3 \gamma_1^2 - 4 \gamma_1 \gamma_2 + 9 \gamma_2^2 \cos^2(2 \phi) + 9 \gamma_3^2 \sin^2(2 \phi) } \nonumber \\
& & \times \frac{2^8 s^2 e}{9 \pi^4} ,
\label{eq:chiForResultSimpleModel}
\end{eqnarray}
requires a high degree of asymmetry because it vanishes when either $\phi \in \{0, \pi/4, \pi/2, \cdots \}$ or $\gamma_2 = \gamma_3$. 

\begin{figure}[tb]
\begin{center}
\includegraphics[width=0.90\linewidth]{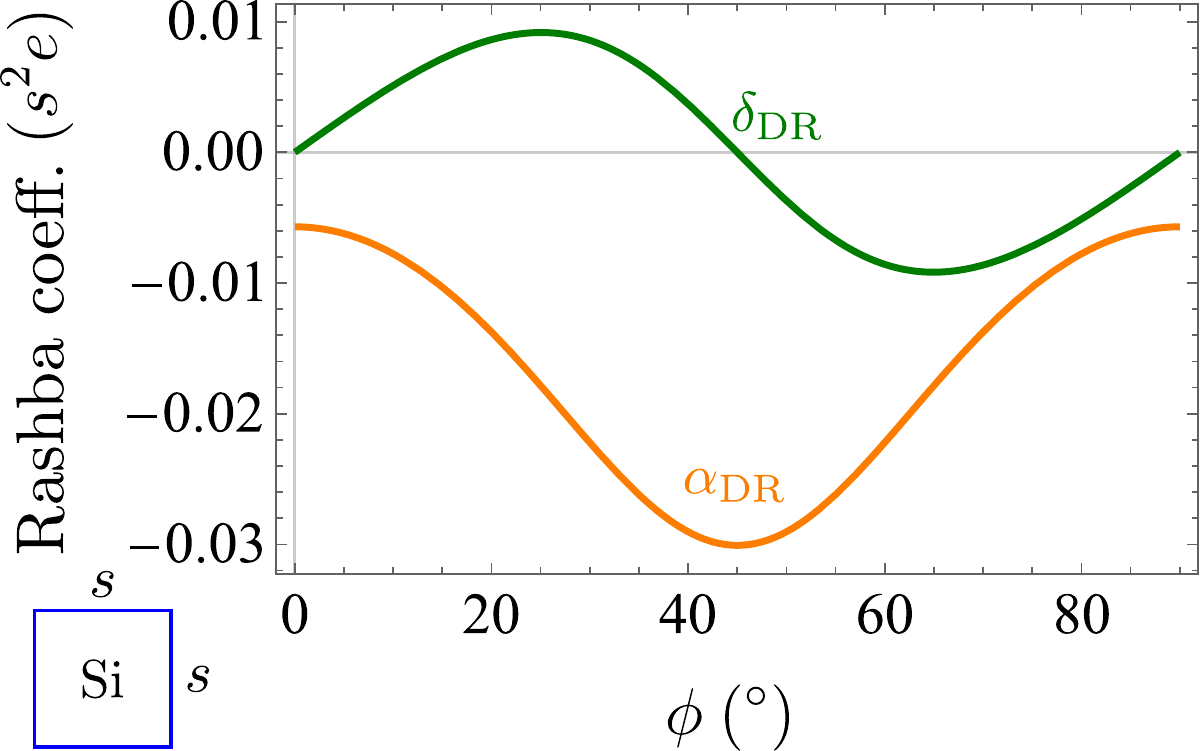}
\caption{Dependence of the derived coefficients $\delta_{\rm DR}(\phi)$ [see Eqs.~(\ref{eq:AnalyticalResultSiDelta}) and (\ref{eq:chiForResultSimpleModel})] and $\alpha_{\rm DR}(\phi)$ [see Eqs.~(\ref{eq:AnalyticalResultSiAlpha}) and (\ref{eq:chiForResultSimpleModel})] on the angle $\phi$. For detailed information, we refer to Sec.~\ref{sec:AnalyticalResultsSiNWs}. We use the Luttinger parameters $\gamma_1 = 4.22$, $\gamma_2 = 0.39$, and $\gamma_3 = 1.44$ for Si (Sec.~\ref{secsub:SiParameters}) \cite{lawaetz:prb71}. The Si NW has a square cross section with side length $s$ and the electric field, applied along the $x$ axis (Fig.~\ref{fig:SetupWireAndQD}), is assumed to be small. }
\label{fig:alphadelta}
\end{center}
\end{figure}

The effective Rashba coefficient of the DRSOI is
\begin{equation}
\alpha_{\rm DR}(\phi) = \chi(\phi) \left[ \gamma_2 + \gamma_3 + ( \gamma_2 - \gamma_3 ) \cos(4 \phi) \right] .
\label{eq:AnalyticalResultSiAlpha}
\end{equation}
For $\phi = 0$ ($x \parallel [100]$), it simplifies to
\begin{equation}
\alpha_{\rm DR}(0) = \frac{2^9 \gamma_2 \gamma_3 s^2 e}{9 \pi^4 \left( 16 \gamma_2^2 - 3 \gamma_1^2 - 4 \gamma_1 \gamma_2 \right)} ,
\end{equation}
whereas
\begin{equation}
\alpha_{\rm DR}(\pi/4) = \frac{2^9 \gamma_3^2 s^2 e}{9 \pi^4 \left( 7 \gamma_2^2 - 3 \gamma_1^2 - 4 \gamma_1 \gamma_2 + 9 \gamma_3^2 \right)} 
\label{eq:alphaDRSimpleModelPhipiOv4}
\end{equation}
for $\phi = \pi/4$ ($x \parallel [110]$). With the Luttinger parameters of Si \cite{lawaetz:prb71, winkler:book, richard:prb04, adachi:properties}, the ratio
\begin{equation}
\frac{\alpha_{\rm DR}(\pi/4)}{\alpha_{\rm DR}(0)} = \frac{\gamma_3 \left( 16 \gamma_2^2 - 3 \gamma_1^2 - 4 \gamma_1 \gamma_2 \right)}{\gamma_2 \left( 7 \gamma_2^2 - 3 \gamma_1^2 - 4 \gamma_1 \gamma_2 + 9 \gamma_3^2 \right)}
\end{equation}
reveals that, compared with $x \parallel [100]$, the effective Rashba coefficient of the DRSOI is more than five times greater when $x \parallel [110]$. This can also be seen in Fig.~\ref{fig:alphadelta}, where we plot the dependence of $\alpha_{\rm DR}$ [Eq.~(\ref{eq:AnalyticalResultSiAlpha})] and $\delta_{\rm DR}$ [Eq.~(\ref{eq:AnalyticalResultSiDelta})] on the angle~$\phi$.

In Eq.~(\ref{eq:AnalyticalSiModel2x2Ham}), we neglected corrections to the effective mass for simplicity. Corrections to the effective mass may easily be accounted for by replacing the $m_{\rm LH}^{\prime}$ in Eq.~(\ref{eq:AnalyticalSiModel2x2Ham}) with the $m_{\rm fit}^{[001]}(\phi)$ introduced in Sec.~\ref{secsub:SiNWsNoFields}. We recall that, for Si NWs, the values $m_{\rm fit}^{[001]}(0) = 0.446 m$ and $m_{\rm fit}^{[001]}(\pi/4) = 0.794 m$ were obtained from the lowest-energy spectra in Fig.~\ref{fig:SiSpectraNoFields}. For a Hamiltonian $\hbar^2 k_z^2/[2 m_{\rm fit}^{[001]}(\phi)] + \alpha_{\rm DR}(\phi) E_x \tilde{\sigma}_y k_z $, the spin-orbit length $l_{\rm SO}$ and the spin-orbit energy $E_{\rm SO}$ are \cite{kloeffel:prb11}
\begin{equation}
l_{\rm SO}(\phi) = \frac{\hbar^2}{m_{\rm fit}^{[001]}(\phi) \left| \alpha_{\rm DR}(\phi) E_x \right|}
\end{equation}
and 
\begin{equation}
E_{\rm SO}(\phi) = \frac{m_{\rm fit}^{[001]}(\phi) \hspace{0.03cm} \alpha_{\rm DR}^2(\phi) E_x^2}{2 \hbar^2} , 
\label{eq:AnalyticalModelSiESOforPlots}
\end{equation}
respectively. Thus, considering a fixed side length $s$, a fixed electric field $E_x$, and provided that $E_x$ is small enough for the perturbative approach in this Sec.~\ref{secsub:AnalyticalModelSiResults} to apply, our analytical results show for Si NWs that by changing from $x \parallel [100]$ to $x \parallel [110]$, the spin-orbit length becomes more than nine times shorter and the spin-orbit energy increases by a factor of about fifty. This agrees very well with our numerical results from Sec.~\ref{sec:Si}. In Fig.~\ref{fig:SiESO}, where Eq.~(\ref{eq:AnalyticalModelSiESOforPlots}) is plotted for the cases $\phi = 0$ (gray dashed line) and $\phi = \pi/4$ (black dashed line) using $s = 10\mbox{ nm}$ and the $\gamma_{1,2,3}$ of Sec.~\ref{secsub:SiParameters}, we find quantitative agreement with the numerical data in the regime of small $E_x$, which is the regime where the abovementioned perturbation theory applies.

\subsection{Validity and remarks}
\label{secsub:AnalyticalModelSiValidity}

We want to conclude this section with several remarks. Our results for $\alpha_{\rm DR}$ [see Eqs.~(\ref{eq:chiForResultSimpleModel}) to (\ref{eq:alphaDRSimpleModelPhipiOv4})] show that $\alpha_{\rm DR} \propto s^2$ increases linearly with the area of the square cross section. This finding is consistent with the model for cylindrical Ge/Si core/shell NWs \cite{kloeffel:prb11}. In the absence of strain (no Si shell), the splitting $\Delta$ in the effective Hamiltonian is proportional to $R^{-2}$, where $R$ is the core radius. Thus, for a bare Ge NW we obtain $\alpha_{\rm DR} = 2 e C U/\Delta \propto R^2$ in the regime of small electric fields, since $C \propto R^{-1}$ and $U \propto R$ (see Sec.~\ref{secsubsub:HolesInNWs}). We wish to emphasize that the remarkable scalings $\alpha_{\rm DR} \propto s^2$ and $\alpha_{\rm DR} \propto R^2$ are only valid within the parameter regimes for which the assumptions and perturbative approaches behind the respective formulas apply. For instance, the off-diagonal coupling $p_1 E_x$ in Eq.~(\ref{eq:Matrix6x6ModelSi}) is proportional to $s$, whereas the splittings $\Delta_{1,2}$ on the diagonal are proportional to $s^{-2}$. Hence, when $s$ is continuously increased, the perturbation theory behind our analytical results in Sec.~\ref{secsub:AnalyticalModelSiResults} will eventually lose validity when $E_x$ is fixed. Also, the validity will eventually be lost when $E_x$ is increased at a fixed $s$, which explains the deviation in Fig.~\ref{fig:SiESO} between the numerical and analytical results beyond the regime of small $E_x$. As a rough estimate, one may use $|p_1 E_x / \Delta_1| \lesssim 0.1$ to identify the regime of small $E_x$ for the model of Eq.~(\ref{eq:Matrix6x6ModelSi}). With $s = 10\mbox{ nm}$ and the Luttinger parameters of Si, this estimate yields $|E_x| \lesssim 2.4\mbox{ V/$\mu$m}$, in good agreement with Fig.~\ref{fig:SiESO}. 

While thicker Si and Ge NWs allow for a stronger SOI when only weak electric fields are present in the system (see the abovementioned $\alpha_{\rm DR} \propto s^2$ and $\alpha_{\rm DR} \propto R^2$), reaching the lowest-energy subband with NWs or NW QDs that have large cross sections may be experimentally challenging because of the small splittings between the subbands. In fact, we find that qualitatively different behaviors can be expected for hole states in different subbands, which is consistent with previous work \cite{csontos:prb09}. 

From a theoretical point of view, Ge/Si core/shell NWs with a core diameter of approximately $4\mbox{--}24\mbox{ nm}$ were found to be most promising (see, e.g., Ref.~\cite{kloeffel:prb13} and its SI). Similar considerations apply to Si NWs, as discussed in Sec.~\ref{sec:Accuracy}. When we use $s = 10\mbox{ nm}$, we already obtain a relatively large $\alpha_{\rm DR}(\pi/4) = -3.0\mbox{ nm}^2 e$ from Eq.~(\ref{eq:alphaDRSimpleModelPhipiOv4}) for holes in Si NWs, which is similar to the calculated Rashba coefficients $\alpha_{\rm el} = 1.2\mbox{ nm}^2 e$ and $\alpha_{\rm el} = 5.2\mbox{ nm}^2 e$ for electrons in InAs and InSb, respectively \cite{winkler:book}. The negative sign obtained with the formulas for $\alpha_{\rm DR}$ in this Sec.~\ref{sec:AnalyticalResultsSiNWs} simply results from our ordering of the two states that the Pauli matrices are based on. If these two states were swapped, the coefficient $\alpha_{\rm DR}$ would change its sign. 

In order to rule out that the term $\delta_{\rm DR}(\phi) E_x \tilde{\sigma}_x k_z$ in Eq.~(\ref{eq:AnalyticalSiModel2x2Ham}) is an artifact of the six-dimensional subspace of Eq.~(\ref{eq:Matrix6x6ModelSi}), we performed numerical calculations for Si NWs with square cross sections, considering a nonzero electric field $\bm{E} = E_x \bm{e}_x$, a nonzero magnetic field $\bm{B} = B_y \bm{e}_y$, and $z \parallel [001]$. Analogously to Sec.~\ref{sec:Si}, we used the method described in Sec.~\ref{sec:Model} to analyze the low-energy hole spectrum. For $\phi = 0$, the two subbands of lowest energy correspond, in good approximation, to two parabolas that are shifted against each other in the $E$-$k_z$ diagram ($E$: Energy). Importantly, the two parabolas cross each other. This is consistent with an effective 2$\times$2 Hamiltonian of the form $\hbar^2 k_z^2/(2 m_{\rm eff}) + \alpha_{\rm DR} E_x \tilde{\sigma}_y k_z + g \mu_B B_y \tilde{\sigma}_y /2$, where $m_{\rm eff}$ is an effective mass and $g$ is a $g$ factor. The same qualitative results are observed at $\phi = \pi/4$ and $\phi = \pi/2$. With $0 < \phi < \pi/4$ or $\pi/4 < \phi < \pi/2$, however, an anticrossing occurs in the spectrum. This anticrossing can only be obtained with an additional term proportional to $\tilde{\sigma}_{x,z}$ in the 2$\times$2 Hamiltonian, in agreement with the term $\delta_{\rm DR} E_x \tilde{\sigma}_x k_z$ in Eq.~(\ref{eq:AnalyticalSiModel2x2Ham}). We recall that, indeed, $\delta_{\rm DR}(\phi)$ vanishes for $0 \leq \phi \leq \pi/2$ only at $\phi = 0$, $\phi = \pi/4$, and $\phi = \pi/2$ (see also Fig.~\ref{fig:alphadelta}).

When we analyze Eq.~(\ref{eq:Matrix6x6ModelSi}) in the regime of very strong electric fields, where $|p_1 E_x / \Delta_1 | \gg 1$, we find that the low-energy eigenstates contain almost equal superpositions of $\ket{1,1,k_z,\pm\frac{1}{2}}$ and $\ket{2,1,k_z,\pm\frac{1}{2}}$. Deriving an effective 2$\times$2 Hamiltonian for the low-energy subbands yields that Eqs.~(\ref{eq:AnalyticalSiModel2x2Ham}), (\ref{eq:AnalyticalResultSiDelta}), and (\ref{eq:AnalyticalResultSiAlpha}) also apply at very strong $E_x$, but with
\begin{equation}
\chi(\phi) = \chi = - \frac{27 \gamma_3 \hbar^4 \pi^4}{8 s^4 m^2 e E_x^2} 
\label{eq:chiRegimeOfLargeEx}
\end{equation}
instead of Eq.~(\ref{eq:chiForResultSimpleModel}). The property $\chi \propto E_x^{-2}$ in Eq.~(\ref{eq:chiRegimeOfLargeEx}) is consistent with the numerical result that the spin-orbit energy decreases at large $E_x$, as illustrated for Si NWs in Fig.~\ref{fig:SiESO}.  

Finally, we want to briefly mention that additional couplings among the basis states $\ket{n_x , n_y, k_z, j_z}$ have, of course, been omitted by considering only the six-dimensional subspace of Eq.~(\ref{eq:Matrix6x6ModelSi}). Nevertheless, the simple analytical results derived in Sec.~\ref{secsub:AnalyticalModelSiResults} apply well to the case of Si NWs. In some other materials, additional couplings may be relatively important as well. For instance, in the case of Ge, where both $\gamma_2/\gamma_1$ and $\gamma_3/\gamma_1$ are greater than in Si, these additional couplings apparently play a larger role than in Si and, depending on the desired accuracy, more basis states should be taken into account, as we do in the numerical calculations of Sec.~\ref{sec:Ge}. However, since $\gamma_2 \simeq \gamma_3$ in Ge, the spherical approximation applies and the effective model of Ref.~\cite{kloeffel:prb11} can be exploited as an analytical alternative [see also Eqs.~(\ref{eq:RecallPRB2011Matrix}), (\ref{eq:H2x2effWithDRSOI}), and (\ref{eq:H2x2effWithDRSOIStrongEx})]. By combining several subsequent transformations, it is sometimes even possible to derive surprisingly simple expressions that remain valid for a relatively wide range of parameters \cite{kloeffel:prb13}.

\section{Accuracy}
\label{sec:Accuracy}

Our numerical results in Secs.~\ref{sec:Ge} and \ref{sec:Si} were calculated by diagonalizing matrices with 36$\times$36 matrix elements, which are obtained from the model and the basis states of Sec.~\ref{sec:Model}. The model focuses on the topmost valence band of Si or Ge ($\Gamma_8^v$ \cite{winkler:book}) and uses hard-wall confinement for the core of the NW. This approach has many advantages. For instance, it allows for very fast calculations without the need for sophisticated numerics. Also, it provides insight into important mechanisms and enables the derivation of analytical results (Sec.~\ref{sec:AnalyticalResultsSiNWs}). On the other hand, it is clear that such an approach is only reliable if the associated requirements are satisfied.

An important requirement is that the considered subspace is well isolated. This may be analyzed by comparing energy scales. When the cross section of the NW core is a square with side length $s$, the energies due to an electric field $E_x$ are usually proportional to $E_x s$ [see, e.g., Eqs.~(\ref{eq:Matrix6x6ModelSi}) and (\ref{eq:p1AnalyticalModelSi})] and those due to the quantum confinement are proportional to $s^{-2}$ [see, e.g., Eqs.~(\ref{eq:Delta1AnalyticalModelSi}) and (\ref{eq:Delta2AnalyticalModelSi})]. Therefore, estimates for the upper bound of $E_x$ in our approach typically lead to a value proportional to $s^{-3}$. The upper bound can be increased by taking more basis states into account, particularly in the case of hard-wall confinement, since the energies associated with hard-wall confinement scale with the squared quantum number [see, e.g., Eqs.~(\ref{eq:HHEnergies1D}) and (\ref{eq:LHEnergies1D})].

In the case of Ge/Si core/shell NWs, the assumption of hard-wall confinement is justified as long as all energies are below the Ge-Si valence band offset of about $0.5\mbox{ eV}$ \cite{lu:pnas05}. In the case of Si NWs, which may be surrounded by materials with a very large band gap (such as SiO$_2$ \cite{voisin:nlt16, prati:nanotech12, weinberg:prb79}), the valence band offset can be even larger than that for the Ge-Si interface. An important boundary in our model for Si NWs is certainly the small splitting of only $44\mbox{ meV}$ \cite{winkler:book, richard:prb04} between the topmost valence band ($\Gamma_8^v$) and the spin-orbit split-off band ($\Gamma_7^v$) at the $\Gamma$ point ($k = 0$) in bulk Si. As a consequence, we estimate that effects of the split-off band may become important in Si NWs with $s < 10\mbox{ nm}$. For Ge, the splitting between $\Gamma_8^v$ and $\Gamma_7^v$ is approximately $0.3\mbox{ eV}$ \cite{winkler:book, richard:prb04} and therefore relatively large. Based on all numbers, we decided to show our results for $s \geq 4\mbox{ nm}$. We also note that $s$ cannot be chosen arbitrarily small because the LK Hamiltonian and the BP Hamiltonian lose validity if the NW has only very few atoms in its cross section. 

While the above explanations lead to a lower bound for $s$ in our model, there are also reasons for an upper bound. If $s$ is chosen very large, a great number of basis states may be necessary for reliable results in the presence of, e.g., strain (can easily exceed $10\mbox{ meV}$ in Ge/Si core/shell NWs \cite{kloeffel:prb11, kloeffel:prb14}) or applied fields (see, e.g., the mentioned proportionality to $s^{-3}$ of our estimated bound for the electric field). Furthermore, since the splittings between subbands scale with $s^{-2}$ in the absence of strain, as evident from Figs.~\ref{fig:GeSpectrumNoStrainComparison} and \ref{fig:SiSpectraNoFields}, it may be challenging to reach the subbands of lowest energy experimentally when $s$ is large.

A rather surprising feature in the numerical results of Sec.~\ref{sec:Ge} and \ref{sec:Si} is the eventual decay of the spin-orbit energy, which is observed when the electric field is continuously increased. For instance, in the example of a Si NW with $s = 10\mbox{ nm}$, $x \parallel \mbox{[110]}$, and $z \parallel \mbox{[001]}$ (blue line in Fig.~\ref{fig:SiESO}), the maximal spin-orbit energy $E_{\rm SO}^{\rm max} = 0.68\mbox{ meV}$ is reached at $E_x = 6.8\mbox{ V/$\mu$m}$, and at stronger $E_x$ the spin-orbit energy $E_{\rm SO}$ decays rapidly. When we recalculate this curve with $n_{x,y} \leq 5$ instead of $n_{x,y} \leq 3$, i.e., with 100 instead of 36 basis states (see Sec.~\ref{secsub:BasisStatesNumDiag}), we find that the maximum now occurs at $E_x = 7.3\mbox{ V/$\mu$m}$, with $E_{\rm SO}^{\rm max} = 0.78\mbox{ meV}$. Again, $E_{\rm SO}$ decays rapidly with increasing $E_x$ once the maximum was reached. In the two calculations, $E_{\rm SO}$ dropped to $E_{\rm SO}^{\rm max} / 2$ at $E_x \simeq 15\mbox{ V/$\mu$m}$ (with $n_{x,y} \leq 3$, Fig.~\ref{fig:SiESO}) and $E_x \simeq 17\mbox{ V/$\mu$m}$ (with $n_{x,y} \leq 5$), respectively. Thus, we obtain here quantitative but not qualitative corrections by changing from $n_{x,y} \leq 3$ to $n_{x,y} \leq 5$. Due to the hard-wall confinement, qualitative corrections to these results from basis states with even larger $n_{x,y}$ are not expected either. This strongly suggests that the eventual decay of $E_{\rm SO}$ at increasing $E_x$ is not an artifact of our finite subspace. We note that a decay of $E_{\rm SO}$ is expected when the simple model of Sec.~\ref{sec:AnalyticalResultsSiNWs} is analyzed in the regime of strong $E_x$ [see Eq.~(\ref{eq:chiRegimeOfLargeEx})]. 

We also recalculated other curves with $n_{x,y} \leq 5$ instead of $n_{x,y} \leq 3$. For a Si NW with $s = 10\mbox{ nm}$, $x \parallel \mbox{[100]}$, and $z \parallel \mbox{[001]}$ (red line in Fig.~\ref{fig:SiESO}), $E_{\rm SO}^{\rm max}$ increases from $0.09\mbox{ meV}$ to $0.13\mbox{ meV}$, and the electric field at which $E_{\rm SO}^{\rm max}$ is reached changes by a factor of two from $E_x = 16\mbox{ V/$\mu$m}$ to $E_x = 32\mbox{ V/$\mu$m}$. When we compare the $E_x$-dependence of $E_{\rm SO}$ with the abovementioned case for $x \parallel \mbox{[110]}$ (blue line in Fig.~\ref{fig:SiESO}), we find that for $x \parallel \mbox{[100]}$, the maximal spin-orbit energy is obtained at a stronger $E_x$ and the range of $E_x$ with $E_{\rm SO} > E_{\rm SO}^{\rm max} /2$ is wider. In the case of $x \parallel \mbox{[100]}$, it turns out that the increase of $E_{\rm SO}$ at small $E_x$ is much steeper than the decrease after $E_{\rm SO} = E_{\rm SO}^{\rm max}$, leading to a plateau-like behavior once $E_{\rm SO} \approx E_{\rm SO}^{\rm max}$ is reached. As evident from Fig.~\ref{fig:GeSiCoreShellESOvsEx}, a relatively slow decay of $E_{\rm SO}$ is also observed for Ge/Si core/shell NWs. When the three curves for $s = 6\mbox{ nm}$, $s = 10\mbox{ nm}$, and $s = 14\mbox{ nm}$ in Fig.~\ref{fig:GeSiCoreShellESOvsEx} are recalculated with $n_{x,y} \leq 5$, $E_{\rm SO}^{\rm max}$ increases by $21\%$, $41\%$, and $74\%$, respectively, and the electric field at which the maximum occurs increases by $40\%$, $54\%$, and $69\%$. Although all these curves decay at strong electric fields, the decay occurs at values for $E_x$ at which we estimate that more basis states must be taken into account for the results to be reliable. We note that a plateau of $E_{\rm SO}$ is obtained when the effective model of Ref.~\cite{kloeffel:prb11} for Ge/Si core/shell NWs is studied in the regime of strong $E_x$ [see Eqs.~(\ref{eq:H2x2effWithDRSOIStrongEx}) and (\ref{eq:ESOGeSiEffModelLimitOfStrongEx})]. However, we wish to emphasize again that the model of Ref.~\cite{kloeffel:prb11}, the simple 6$\times$6 model of Sec.~\ref{sec:AnalyticalResultsSiNWs}, and our numerical approach discussed in Sec.~\ref{secsub:BasisStatesNumDiag} are all based on finite subspaces, which are no longer well isolated once the electric-field-induced couplings to omitted basis states become relatively strong. Consequently, the detailed behavior of $E_{\rm SO}$ in the regime of large $E_x$ is currently an open problem and requires further research.         

In summary, the model and the numerical approach of Sec.~\ref{sec:Model} have several advantages because of their simplicity. It is clear, however, that results with a high quantitative precision will require an extended model. Nevertheless, based on our estimates and calculations described above, we are convinced that our quantitative results are within the right order of magnitude for the range of $s$ considered in Secs.~\ref{sec:Ge} and \ref{sec:Si}, and most importantly, that the qualitative findings are reliable. The detailed behavior of the SOI at strong electric fields, however, requires further research. We identified cases with a clear and relatively fast decay of $E_{\rm SO}$, and cases with a plateau-like behavior where the decay is slow. In the latter cases, the decay occurs at electric fields outside of our estimated parameter range within which the model assumptions (isolated subspace) are well satisfied.

\section{Conclusion}
\label{sec:Conclusion}

We studied low-energy hole states in Si- and Ge-based NWs whose cores have rectangular cross sections. In particular, we analyzed the case where the cross section is a square with side length $s$. It turned out that the DRSOI is the dominant contribution to the SOI and that the shortest achievable spin-orbit length is approximately proportional to $s$. 

For Ge and Ge/Si core/shell NWs, we found that the orientation of the crystallographic axes has relatively small effects on the low-energy hole spectrum. Furthermore, we obtained very good agreement with the results of Ref.~\cite{kloeffel:prb11}, where Ge and Ge/Si core/shell NWs with cylindrical symmetry were considered and an effective model for the low-energy subbands was developed. Thus, our work strongly supports recent calculations \cite{maier:prb13, kloeffel:prb13, maier:prb14a, maier:prb14b, nigg:prl17} that make use of the effective Hamiltonian of Ref.~\cite{kloeffel:prb11}. In addition, the agreement suggests that the exact shape of the Ge core is not important for the low-energy hole states, as long as the confinement in the transverse directions is approximately similar. As a consequence, our results of Sec.~\ref{sec:Ge} and those of Refs.~\cite{kloeffel:prb11, maier:prb13, kloeffel:prb13, maier:prb14a, maier:prb14b, nigg:prl17} may equally be used for NWs with, e.g., circular, square, or hexagonal cross sections.         

The orientation of the crystallographic axes is very important for the hole states in Si NWs. By comparing the results for different orientations, we found significant differences among the effective masses of the lowest-energy subbands and among the strengths of the SOI. Considering a perpendicularly applied electric field along $x$, a particularly strong SOI was obtained for Si NWs with $x \parallel \mbox{[110]}$ and $z \parallel \mbox{[001]}$ (see Fig.~\ref{fig:SetupWireAndQD} for a sketch of the axes and the NW), in agreement with our analytical results of Sec.~\ref{sec:AnalyticalResultsSiNWs}. For these NWs with $x \parallel \mbox{[110]}$ and $z \parallel \mbox{[001]}$, an additional enhancement of the achievable spin-orbit energy was observed when the electric field was applied parallel to the diagonal of the square cross section. Including magnetic fields in our model showed that a helical gap at $k_z = 0$ can be opened.  

We found that the preferable choice of the side length $s$ depends on the setup and the application. If only relatively weak electric fields are feasible, a stronger SOI may be achieved by using a larger $s$, since $\alpha_{\rm DR} \propto s^2$ within the regime of small electric fields and in the absence of strain. (Details are provided in Sec.~\ref{sec:AnalyticalResultsSiNWs}, and we note that this feature is not observed in Fig.~\ref{fig:GeSiCoreShellESOvsEx} for Ge/Si core/shell NWs because of the strain.) If there is practically no limitation on the electric field, the strongest achievable SOI is increased when $s$ is decreased. 

Our work also points out some currently open questions. For instance, as explained in Sec.~\ref{sec:Accuracy}, a detailed analysis of the SOI and its electric-field-dependence in the regime of strong fields would be desirable. Furthermore, predictions for holes in Si NWs with circular cross sections \cite{barraud:edl12} would be useful. Given our results for square cross sections, particularly the strong dependence of the SOI on the orientation of the crystallographic axes, we expect that the SOI in Si NWs with circular cross sections depends on both the growth direction and the orientation of the electric field. This may be analyzed with an approach similar to that of Sec.~\ref{sec:Model}, using cylindrical confinement \cite{sercel:prb90, csontos:prb09}, which is beyond the scope of the present work.    

Although we primarily considered square cross sections, we also studied rectangular cross sections with different aspect ratios $L_x / L_y$. As expected \cite{harada:prb06, watzinger:nlt16}, we found that the HH-LH mixing decreases when $L_x / L_y$ or $L_y / L_x$ is changed from 1 (square cross section) toward 0. When the HH-LH mixing is reduced, the DRSOI becomes less pronounced, as explained in Sec.~\ref{sec:DRSOI}.   

In conclusion, our calculations show that holes in \mbox{Si- and} Ge-based NWs are promising platforms for applications which require a strong and/or electrically tunable SOI. Spin-orbit energies of several millielectronvolts can be achieved, and the SOI can be switched on and off via the electric field. In Si NWs, the orientation of the crystallographic axes strongly affects the properties of the low-energy hole states.

\begin{acknowledgments} 
We thank M.~Brauns, S.~De Franceschi, S.~Hoffman, R.~Maurand, M.~Sanquer, P.~Stano, M.~Vinet, J.~R.~Wootton, and F.~A.~Zwanenburg for helpful discussions and acknowledge support from  the Swiss National Science Foundation, NCCR QSIT, and SiSPIN.
\end{acknowledgments}

\appendix

\section{Orbital contributions of a magnetic field}
\label{app:OrbitalContribMagnField}

As pointed out in Sec.~\ref{secsubsub:ModelHamLK}, the $\hbar k_i$ in the LK Hamiltonian correspond to the kinetic electron momenta, i.e., 
\begin{equation}
\bm{k} = - i \nabla + \frac{e}{\hbar} \bm{A}, 
\label{eq:KineticMomentumForOrbContrAppendix}
\end{equation}  
where $e$ is the elementary positive charge and $\bm{A}$ is the vector potential with $\bm{B} = \nabla \times \bm{A}$ \cite{luttinger:pr56}. Considering a homogeneous magnetic field $\bm{B} = B_x \bm{e}_x + B_y \bm{e}_y + B_z \bm{e}_z$ with arbitrary strength and direction, we choose the vector potential
\begin{equation}
\bm{A} = - \frac{1}{2} B_z y \bm{e}_x + \frac{1}{2} B_z x \bm{e}_y + \left( B_x y - B_y x \right) \bm{e}_z .
\label{eq:VectorPotentialForOurBField}
\end{equation}
That is, we choose a symmetric gauge for the magnetic field $B_z$ along the wire and a Landau gauge for the components $B_x$ and $B_y$ perpendicular to the wire. The vector potential in Eq.~(\ref{eq:VectorPotentialForOurBField}) was also used in our previous works on Ge/Si NWs \cite{kloeffel:prb11, maier:prb13, kloeffel:prb13}, and it may easily be verified that the relation $\bm{B} = \nabla \times \bm{A}$ is satisfied. From Eqs.~(\ref{eq:KineticMomentumForOrbContrAppendix}) and (\ref{eq:VectorPotentialForOurBField}), we obtain
\begin{eqnarray}
k_x &=& -i \partial_x - \frac{e B_z}{2 \hbar} y  , \\
k_y &=& -i \partial_y + \frac{e B_z}{2 \hbar} x , \\
k_z &=& -i \partial_z + \frac{e B_x}{\hbar} y - \frac{e B_y}{\hbar} x .
\end{eqnarray}
Furthermore,
\begin{eqnarray}
k_x^2 &=&  - \partial_x^2 + i \frac{e B_z}{\hbar} y \partial_x + \frac{e^2 B_z^2}{4 \hbar^2} y^2  , \label{eq:kxSquaredAppendix} \\
k_y^2 &=&  - \partial_y^2 - i \frac{e B_z}{\hbar} x \partial_y + \frac{e^2 B_z^2}{4 \hbar^2} x^2  , \\
k_z^2 &=& 
- \partial_z^2
- 2 i \frac{e}{\hbar} \left( B_x y \partial_z - B_y x \partial_z \right) \nonumber \\ 
& & + \frac{e^2}{\hbar^2} \left( B_x^2 y^2 - 2 B_x B_y x y + B_y^2 x^2 \right) . \label{eq:kzSquaredAppendix} 
\end{eqnarray}
In addition, we find
\begin{eqnarray}
k_x k_y &=&  - \partial_x \partial_y - i \frac{e B_z}{2 \hbar} \left(1 + x \partial_x - y \partial_y  \right) - \frac{e^2 B_z^2}{4 \hbar^2} x y  , \\
k_y k_x &=&  - \partial_x \partial_y - i \frac{e B_z}{2 \hbar} \left( -1 + x \partial_x - y \partial_y  \right) - \frac{e^2 B_z^2}{4 \hbar^2} x y  , \mbox{ } 
\end{eqnarray}
and
\begin{eqnarray}
k_x k_z &=&
- \partial_x \partial_z 
- i \frac{e}{\hbar} \left( B_x y \partial_x
- B_y - B_y x \partial_x
- \frac{B_z}{2} y \partial_z \right)  \nonumber \\
& & + \frac{e^2 B_z}{2 \hbar^2} \left( B_y x y
- B_x y^2 \right) , \\
k_z k_x &=& 
- \partial_x \partial_z 
- i \frac{e}{\hbar} \left( B_x y \partial_x
- B_y x \partial_x
- \frac{B_z}{2} y \partial_z \right)  \nonumber \\
& & + \frac{e^2 B_z}{2 \hbar^2} \left( B_y x y
- B_x y^2 \right) , 
\end{eqnarray}
and
\begin{eqnarray}
k_y k_z &=& 
- \partial_y \partial_z 
- i \frac{e}{\hbar} \left(B_x + B_x y \partial_y
- B_y x \partial_y
+ \frac{B_z}{2} x \partial_z \right)  \nonumber \\
& & + \frac{e^2 B_z}{2 \hbar^2} \left( B_x x y
- B_y x^2 \right) , \\
k_z k_y &=& 
- \partial_y \partial_z 
- i \frac{e}{\hbar} \left( B_x y \partial_y
- B_y x \partial_y
+ \frac{B_z}{2} x \partial_z \right)  \nonumber \\
& & + \frac{e^2 B_z}{2 \hbar^2} \left( B_x x y
- B_y x^2 \right) .  \label{eq:kzkyAppendix}
\end{eqnarray}
We note that 
\begin{eqnarray}
k_y k_z - k_z k_y &=& - i \frac{e}{\hbar} B_x , \\
k_z k_x - k_x k_z &=& - i \frac{e}{\hbar} B_y , \\ 
k_x k_y - k_y k_x &=&  - i \frac{e}{\hbar} B_z ,
\end{eqnarray}
and so $\bm{k}\times\bm{k} = - i e \bm{B}/\hbar$ is indeed satisfied \cite{winkler:book}. 

Equations~(\ref{eq:kxSquaredAppendix}) to (\ref{eq:kzkyAppendix}) correspond to the chosen representations of the operators $k_i k_j$ in the LK Hamiltonian. However, when the magnetic field is relatively weak, terms in $k_i k_j$ that are quadratic in the magnetic field may be neglected, analogous to previous theoretical studies on Ge/Si NWs \cite{kloeffel:prb11, maier:prb13, kloeffel:prb13}. We verified numerically that these quadratic terms are indeed negligible for the considered parameter range. Hence, only the terms in Eqs.~(\ref{eq:kxSquaredAppendix}) to (\ref{eq:kzkyAppendix}) that are either independent of or linear in the magnetic field are important for the plots shown in this work.

\section{Transformation of the Luttinger-Kohn Hamiltonian}
\label{app:Transformation}

Since the spherical approximation does not apply to Si, the Hamiltonian of our model depends on the details of the NW fabrication. That is, the relations between the coordinate systems $\Sigma$ for the NW $(x,y,z)$ and $\Sigma^\prime$ for the main crystallographic axes $(x^\prime,y^\prime,z^\prime)$ must be taken into account, see Sec.~\ref{secsub:ModelCoordinateSystems}. The coordinate system $\Sigma^\prime$ is based on the orthonormal vectors $\bm{e}_{x^\prime}$, $\bm{e}_{y^\prime}$, and $\bm{e}_{z^\prime} = \bm{e}_{x^\prime} \times \bm{e}_{y^\prime}$, which correspond to the crystallographic directions [100], [010], and [001], respectively. Analogously, the basis vectors of $\Sigma$ are $\bm{e}_{x}$, $\bm{e}_{y}$, and $\bm{e}_{z} = \bm{e}_{x} \times \bm{e}_{y}$ and point along the axes $x$ (``height''), $y$ (``width''), and $z$ (``length'') of the NW, as illustrated in Fig.~\ref{fig:SetupWireAndQD}. In this appendix, we briefly explain how the $\Sigma^\prime$-based LK Hamiltonian of Eq.~(\ref{eq:LKfullMainaxes}) is rewritten in terms of $\Sigma$. All details are provided in the SI \cite{supplement}, including information on rewriting the BP Hamiltonian of Eq.~(\ref{eq:BirPikusFull}).

\subsection{Nanowire axis along [001]}
\label{appsub:TransformationFor001}

For the momentum operator $\hbar \bm{k}$ and its components, the equality
\begin{equation}
\bm{k} = \bm{e}_{x} k_x + \bm{e}_{y} k_y  + \bm{e}_{z} k_z = \bm{e}_{x^\prime} k_{x^\prime} + \bm{e}_{y^\prime} k_{y^\prime} + \bm{e}_{z^\prime} k_{z^\prime}
\label{eq:vectorkDifferentBasisVectors} 
\end{equation}
applies. Equations~(\ref{eq:RelationsAxesMainTextFor001start}) to (\ref{eq:RelationsAxesMainTextFor001end}) therefore imply that 
\begin{eqnarray}
k_{x^\prime} &=& k_x \cos\phi - k_y \sin\phi , \label{eq:kxprimeFor001} \\
k_{y^\prime} &=& k_x \sin\phi + k_y \cos\phi , \\
k_{z^\prime} &=& k_z .   \label{eq:kzprimeFor001}
\end{eqnarray}
The relations between $J_{x^\prime, y^\prime, z^\prime}$ and $J_{x,y,z}$ for the spin are formally equivalent to those for the momentum and can be derived analogously via 
\begin{equation}
\bm{J} = \bm{e}_{x} J_x + \bm{e}_{y} J_y  + \bm{e}_{z} J_z = \bm{e}_{x^\prime} J_{x^\prime} + \bm{e}_{y^\prime} J_{y^\prime} + \bm{e}_{z^\prime} J_{z^\prime} .
\label{eq:vectorJDifferentBasisVectors} 
\end{equation}
Insertion of the expressions for $k_{x^\prime, y^\prime, z^\prime}$ and $J_{x^\prime, y^\prime, z^\prime}$ into Eq.~(\ref{eq:LKfullMainaxes}), followed by algebraic simplification, yields Eq.~(\ref{eq:LKfull001}). We want to mention that the inverse relations for Eqs.~(\ref{eq:RelationsAxesMainTextFor001start}) to (\ref{eq:RelationsAxesMainTextFor001end}) are 
\begin{eqnarray}
\bm{e}_{x^\prime} &=& \bm{e}_{x} \cos\phi - \bm{e}_{y} \sin\phi , \\
\bm{e}_{y^\prime} &=& \bm{e}_{x} \sin\phi + \bm{e}_{y} \cos\phi , \\
\bm{e}_{z^\prime} &=& \bm{e}_{z} 
\end{eqnarray}  
and resemble Eqs.~(\ref{eq:kxprimeFor001}) to (\ref{eq:kzprimeFor001}).

\subsection{Nanowire axis along [110]}
\label{appsub:TransformationFor110}

We proceed analogously to Sec.~\ref{appsub:TransformationFor001}. Using Eqs.~(\ref{eq:vectorkDifferentBasisVectors}), (\ref{eq:vectorJDifferentBasisVectors}), and (\ref{eq:RelationsAxesMainTextFor110start}) to (\ref{eq:RelationsAxesMainTextFor110end}), one finds
\begin{eqnarray}
k_{x^\prime} &=&  k_x \frac{\sin\xi}{\sqrt{2}} + k_y \frac{\cos\xi}{\sqrt{2}} + k_z \frac{1}{\sqrt{2}} , \\
k_{y^\prime} &=& - k_x \frac{\sin\xi}{\sqrt{2}} - k_y \frac{\cos\xi}{\sqrt{2}} + k_z \frac{1}{\sqrt{2}} , \\
k_{z^\prime} &=& k_x \cos\xi - k_y \sin\xi ,
\end{eqnarray}
and the formally equivalent relations between $J_{x^\prime, y^\prime, z^\prime}$ and $J_{x,y,z}$. Again, we briefly mention that these relations resemble the inverse relations
\begin{eqnarray}
\bm{e}_{x^\prime} &=& \bm{e}_{x} \frac{\sin\xi}{\sqrt{2}} + \bm{e}_{y} \frac{\cos\xi}{\sqrt{2}} + \bm{e}_{z} \frac{1}{\sqrt{2}} , \\
\bm{e}_{y^\prime} &=& - \bm{e}_{x} \frac{\sin\xi}{\sqrt{2}} - \bm{e}_{y} \frac{\cos\xi}{\sqrt{2}} + \bm{e}_{z} \frac{1}{\sqrt{2}} , \\
\bm{e}_{z^\prime} &=& \bm{e}_{x} \cos\xi  - \bm{e}_{y} \sin\xi 
\end{eqnarray} 
for Eqs.~(\ref{eq:RelationsAxesMainTextFor110start}) to (\ref{eq:RelationsAxesMainTextFor110end}). By inserting the expressions for  $k_{x^\prime, y^\prime, z^\prime}$ and $J_{x^\prime, y^\prime, z^\prime}$ into Eq.~(\ref{eq:LKfullMainaxes}), we obtain the $\xi$-dependent LK Hamiltonian displayed in the SI \cite{supplement} after algebraic simplification. The special case with $\xi = 0$ (or $\xi = \pi$ because of the symmetry) is shown in Eq.~(\ref{eq:LKx001y1M10z110}).

\section{Terms caused by electric fields}
\label{app:TermsCausedByEFields}

In this appendix, we provide information about the two electric-field-dependent terms $H_{\rm dir}^h$ and $H_R^h$ in our model Hamiltonian for low-energy hole states in NWs [Sec.~\ref{sec:Model}, Eq.~(\ref{eq:HamiltonianOurModel})]. When an effective electric field $\bm{E}$ is present inside the NW core, the Hamiltonian \cite{kloeffel:prb11, winkler:book}
\begin{equation}
H_{\rm dir}^h = - e \bm{E} \cdot \bm{r} = - e \left( E_x x  + E_y y + E_z z \right)
\end{equation}
describes the direct coupling between the hole and the electric field. Additional corrections can, e.g., be derived via $\bm{k}\cdot\bm{p}$ theory \cite{winkler:book}. For holes in the valence band $\Gamma_8^v$, which is the topmost valence band of Si and Ge, the most prominent correction is the standard Rashba SOI
\begin{equation}
H_R^h = \alpha_h \bm{E} \cdot \left( \bm{k} \times \bm{J} \right) .
\end{equation}
By means of third-order perturbation theory, starting with the extended Kane model, one obtains \cite{winkler:book}
\begin{equation}
\alpha_h \simeq - \frac{e P^2}{3 E_0^2} + \frac{e Q^2}{9} \left[\frac{10}{E_0^{\prime 2}} - \frac{7}{\left( E_0^\prime + \Delta_0^\prime \right)^2 } \right] 
\end{equation}
for the Rashba coefficient, where the energies $E_0$, $E_0^\prime$, and $\Delta_0^\prime$ quantify the splittings between the considered bands and $P$ and $Q$ are the parameters for the momentum matrix elements. Taking the Si values $E_0 = 4.19\mbox{ eV}$, $E_0^\prime = 3.40\mbox{ eV}$, $\Delta_0^\prime = 0$, $P = 8.72\mbox{ eV \AA}$, and $Q = 7.51\mbox{ eV \AA}$ from Ref.~\cite{richard:prb04}, we find $\alpha_h \approx 0.002\mbox{ nm$^2 e$}$ for Si, which is much smaller than the Rashba coefficient $\alpha_h \approx - 0.4\mbox{ nm$^2 e$}$ obtained for Ge \cite{kloeffel:prb11}. 

Although $H_R^h$ is fully taken into account in our numerical calculations, we have verified that $H_R^h$ is negligible for all results plotted here, both in the case of Ge and Si, because the DRSOI in the studied systems clearly dominates.


\onecolumngrid
\clearpage
\begin{center}
\textbf{\large Supplementary information }
\end{center}

\section{Change of basis: Luttinger-Kohn Hamiltonian}
\label{secSI:LKHamiltonian}

The Luttinger-Kohn Hamiltonian for holes is \cite{luttinger:pr56}
\begin{equation}
H_{\rm LK} = 
\frac{\hbar^2}{2m}\left[ 
\left(\gamma_1 + \frac{5 \gamma_2}{2}\right)k^2 
- 2 \gamma_2 \left( k_{x^\prime}^2 J_{x^\prime}^2 + k_{y^\prime}^2 J_{y^\prime}^2 + k_{z^\prime}^2 J_{z^\prime}^2 \right)
- 4 \gamma_3 \left( \{k_{x^\prime}, k_{y^\prime}\}\{J_{x^\prime}, J_{y^\prime}\} \mbox{ + c.p.} \right)
\right] , 
\label{eqSI:LKMainaxes} 
\end{equation}
where ``c.p.'' stands for cyclic permutations and $\{A, B\} = (AB + BA)/2$. The isotropic and anisotropic Zeeman terms, which are proportional to the parameters $\kappa$ and $q$, respectively, are omitted in Eq.~(\ref{eqSI:LKMainaxes}).

\subsection{Nanowire axis along [001]}
\label{secsubSI:LKz001}

When the $z$ axis, which points along the nanowire, coincides with the [001] direction, the relations between the basis vectors $\bm{e}_{x}$, $\bm{e}_{y}$, $\bm{e}_{z}$ (related to the rectangular cross section and the nanowire axis) and $\bm{e}_{x^\prime}$, $\bm{e}_{y^\prime}$, $\bm{e}_{z^\prime}$ (main crystallographic directions [100], [010], [001]) are
\begin{eqnarray}
\bm{e}_x &=& \bm{e}_{x^\prime} \cos\phi + \bm{e}_{y^\prime} \sin\phi , \\
\bm{e}_y &=& - \bm{e}_{x^\prime} \sin\phi + \bm{e}_{y^\prime} \cos\phi , \\
\bm{e}_z &=& \bm{e}_{z^\prime} .
\end{eqnarray}     
We note that the inverse relations are 
\begin{eqnarray}
\bm{e}_{x^\prime} &=& \bm{e}_{x} \cos\phi - \bm{e}_{y} \sin\phi , \\
\bm{e}_{y^\prime} &=& \bm{e}_{x} \sin\phi + \bm{e}_{y} \cos\phi , \\
\bm{e}_{z^\prime} &=& \bm{e}_{z} .
\end{eqnarray}  
For the momentum operators, one obtains
\begin{eqnarray}
\bm{k} &=& \bm{e}_{x} k_x + \bm{e}_{y} k_y  + \bm{e}_{z} k_z \nonumber \\
&=& \left(\bm{e}_{x^\prime} \cos\phi + \bm{e}_{y^\prime} \sin\phi \right) k_x + \left( - \bm{e}_{x^\prime} \sin\phi + \bm{e}_{y^\prime} \cos\phi \right) k_y  + \bm{e}_{z^\prime} k_z \nonumber \\
&=& \bm{e}_{x^\prime} \left( k_x \cos\phi - k_y \sin\phi \right) + \bm{e}_{y^\prime} \left( k_x \sin\phi + k_y \cos\phi \right) + \bm{e}_{z^\prime} k_z \nonumber \\
&=& \bm{e}_{x^\prime} k_{x^\prime} + \bm{e}_{y^\prime} k_{y^\prime} + \bm{e}_{z^\prime} k_{z^\prime} ,
\end{eqnarray}
and so
\begin{eqnarray}
k_{x^\prime} &=& k_x \cos\phi - k_y \sin\phi , 
\label{eqSI:kxprimeForz001} \\
k_{y^\prime} &=& k_x \sin\phi + k_y \cos\phi , \\
k_{z^\prime} &=& k_z .
\label{eqSI:kzprimeForz001}
\end{eqnarray}
Analogously, one can derive the formally equivalent relations
\begin{eqnarray}
J_{x^\prime} &=& J_x \cos\phi - J_y \sin\phi , 
\label{eqSI:JxprimeForz001} \\
J_{y^\prime} &=& J_x \sin\phi + J_y \cos\phi , \\
J_{z^\prime} &=& J_z 
\label{eqSI:JzprimeForz001}
\end{eqnarray}
for the spin operators.

Insertion of Eqs.~(\ref{eqSI:kxprimeForz001}) to (\ref{eqSI:JzprimeForz001}) into Eq.~(\ref{eqSI:LKMainaxes}) yields  
\begin{eqnarray}
H_{\rm LK}^{\rm [001]}(\phi) = \frac{\hbar^2}{2m} &\Biggl[& 
\left(\gamma_1 + \frac{5 \gamma_2}{2}\right)k^2 - 2 \gamma_2 \left( k_x^2 J_x^2 + k_y^2 J_y^2 \right) \left(\cos^4\phi + \sin^4\phi \right) - 2 \gamma_2 k_z^2 J_z^2 \nonumber  \\
& & - \gamma_2 \left[ \left( k_y^2 - k_x^2 \right) (J_x J_y + J_y J_x) + (k_x k_y + k_y k_x) \left(J_y^2 - J_x^2 \right) \right] \sin(2\phi) \cos(2\phi) \nonumber \\
& & - \gamma_2 \left[ k_x^2 J_y^2 + k_y^2 J_x^2 + (k_x k_y + k_y k_x) (J_x J_y + J_y J_x) \right] \sin^2(2\phi) \nonumber \\
& & - \gamma_3 \left[ \left(k_x^2 -k_y^2 \right) \sin(2\phi) + \left( k_x k_y + k_y k_x \right) \cos(2\phi) \right] \left[ \left(J_x^2 - J_y^2 \right) \sin(2\phi) + \left( J_x J_y + J_y J_x \right) \cos(2\phi) \right] \nonumber \\
& & - \gamma_3 \left[ \left( k_y k_z + k_z k_y \right) \left( J_y J_z + J_z J_y \right) + \left( k_z k_x + k_x k_z \right) \left( J_z J_x + J_x J_z \right) \right] \Biggr] ,
\label{eqSI:LKz001}
\end{eqnarray}
which is equivalent to Eq.~(\ref{eq:LKfull001}) of the main text. We note that the trigonometric identities $\cos^2\phi - \sin^2\phi = \cos(2\phi)$, $2 \sin\phi \cos\phi = \sin(2\phi)$, and $\sin^2\phi + \cos^2\phi = 1$ were used when Eq.~(\ref{eqSI:LKz001}) was calculated.

\subsection{Nanowire axis along [110]}
\label{secsubSI:LKz110}

When the $z$ axis corresponds to the [110] direction, the relations between the basis vectors are
\begin{eqnarray}
\bm{e}_x &=& \bm{e}_{x^\prime} \frac{\sin\xi}{\sqrt{2}} - \bm{e}_{y^\prime} \frac{\sin\xi}{\sqrt{2}} + \bm{e}_{z^\prime} \cos\xi , \\
\bm{e}_y &=& \bm{e}_{x^\prime} \frac{\cos\xi}{\sqrt{2}} - \bm{e}_{y^\prime} \frac{\cos\xi}{\sqrt{2}} - \bm{e}_{z^\prime} \sin\xi , \\
\bm{e}_z &=& \bm{e}_{x^\prime} \frac{1}{\sqrt{2}} + \bm{e}_{y^\prime} \frac{1}{\sqrt{2}} ,
\end{eqnarray}     
and the inverse relations read 
\begin{eqnarray}
\bm{e}_{x^\prime} &=& \bm{e}_{x} \frac{\sin\xi}{\sqrt{2}} + \bm{e}_{y} \frac{\cos\xi}{\sqrt{2}} + \bm{e}_{z} \frac{1}{\sqrt{2}} , \\
\bm{e}_{y^\prime} &=& - \bm{e}_{x} \frac{\sin\xi}{\sqrt{2}} - \bm{e}_{y} \frac{\cos\xi}{\sqrt{2}} + \bm{e}_{z} \frac{1}{\sqrt{2}} , \\
\bm{e}_{z^\prime} &=& \bm{e}_{x} \cos\xi  - \bm{e}_{y} \sin\xi .
\end{eqnarray}  
For the momentum operators, one finds
\begin{eqnarray}
\bm{k} &=& \bm{e}_{x} k_x + \bm{e}_{y} k_y  + \bm{e}_{z} k_z \nonumber \\
&=& \left( \bm{e}_{x^\prime} \frac{\sin\xi}{\sqrt{2}} - \bm{e}_{y^\prime} \frac{\sin\xi}{\sqrt{2}} + \bm{e}_{z^\prime} \cos\xi \right) k_x + \left( \bm{e}_{x^\prime} \frac{\cos\xi}{\sqrt{2}} - \bm{e}_{y^\prime} \frac{\cos\xi}{\sqrt{2}} - \bm{e}_{z^\prime} \sin\xi \right) k_y  + \left( \bm{e}_{x^\prime} \frac{1}{\sqrt{2}} + \bm{e}_{y^\prime} \frac{1}{\sqrt{2}} \right) k_z \nonumber \\
&=& \bm{e}_{x^\prime} \left( k_x \frac{\sin\xi}{\sqrt{2}} + k_y \frac{\cos\xi}{\sqrt{2}} + k_z \frac{1}{\sqrt{2}} \right) + \bm{e}_{y^\prime} \left( - k_x \frac{\sin\xi}{\sqrt{2}} - k_y \frac{\cos\xi}{\sqrt{2}} + k_z \frac{1}{\sqrt{2}} \right) + \bm{e}_{z^\prime} \left( k_x \cos\xi - k_y \sin\xi \right) \nonumber \\
&=& \bm{e}_{x^\prime} k_{x^\prime} + \bm{e}_{y^\prime} k_{y^\prime} + \bm{e}_{z^\prime} k_{z^\prime} ,
\end{eqnarray}
and so
\begin{eqnarray}
k_{x^\prime} &=&  k_x \frac{\sin\xi}{\sqrt{2}} + k_y \frac{\cos\xi}{\sqrt{2}} + k_z \frac{1}{\sqrt{2}} , \label{eqSI:kxprimeForz110} \\
k_{y^\prime} &=& - k_x \frac{\sin\xi}{\sqrt{2}} - k_y \frac{\cos\xi}{\sqrt{2}} + k_z \frac{1}{\sqrt{2}} , \\
k_{z^\prime} &=& k_x \cos\xi - k_y \sin\xi .  \label{eqSI:kzprimeForz110} 
\end{eqnarray}
Analogously, the formally equivalent relations
\begin{eqnarray}
J_{x^\prime} &=&  J_x \frac{\sin\xi}{\sqrt{2}} + J_y \frac{\cos\xi}{\sqrt{2}} + J_z \frac{1}{\sqrt{2}} , \label{eqSI:JxprimeForz110} \\
J_{y^\prime} &=& - J_x \frac{\sin\xi}{\sqrt{2}} - J_y \frac{\cos\xi}{\sqrt{2}} + J_z \frac{1}{\sqrt{2}} , \\
J_{z^\prime} &=& J_x \cos\xi - J_y \sin\xi  \label{eqSI:JzprimeForz110} 
\end{eqnarray}
are obtained for the spin operators.

Insertion of Eqs.~(\ref{eqSI:kxprimeForz110}) to (\ref{eqSI:JzprimeForz110}) into Eq.~(\ref{eqSI:LKMainaxes}) yields
\begin{eqnarray}
H_{\rm LK}^{\rm [110]}(\xi)  
= \frac{\hbar^2}{2m} &\Biggl[& 
\left(\gamma_1 + \frac{5 \gamma_2}{2}\right)k^2 
- \gamma_2 k_x^2 J_x^2 \left( \sin^4\xi + 2 \cos^4\xi \right) 
- \gamma_2  k_y^2 J_y^2 \left( \cos^4\xi + 2 \sin^4\xi \right) 
- \gamma_2  k_z^2 J_z^2    \nonumber \\ & & 
- 3 \gamma_2 \left[ k_x^2 J_y^2 + k_y^2 J_x^2 + \left( k_x k_y + k_y k_x \right) \left( J_x J_y + J_y J_x \right) \right] \frac{\sin^2(2\xi) }{4}   \nonumber \\ & & 
- \gamma_2  \left( k_x^2 J_z^2 + k_z^2 J_x^2 \right) \sin^2\xi 
- \gamma_2  \left( k_y^2 J_z^2 + k_z^2 J_y^2 \right) \cos^2\xi    \nonumber \\ & & 
- \gamma_2  \left[ k_x^2 \left( J_x J_y + J_y J_x \right) + \left( k_x k_y + k_y k_x \right) J_x^2 \right] \frac{\sin(2\xi) }{2} \left( \sin^2\xi - 2 \cos^2\xi \right)   \nonumber \\ & & 
- \gamma_2  \left[ k_y^2 \left( J_x J_y + J_y J_x \right) + \left( k_x k_y + k_y k_x \right) J_y^2 \right] \frac{\sin(2\xi) }{2} \left( \cos^2\xi - 2 \sin^2\xi \right)  \nonumber \\ & &
- \gamma_2  \left[ k_z^2 \left( J_x J_y + J_y J_x \right) + \left( k_x k_y + k_y k_x \right) J_z^2 \right] \frac{\sin(2\xi) }{2}   \nonumber \\ & &  
- \gamma_2  \left[ \left( k_x k_z + k_z k_x \right) \left( J_y J_z + J_z J_y \right) + \left( k_y k_z + k_z k_y \right) \left( J_x J_z + J_z J_x \right) \right] \frac{\sin(2\xi) }{2}    \nonumber \\ & &
- \gamma_2  \left( k_x k_z + k_z k_x \right) \left( J_x J_z + J_z J_x \right) \sin^2\xi 
- \gamma_2  \left( k_y k_z + k_z k_y \right) \left( J_y J_z + J_z J_y \right) \cos^2\xi  \nonumber \\ & &
- \gamma_3 \left[ k_x^2 \sin^2\xi + k_y^2 \cos^2\xi - k_z^2
+ \left( k_x k_y + k_y k_x \right) \frac{\sin(2\xi )}{2} \right]   \nonumber \\ & &
 \mbox{ } \mbox{ } \mbox{ } \times \left[ J_x^2 \sin^2\xi + J_y^2 \cos^2\xi - J_z^2
+ \left( J_x J_y + J_y J_x \right) \frac{\sin(2\xi )}{2} \right]  \nonumber \\ & & 
- \gamma_3 \left[ \left( k_x^2 - k_y^2 \right) \sin(2\xi ) + \left( k_x k_y + k_y k_x \right) \cos(2\xi ) \right] 
\left[ \left( J_x^2 - J_y^2 \right) \sin(2\xi ) + \left( J_x J_y + J_y J_x \right) \cos(2\xi ) \right]  \nonumber \\ & & 
- \gamma_3 \left[ \left( k_x k_z + k_z k_x \right) \cos\xi - \left( k_y k_z + k_z k_y \right)  \sin\xi  \right]
\left[ \left( J_x J_z + J_z J_x \right) \cos\xi - \left( J_y J_z + J_z J_y \right)  \sin\xi  \right]
\Biggr]  .
\label{eqSI:LKz110}
\end{eqnarray}
As in Appendix~\ref{secsubSI:LKz001}, we used the trigonometric identities $\cos^2\xi - \sin^2\xi = \cos(2\xi)$, $2 \sin\xi \cos\xi = \sin(2\xi)$, and $\sin^2\xi + \cos^2\xi = 1$. In the case of $\xi = 0$ (or $\xi = \pi$ because of the symmetry), Eq.~(\ref{eqSI:LKz110}) simplifies to Eq.~(\ref{eq:LKx001y1M10z110}) of the main text.

\section{Change of basis: Bir-Pikus Hamiltonian}
\label{secSI:BPHamiltonian}

The Bir-Pikus Hamiltonian for holes reads \cite{birpikus:book}
\begin{equation}
H_{\rm BP} = b \left( \varepsilon_{x'x'} J_{x'}^2 + \varepsilon_{y'y'} J_{y'}^2 + \varepsilon_{z'z'} J_{z'}^2 \right)
+ \frac{2 d}{\sqrt{3}} \left( \varepsilon_{x'y'} \left\{ J_{x'} , J_{y'} \right\} + \varepsilon_{y'z'} \left\{ J_{y'} , J_{z'} \right\} + \varepsilon_{z'x'} \left\{ J_{z'} , J_{x'} \right\}  \right) .
\label{eqSI:BPHamiltonian}
\end{equation}
We omitted here all spin-independent terms, because they only lead to a global energy shift in our model and therefore do not affect the results. We recall that $\varepsilon_{ij} = \varepsilon_{ji}$ are the strain tensor elements, $x^\prime$, $y^\prime$, and $z^\prime$ refer to the main crystallographic axes, and $b$ and $d$ are deformation potentials.

\subsection{Nanowire axis along [001]}
\label{secsubSI:BPz001}

We use the relations listed in Appendix~\ref{secsubSI:LKz001}. Following Appendix~B~2 of Ref.~\cite{kloeffel:prb14}, we obtain 
\begin{equation}
R = \begin{pmatrix} 
\cos\phi   & \sin\phi & 0 \\
- \sin\phi & \cos\phi & 0 \\
0 & 0 & 1 
\end{pmatrix}
\label{eqSI:matrixRforz001}
\end{equation}
and
\begin{equation}
\begin{pmatrix}
\varepsilon_{x'x'} & \varepsilon_{x'y'} & \varepsilon_{x'z'} \\
\varepsilon_{y'x'} & \varepsilon_{y'y'} & \varepsilon_{y'z'} \\
\varepsilon_{z'x'} & \varepsilon_{z'y'} & \varepsilon_{z'z'} 
\end{pmatrix} = R^{\rm T} \begin{pmatrix}
\varepsilon_{xx} & \varepsilon_{xy} & \varepsilon_{xz} \\
\varepsilon_{yx} & \varepsilon_{yy} & \varepsilon_{yz} \\
\varepsilon_{zx} & \varepsilon_{zy} & \varepsilon_{zz} 
\end{pmatrix} R , 
\label{eqSI:mainEquForStrTensElemsGivenz001}
\end{equation}
where $R^{\rm T}$ is the transpose of the matrix $R$ in Eq.~(\ref{eqSI:matrixRforz001}). By exploiting trigonometric identities and $\varepsilon_{ij} = \varepsilon_{ji}$, we find
\begin{gather}
\varepsilon_{x'x'} = \varepsilon_{xx} \cos^2\phi + \varepsilon_{yy} \sin^2\phi - \varepsilon_{xy} \sin(2 \phi) , \label{eqSI:epsxprxprForz001} \\
\varepsilon_{y'y'} = \varepsilon_{xx} \sin^2\phi + \varepsilon_{yy} \cos^2\phi + \varepsilon_{xy} \sin(2 \phi) , \\
\varepsilon_{z'z'} = \varepsilon_{zz} , \\
\varepsilon_{x'y'} = \varepsilon_{xy} \cos(2\phi) + (\varepsilon_{xx} - \varepsilon_{yy}) \sin\phi \cos\phi , \\
\varepsilon_{x'z'} = \varepsilon_{xz} \cos\phi - \varepsilon_{y z} \sin\phi , \\
\varepsilon_{y'z'} = \varepsilon_{yz} \cos\phi + \varepsilon_{xz} \sin\phi  
\label{eqSI:epsyprzprForz001}
\end{gather} 
from Eq.~(\ref{eqSI:mainEquForStrTensElemsGivenz001}). Finally, we obtain the Bir-Pikus Hamiltonian $H_{\rm BP}^{\rm [001]}(\phi)$ in the unprimed basis by inserting Eqs.~(\ref{eqSI:JxprimeForz001}) to~(\ref{eqSI:JzprimeForz001}) and Eqs.~(\ref{eqSI:epsxprxprForz001}) to~(\ref{eqSI:epsyprzprForz001}) into Eq.~(\ref{eqSI:BPHamiltonian}).

\subsection{Nanowire axis along [110]}
\label{secsubSI:BPz110}

Now we use the relations listed in Appendix~\ref{secsubSI:LKz110} and proceed analogously to Appendix~\ref{secsubSI:BPz001}. Following Appendix~B~2 of Ref.~\cite{kloeffel:prb14}, we find 
\begin{equation}
R = \begin{pmatrix} 
\frac{\sin\xi}{\sqrt{2}} & - \frac{\sin\xi}{\sqrt{2}} & \cos\xi \\
\frac{\cos\xi}{\sqrt{2}} & - \frac{\cos\xi}{\sqrt{2}} & - \sin\xi \\
\frac{1}{\sqrt{2}} & \frac{1}{\sqrt{2}} & 0 
\end{pmatrix} ,
\end{equation}
which results in
\begin{gather}
\varepsilon_{x'x'} = \frac{1}{2} \left[ \varepsilon_{zz} + \varepsilon_{xx} \sin^2\xi + \varepsilon_{yy} \cos^2\xi + \varepsilon_{xy} \sin(2 \xi)   \right] + \varepsilon_{xz} \sin\xi + \varepsilon_{yz} \cos\xi , 
\label{eqSI:epsxprxprForz110} \\
\varepsilon_{y'y'} = \frac{1}{2} \left[ \varepsilon_{zz} + \varepsilon_{xx} \sin^2\xi + \varepsilon_{yy} \cos^2\xi + \varepsilon_{xy} \sin(2 \xi)   \right] - \varepsilon_{xz} \sin\xi - \varepsilon_{yz} \cos\xi , \\
\varepsilon_{z'z'} = \varepsilon_{xx} \cos^2\xi + \varepsilon_{yy} \sin^2\xi - \varepsilon_{xy} \sin(2 \xi) , \\
\varepsilon_{x'y'} = \frac{1}{2} \left[ \varepsilon_{zz} - \varepsilon_{xx} \sin^2\xi - \varepsilon_{yy} \cos^2\xi - \varepsilon_{xy} \sin(2 \xi) \right] , \\
\varepsilon_{x'z'} = \frac{1}{\sqrt{2}} \left[ \varepsilon_{xz} \cos\xi - \varepsilon_{yz} \sin\xi + \varepsilon_{xy} \cos(2 \xi) + (\varepsilon_{xx} - \varepsilon_{yy}) \sin\xi \cos\xi \right] , \\
\varepsilon_{y'z'} = \frac{1}{\sqrt{2}} \left[ \varepsilon_{xz} \cos\xi - \varepsilon_{yz} \sin\xi - \varepsilon_{xy} \cos(2 \xi) + (\varepsilon_{yy} - \varepsilon_{xx}) \sin\xi \cos\xi \right]  . 
\label{eqSI:epsyprzprForz110}
\end{gather} 
By inserting Eqs.~(\ref{eqSI:JxprimeForz110}) to~(\ref{eqSI:JzprimeForz110}) and Eqs.~(\ref{eqSI:epsxprxprForz110}) to~(\ref{eqSI:epsyprzprForz110}) into Eq.~(\ref{eqSI:BPHamiltonian}), one obtains the Bir-Pikus Hamiltonian $H_{\rm BP}^{\rm [110]}(\xi)$ in the unprimed basis.

\section{Strain-dependent hole spectrum in a cylindrical Ge/Si core/shell nanowire}
\label{secSI:SpectrumGeSiCoreShellNW}

\begin{figure}[tb]
\vspace{0.0cm}                                 
\centering                                     
\includegraphics[width=0.85\linewidth]{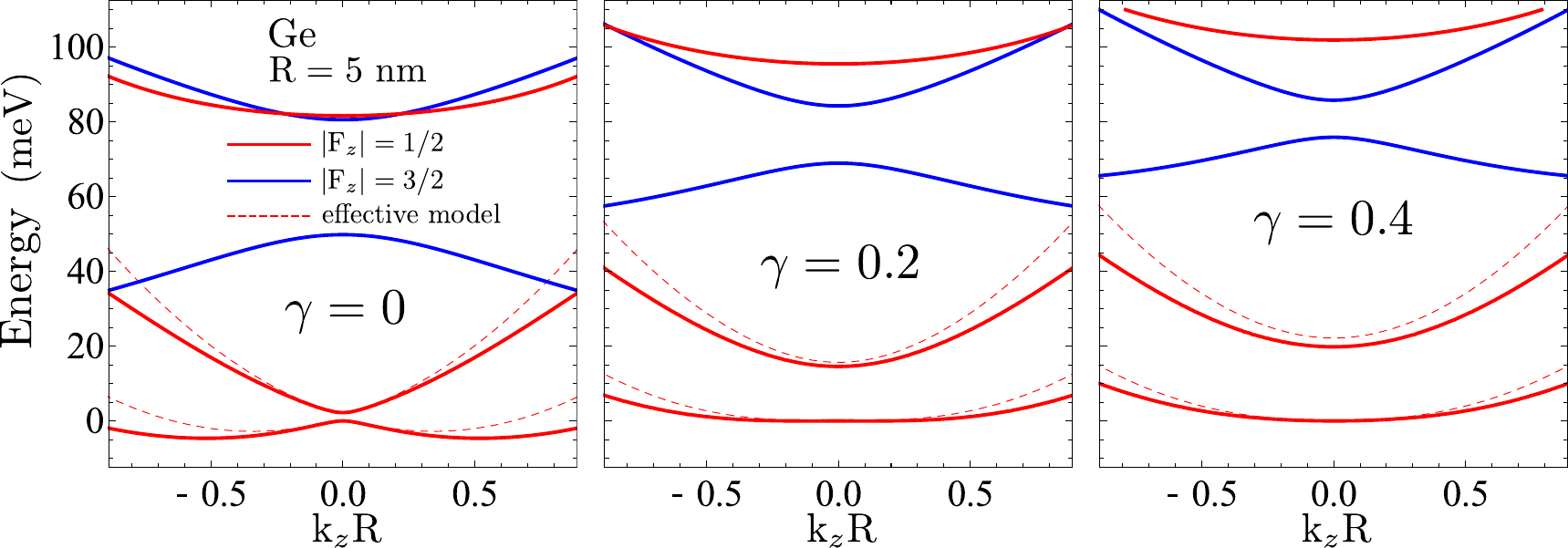}   
\vspace{0.0cm} 
\caption[]{Low-energy hole spectra in Ge/Si core/shell nanowires with a fixed core radius $R = 5\mbox{ nm}$ and different shell thicknesses. The relative shell thickness is defined as $\gamma = (R_s - R)/R$, where $R_s$ is the shell radius, i.e., the outer radius of the Ge/Si core/shell nanowire. The solid lines show the exact eigenenergies of the model Hamiltonian (see Ref.~\cite{kloeffel:prb11} for details), which comprises the Luttinger-Kohn Hamiltonian \cite{luttinger:pr56} in the spherical approximation, the Bir-Pikus Hamiltonian \cite{birpikus:book} in the spherical approximation, and cylindrical hard-wall confinement. In each of the three panels, an offset was chosen such that the ground state energy at $k_z = 0$ is zero. In order to calculate the eigenenergies, we extended the approach described in Ref.~\cite{sercel:prb90} to the case where shell-induced static strain is present. We note that each plotted line represents two degenerate subbands. The color red (blue) indicates that the total angular momentum along the $z$ axis is $F_z = \pm 1/2$ ($F_z = \pm 3/2$). All hole states with $F_z \in \{ \pm 5/2, \pm 7/2, \cdots \}$ are relatively high in energy and lie outside of the displayed range. Even at large $\gamma$, the two energetically lowest doublets with $F_z = \pm 1/2$ are well separated from other subbands, which justifies the perturbative approach (projection onto subspace) used in Ref.~\cite{kloeffel:prb11} to obtain an effective model for the four low-energy subbands at $|k_z| R < 1$. The dashed lines show the spectra that were calculated with this effective model, i.e., with the 4$\times$4 Hamiltonian developed in Ref.~\cite{kloeffel:prb11}. Importantly, both the exact calculation and the effective model show that the compressive strain in the Ge core, which is caused by the Si shell, lifts the additional quasi-degeneracy at $k_z = 0$ between the two degenerate ground states and the two degenerate states with second-lowest energy. As a consequence, when $\gamma$ is continuously increased, the effective mass of the lowest-energy subbands changes from a negative value (no strain, $\gamma = 0$) towards minus infinity, where a change of sign occurs, and then decreases to a positive value. For details and explanations, see Refs.~\cite{kloeffel:prb11, kloeffel:prb14} and the main text.}  
\label{figSI:GeSiCoreShellNWSpectrum}                                       
\end{figure}

Figure~\ref{figSI:GeSiCoreShellNWSpectrum} shows the low-energy hole spectra of three different Ge/Si core/shell NWs with cylindrical symmetry. The spectra were calculated as described in Ref.~\cite{kloeffel:prb11}, the radius of the Ge core is always $R = 5\mbox{ nm}$. In the left panel, the relative shell thickness is $\gamma = 0$, i.e., there is no Si shell and the spectrum corresponds to that of a bare Ge NW. In the middle and right panel, the relative shell thickness is $\gamma = 0.2$ and $\gamma = 0.4$, i.e., the Si shell is $1\mbox{ nm}$ and $2\mbox{ nm}$ thick, respectively. In each panel, the solid lines show the exact eigenenergies of the Hamiltonian of Ref.~\cite{kloeffel:prb11}, which comprises the Luttinger-Kohn Hamiltonian $H_{\rm LK}$, the Bir-Pikus Hamiltonian $H_{\rm BP}$, and cylindrical hard-wall confinement. We note that every plotted line represents two degenerate subbands. Furthermore, because of the cylindrical symmetry, all eigenstates and therefore the subbands can be classified regarding their total angular momentum $F_z$ \cite{sercel:prb90, csontos:prb09} along the $z$~axis, which is the nanowire axis.

Besides the abovementioned exact eigenenergies, Fig.~\ref{figSI:GeSiCoreShellNWSpectrum} also shows the results from the effective model (4$\times$4 Hamiltonian) that was developed in Ref.~\cite{kloeffel:prb11} in order to describe the states of lowest energy. When a Si shell is present (middle and right panel), i.e., when the Ge core is strained, the effective model does not fully coincide with the exact calculation at $k_z = 0$, in contrast to the case without core strain (left panel), where the effective model and the exact calculation exactly coincide at $k_z = 0$. Given the details of the derivation \cite{kloeffel:prb11} of the effective model, this observation is not surprising, since $[H_{\rm BP}, H_{\rm LK}] = 0$ is only satisfied when there is no strain, i.e., no Si shell ($\gamma = 0$). Nevertheless, as evident from Fig.~\ref{figSI:GeSiCoreShellNWSpectrum}, we find that the energy splitting at $k_z = 0$ is always reproduced with good accuracy, even for highly strained Ge/Si core/shell NWs (large~$\gamma$). Moreover, even at large~$\gamma$, the effective model closely resembles the exact spectrum when $|k_z| R < 1$. Furthermore, it is important to note that the energy gap at $k_z = 0$ that separates the four-dimensional low-energy subspace from the energetically lowest states with $F_z = \pm 3/2$ increases when the relative shell thickness $\gamma$ is increased. Hence, the low-energy subspace described by the effective model remains energetically well isolated, even in the presence of a thick Si shell.  

The three panels of Fig.~\ref{figSI:GeSiCoreShellNWSpectrum} illustrate a remarkable effect of the Si shell on the low-energy hole spectrum. When the relative shell thickness $\gamma$ and therefore the strain in the Ge core is continuously increased, the effective mass for the lowest-energy subbands changes from a negative value towards minus infinity, where its sign changes, and then decreases from plus infinity to a positive value, leading to electron-like parabolas (in good approximation) with a positive effective mass. For explanations, we refer to the main text.


\begin{thebibliography}{}
\bibitem{footnote:GeLKparams}{The parameters $\gamma_1 = 13.35$, $\gamma_2 = 4.25$, and $\gamma_3 = 5.69$ for Ge are taken from Ref.~\cite{lawaetz:prb71}. Considering the spherical approximation \cite{lipari:prl70, winkler:sst08}, we use $\gamma_s = \left( 2 \gamma_2 + 3 \gamma_3 \right)/5 = 5.114$.}
\bibitem{footnote:gfactor}{We use a standard definition for the effective $g$~factor. When $\Delta_Z$ is the Zeeman splitting caused by the magnetic field $\bm{B}$, we calculate the associated $g$~factor via $g = \Delta_Z/(\mu_B |\bm{B}|)$. We note that $g \geq 0$ since we consider the Zeeman splitting $\Delta_Z$ as positive. }
\bibitem{footnote:estimateStrongEx}{The two inequalities $|e U E_x / \Delta | > 1$ and $|E_x| < (10\mbox{ nm}/R)^3 \times 5\mbox{ V/$\mu$m}$ can be satisfied when \mbox{$\sqrt{\Delta_{\rm BP}/(694\mbox{ meV})} \times R/\mbox{nm} < 1$}, where the strain-based splitting $\Delta_{\rm BP} = \Delta_{\rm BP}(\gamma)$ depends on the relative shell thickness. Using the upper limit $\Delta_{\rm BP} \approx 30\mbox{ meV}$ \cite{kloeffel:prb11, kloeffel:prb14}, which is caused by a very thick Si shell ($\gamma \to \infty$), we obtain our estimate $R \lesssim 5\mbox{ nm}$ for the core radius. }
\end{thebibliography}
\end{document}